\documentclass[12pt,openright,oneside,a4paper,english]{book}
\usepackage{subfig}
\usepackage{subcaption}
\usepackage{braket}
\newcommand{\cmmnt}[1]{}
\usepackage{booktabs}
\usepackage{slashed}
\usepackage{dsfont}
\usepackage{amsmath, amsthm, amssymb, amsfonts, amstext}
\usepackage{graphicx}
\usepackage[nottoc]{tocbibind} 
\makeatletter
\def\@bracketheight{2pt}

\def\overbracket{\@ifnextchar[{\@overbracket}{\@overbracket[\@bracketheight]}}
\def\@overbracket[#1]{\@ifnextchar[{\@over@bracket[#1]}{\@over@bracket[#1][0.3em]}}
\def\@over@bracket[#1][#2]#3{
	\mathop {\vbox {\m@th \ialign {##\crcr \noalign {\kern 3\p@
					\nointerlineskip }\downbracketfill {#1}{#2}
				\crcr \noalign {\kern 3\p@ }
				\crcr $\hfil \displaystyle {#3}\hfil $%
				\crcr} }}\limits}
\def\downbracketfill#1#2{$\m@th \setbox \z@ \hbox {$\braceld$}
	\edef\@bracketheight{\the\ht\z@}\downbracketend{#1}{#2}
	\leaders \vrule \@height #1 \@depth \z@ \hfill
	\leaders \vrule \@height #1 \@depth \z@ \hfill
	\downbracketend{#1}{#2}$}
\def\downbracketend#1#2{\vrule depth #2 width #1\relax}
\makeatother

\usepackage{graphicx}
\usepackage{animate}
\usepackage[english]{babel}
\usepackage[left=30mm, right=20mm, top=30mm, bottom=20mm]{geometry}
\usepackage{multirow}
\usepackage[utf8]{inputenc}
\usepackage{pifont}
\graphicspath{ {./Figuras} }
\usepackage{mathrsfs}
\usepackage[dvipsnames]{xcolor}
\usepackage[numbers]{natbib}
\usepackage{notoccite}
\usepackage{authblk}
\usepackage{indentfirst}
\usepackage{tensor}
\usepackage[T1]{fontenc}
\usepackage{fancyhdr}
\usepackage{url}
\usepackage{setspace} 
\usepackage{pdfpages}
\usepackage{microtype}
\usepackage{footnotebackref}
\usepackage{hyperref}
\usepackage{afterpage}

\hypersetup{
	colorlinks=true,
	linkcolor=blue,
	filecolor=blue,      
	urlcolor=black,
	citecolor=blue,
	pdftitle={Lucas – Gravitational Waves from the Big Bang},
	pdfauthor={Lucas Martins Barreto Alves},
	pdfsubject={Physics, Cosmology},
	pdfkeywords={Gravitation, Cosmology, Gravitational Waves, Cosmic Inflation, Lucas Martins Barreto Alves},
}
\usepackage{accents}
\theoremstyle{definition}

\providecommand{\headheight}{\baselineskip}

\fancypagestyle{plain}{
	\fancyhf{} 
	\fancyfoot{}

	\fancyhead[R]{\bfseries\thepage}
}

\numberwithin{equation}{section}

\begin{document}
\onehalfspacing	
\pagenumbering{arabic} 
\setcounter{page}{0}
\begin{titlepage}
\begin{center}

\textbf{UNIVERSIDADE FEDERAL DE MINAS GERAIS\\}
\textbf{Instituto de Ciências Exatas\\}
\textbf{Programa de Pós-Graduação em Física}

\vspace{5cm}

\href{https://scholar.google.com/citations?user=3ss5qxYAAAAJ&hl=en}{Lucas Martins Barreto Alves}

\vspace{3cm}
\textbf{GRAVITATIONAL WAVES FROM THE BIG BANG}
\vfill
{Belo Horizonte \\ 2025}
\end{center}
\end{titlepage}

\newpage

\thispagestyle{empty}

	\begin{center}

		\href{https://scholar.google.com/citations?user=3ss5qxYAAAAJ&hl=en}{Lucas Martins Barreto Alves}

		\vfill
		\textbf{GRAVITATIONAL WAVES FROM THE BIG BANG}
		\vfill

		\begin{flushright}
			\begin{minipage}{8cm}
				Dissertation submitted to the Graduate Program in Physics of Universidade Federal de Minas Gerais in partial fulfillment of the requirements for obtaining the degree of Master of Science in Physics.
				
				\vspace{0.8cm}
				
				Supervisor: Prof. Dr. Gláuber Carvalho Dorsch

				\vspace{0.8cm}
	
			\end{minipage}
		\end{flushright}
	\end{center}

\vspace{1.5cm}
		
\begin{center}
		{ Belo Horizonte\\ 2025}
\end{center}

\newpage



**Processo:** 23072.251031/2025-45  **Documento:** 4479797

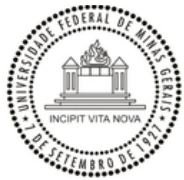

UNIVERSIDADE FEDERAL DE MINAS GERAIS

## ATA

**ATA DA SESSÃO DE ARGUIÇÃO DA 776ª DISSERTAÇÃO DO PROGRAMA DE PÓS-GRADUAÇÃO EM FÍSICA, DEFENDIDA POR LUCAS MARTINS BARRETO ALVES** orientado pelo professor Gláuber Carvalho Dorsch para obtenção do grau de **MESTRE EM FÍSICA, área de concentração Física.** Às dezesseis horas do dia vinte e um de agosto de dois mil e vinte e cinco reuniu-se, presencialmente, a Comissão Examinadora composta pelos professores **Gláuber Carvalho Dorsch** (Orientador - Departamento de Física/UFMG), **Lucas Lages Wardil** (Departamento de Física/UFMG) e Nelson de Oliveira Yokomizo (Departamento de Física/UFMG), para dar cumprimento ao Artigo 37 do Regimento Geral da UFMG, submetendo **LUCAS MARTINS BARRETO ALVES** à arguição de seu trabalho de dissertação, que recebeu o título de **"Gravitational Waves from the Big Bang".** O candidato fez uma exposição oral de seu trabalho durante aproximadamente 50 minutos. Após esta, os membros da comissão prosseguiram com a sua arguição, e apresentaram seus pareceres individuais sobre o trabalho, concluindo pela aprovação do candidato.

Belo Horizonte, 21 de agosto de 2025.

Prof. Gláuber Carvalho Dorsch
Orientador do estudante
Departamento de Física/UFMG

Prof. Lucas Lages Wardil
Departamento de Física/UFMG

Prof. Nelson de Oliveira Yokomizo
Departamento de Física/UFMG

**Candidato:** Lucas Martins Barreto Alves

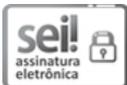 Documento assinado eletronicamente por **Lucas Lages Wardil**, **Professor do Magistério Superior**, em 22/08/2025, às 10:14, conforme horário oficial de Brasília, com fundamento no art. 5º do Decreto nº 10.543, de 13 de novembro de 2020.

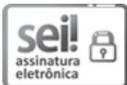 Documento assinado eletronicamente por **Nelson de Oliveira Yokomizo**, **Professor do Magistério Superior**, em 22/08/2025, às 10:43, conforme horário oficial de Brasília, com fundamento no art. 5º do Decreto nº 10.543, de 13 de novembro de 2020.

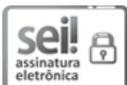 Documento assinado eletronicamente por **Gláuber Carvalho Dorsch**, **Professor do Magistério Superior**, em 22/08/2025, às 10:47, conforme horário oficial de Brasília, com fundamento no art. 5º do Decreto nº 10.543, de 13 de novembro de 2020.

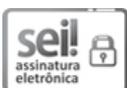 Documento assinado eletronicamente por **Lucas Martins Barreto Alves**, **Usuário Externo**, em 22/08/2025, às 19:05, conforme horário oficial de Brasília, com fundamento no art. 5º do Decreto nº 10.543, de 13 de novembro de 2020.

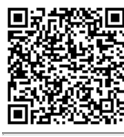 A autenticidade deste documento pode ser conferida no site https://sei.ufmg.br/sei/controlador_externo.php?acao=documento_conferir&id_orgao_acesso_externo=0, informando o código verificador **4479797** e o código CRC **D8394F3D**.

**Referência:** Processo nº 23072.251031/2025-45   SEI nº 4479797




\thispagestyle{empty}
\begin{center}

		\vspace*{\fill}
		\begin{flushright}
	\begin{minipage}{8cm}

		\begin{flushright}

			\textit{To my eternally dear friend Vinicius, one of the earliest, most enthusiastic, and most important supporters of my journey in physics.}

		\end{flushright}
	\end{minipage}
		\end{flushright}

\end{center}

\chapter*{\centering ACKNOWLEDGMENTS}
\thispagestyle{empty}
\noindent

I thank my parents for providing most of the support of all kinds I have ever received. I thank my father for shaping me into a philosopher like himself, deeply kindling my motivation. I thank my mother for using chalk on our old home's doors to teach me how to read and write early on, and for everything that came about as a consequence. I am grateful because, together with my parents, my grandparents, uncles, aunts, and cousins have raised me under remarkable doses of love, filling my soul with positivity. I am especially grateful for having a great friend in my grandfather Suelio, the other STEM mind in the family—progressing in my studies always playfully has the taste of trying to outsmart you!—, who cares deeply to learn about every detail of every step I take in my life. I also thank my dear grandaunt Nena for making me feel absolutely at home in her apartment, including during crucial months of my Master's—I deeply love being there. And I could not single out these two without thanking my grandmother Ana, a source of some of the purest and most intense love I have received.

I thank all my friends deeply for guiding me, encouraging me, and making me so happy just by being around. I love it that prioritizing spending time with you turns out to be a good decision every time. Thank you, Rafa, for bringing about such a deep and positive impact in my life as to allow me to understand so early on that you are the one sibling I have. Thank you, Turi, for being my oldest friend who resonates scientifically with me—I am delighted by the diversity of contexts in which we thrive in each other's company. Thank you, Walter, for our unsurpassable connection and for the effort to be around even in this challenging time. Thank you, Dani, for wisely suggesting early on that a Master's at UFMG could be a good idea for me—you could not have been more right. Thank you, Clarinha, Júlia, Iago, Hércules, Rodrigo, Endrik, Pituca, Felipe, André, Gustavo, Victor L., Maressa, João, Gabi D., Gabi Y., Aninha, Pedro, Cerol, Rapha, Carlos, Daniel, Victor C., JP, Gabe, Avishi, Shloka, Juliana, and Elias. It pains me to bring up a limited list of names—I also thank all my other friends whom I have shamefully not cited.

I thank Carol for having her sweet presence mingled with many of the memories I have of studying modern physics all over Brazil and the United States. I also thank her for so many years of genuine support and patience regarding my academic endeavors.

I thank Andrew for being my first research mentor, for such a treasurable friendship, and for our synergy doing science. I thank Prof. Szabi Márka and Dr. Zsuzsa Márka for welcoming me with open arms so early into their research group and for teaching me so many invaluable lessons about astrophysics, science, and life. I also thank them for all the freedom they have given me in several respects and for their truly unwavering support. I have cherished and grown from being a part of the Columbia Experimental Gravity Group; I thank all my colleagues for contributing to my great experience and hope our scientific collaboration and friendship are far from over.

I thank Prof. Gláuber Dorsch for his welcoming stance toward me since the day we met. I appreciate the similarities in our life stories and in the way we see life, and am grateful you could teach me so much about the career as a physicist and about physics itself—and the topics you are a specialist in are among the most fascinating for me. I also thank you for the combination of easygoingness and excellence you foster—I think it should be a standard for any professional, and it certainly is for me. You were the perfect Master's advisor. I thank Luiz and Beatriz, without whose help this dissertation would be incomplete. I appreciate the bond the four of us have formed working on this project and hope that, together, we will discover something exciting about gravitational waves from the early universe. I also thank all my other colleagues at the UFMG Particle Physics and Cosmology Group. I hope we will all be leading the scientific enterprise in Brazil and worldwide together in the future.

I thank Prof. Alan Guth for his deeply inspiring humility, which reminds me of a lesson from Plato: being a philosopher should be primarily worthwhile to satiate one's thirsty curiosity, and not for honor, fame, or respect—although these things can help one lead a happy life. Being a student of one of the most prominent humans ever involved in the study of the origin of the cosmos is among the most meaningful chapters of my life story. I thank Prof. David Kaiser for being inspiring to me in so many ways: for showing that excelling as a multidisciplinary scholar is possible; for the most refined ability to combine theoretical and observational tools to tackle fundamental physics questions I have ever seen; and for being so remarkably friendly, a presence of unparalleled warmth. You two amaze me at the personal, professional, and academic levels, making me unable to imagine better people in a better place to advise my doctoral studies. I also thank Dr. Josu Aurrekoetxea and Elba Alonso-Monsalve for inspiring me with their excellence and the captivating science they do. You all make me confident that, at MIT, I can continue to lead a happy life, striving to steer a lifelong intellectual pursuit. 

Finally, I would like to thank all teachers I have had, whose positive impact I continue to feel to this day.

\thispagestyle{empty}
\begin{center}

		\vspace*{\fill}
		\begin{flushright}
	\begin{minipage}{8cm}

			\textit{``P.S. As far as the white rabbit is concerned, it might be better to compare it with the whole universe. We who live here are microscopic insects existing deep down in the rabbit’s fur. But philosophers are always trying to climb up the fine hairs of the fur in order to stare right into the magician’s eyes.''}
\begin{flushright}
            \textit{(Jostein Gaarder)}
\end{flushright}

	\end{minipage}
		\end{flushright}

\end{center}

\chapter*{\centering RESUMO}

\noindent
\thispagestyle{empty}

\noindent
Por milênios, a humanidade dependeu exclusivamente da luz — inicialmente da luz visível e, mais tarde, de porções cada vez mais amplas do espectro eletromagnético — para observar o universo. Na última década, um notável capítulo na extensão da astronomia para além de antenas eletromagnéticas foi concretizado: o raiar da astronomia de ondas gravitacionais abriu uma nova janela de observação para o cosmos. Dentre as muitas novas fontes astronômicas que agora podemos buscar e estudar através de seus sinais de ondas gravitacionais, o Big Bang está certamente entre as mais fascinantes. Ondas gravitacionais nos dão esperanças concretas de observar diretamente o universo primordial, cuja luz, emitida há mais de 13,5 bilhões de anos atrás, tem sua passagem bloqueada e não consegue atingir nossos telescópios. Esta dissertação objetiva estudar ondas gravitacionais da inflação cósmica, o principal paradigma científico sobre o universo primordial. Portanto, o texto está dividido em capítulos sobre ondas gravitacionais, cosmologia inflacionária e ondas gravitacionais inflacionárias. Mais especificamente, nossa discussão será guiada pela tentativa de explicar como o sinal de ondas gravitacionais buscado pelo observatório NANOGrav poderia ter se originado no universo primordial.

\vskip 2cm

\noindent
\textbf{Palavras-chave:} gravitação; cosmologia; ondas gravitacionais; inflação cósmica.

\chapter*{\centering ABSTRACT}

\noindent
\thispagestyle{empty}

\noindent
For millennia, humanity has relied exclusively on light—initially visible light and, later, broader and broader portions of the electromagnetic spectrum—to observe the universe. In the past decade, a remarkable chapter in extending astronomy beyond electromagnetic antennas has been concretized: the dawn of gravitational-wave astronomy has opened a new observational window into the cosmos. Among the many new astronomical sources we may now look for and study through their gravitational-wave signals, the Big Bang is surely among the most fascinating. Gravitational waves give us concrete hope of directly observing the primordial universe, whose light, emitted more than 13.7 billion years ago, is blocked from reaching our telescopes. This dissertation is aimed at the study of gravitational waves from cosmic inflation, the main scientific paradigm for the very early universe. Therefore, the text is divided into chapters on gravitational waves, inflationary cosmology, and inflationary gravitational waves. More specifically, our discussion will be steered by the endeavor to explain how the gravitational-wave signal sought by the NANOGrav observatory could have originated in the primordial universe.

\vskip 2cm

\noindent
\textbf{Keywords}: gravitation; cosmology; gravitational waves; cosmic inflation. 


\addtocontents{toc}{\protect\thispagestyle{empty}}
\renewcommand*\contentsname{\centering SUMMARY}

\tableofcontents\thispagestyle{empty}


\pagebreak

\fancyhead[r]{\bfseries\thepage}
\fancyhead[l]{\bfseries \leftmark}

\chapter{\textsc{Introduction}}
Cosmology, the branch of physics and astronomy in which the history of the universe as a whole is studied, has become a data-driven science, and theoretical models are observationally constrained in detail. At first, one may find this surprising and counterintuitive: how could scientists know with precision what was going on in the cosmos billions of years ago? How contrived and powerful must these scientific models be to predict the details of epochs separated by eons from our own? Nonetheless, just by keeping in mind that the speed of light is finite, one should easily convince oneself that studying the entire history of the universe is, in principle, as straightforward as any other endeavor in astronomy. Observing the sky is looking at the past: we see the Sun not as it is instantaneously, but as it had been 8 minutes before; we see the stars in the Southern Cross not as they are instantaneously, but as they had been hundreds of years before; and so on. This is because light needs these non-negligible times to traverse the astronomical distances that separate us from such sources. So, the farther out we observe, the earlier in the history of the universe we see.

Of course, things are not as simple as they could have been. The universe only became transparent to light around 13.7 billion years ago. Before then, the cosmos was so densely filled with ionized particles as to be opaque to photons: trying to observe the astronomical objects from back then is like looking for a far-away building on a foggy day. For this reason, we have mainly relied on indirect—although robust and remarkably clever—evidence to study the earliest epochs in the history of the universe. But, with the dawn of gravitational-wave astronomy, this might soon no longer be the case.

According to the contemporary understanding of gravity—general relativity—space and time form a single structure, spacetime, whose curvature is brought about by the presence of masses. Space is hence no longer understood as the inert stage where the histories of physical entities, like particles, fields, moons, and people, unfold. Rather, space—and time—are as dynamical as its matter content: its curvature is molded by the distribution of masses in it and, in turn, the motion of masses is impacted by the geometry of the region of spacetime they traverse, in a symbiotic interplay. 

Malleable as it is, spacetime can have ripples sent across its structure by masses that move sufficiently violently, in much the same way as a heavy enough rock will send ripples across the surface of a pond onto which it is thrown. These undulations of spacetime are called gravitational waves. In the real world, they are very faint: more than a century after their proposal in the early days of general relativity, a decades-long, perhaps unprecedentedly sophisticated technological endeavor was required before the first direct detection of gravitational waves in 2015. And now, a decade later, the field of gravitational-wave astronomy is skyrocketing, with concretized triumphs and an excitingly huge potential for discovery for astrophysics and cosmology.

Despite its opaqueness to electromagnetic waves early on, the universe has always been transparent to gravitational waves. As a result, these may serve as direct messengers from the early universe: they can freely stream, with no impediments, for billions of years from the early universe up to the present, when they reach our observatories. This renders the direct observation of the infancy of the cosmos, impossible through light, now an exciting possibility. Although the detection of gravitational waves from the primordial universe has not been achieved yet, efforts have been growing for a few decades now on the observational and theoretical fronts. This dissertation is inserted in this context.

Well-understood gravitational-wave detections to date have solely involved astrophysical events: the inspiral and merger of binaries of black holes and/or neutron stars. However, the NANOGrav detector, which has been collecting data for over 15 years, has shown increasingly statistically significant evidence for a gravitational-wave signal, whose origin remains a mystery. Although it may also have had an astrophysical origin, from binary black holes, NANOGrav could be seeing primordial gravitational waves: oscillations produced everywhere in the early universe—by a diverse set of possible sources. It could be, that is, actualizing humanity's first direct look at the primordial cosmos.

In this dissertation, we are interested in primordial gravitational waves produced by cosmic inflation, the main paradigm for the history of the very early universe. Our end goal is to understand an example theoretical scenario in which inflationary gravitational waves could explain NANOGrav's results while evading other existing observational constraints. With this goal in mind, this text follows a straightforward structure: Section \ref{GWchapter} is about gravitational waves, Section \ref{cosmochapter}, about inflationary cosmology, and Section \ref{IGWchapter}, about inflationary gravitational waves.

More specifically, in Section \ref{GWchapter}, after a brief review of Einstein's theory of gravity, we start by working out in general relativity the solution for gravitational waves propagating on an empty and flat spacetime—flat as opposed to curved, i.e. in a context where the wave is the only nontrivial gravitational phenomenon. Then, we discuss the two most well-established techniques to detect gravitational waves. The first is interferometry, the one that has achieved repeated indisputable success since the first detection in 2015, which we discuss qualitatively. The second is pulsar timing, used by NANOGrav and hence more directly related to our problem of interest, which we therefore overview quantitatively. We provide a detailed discussion of gravitational-wave signal analysis, clarifying the definitions of the various quantities used in the literature to report the intensity of gravitational waves and how they are related. To wrap up, we also discuss the properties of a stochastic background of gravitational waves produced by a cosmological collection of binaries, arguably the main explanation under consideration for the NANOGrav signal, to which a primordial, inflationary origin provides a serious alternative.

Section \ref{cosmochapter} also starts in an applied-general-relativity spirit, working out the dynamical equations and solutions that describe a universe like ours. We establish the main features of the standard Hot Big Bang cosmological model and, then, study its shortcomings that led to the development of the theory of cosmic inflation. This section ends with a discussion of the essential features of inflationary dynamics.

Section \ref{IGWchapter} introduces the perturbative formalism for inflation and works out the dynamics of tensorial metric perturbations, which are the inflationary gravitational waves. Applying the concepts from Section \ref{signal_analysis}, we study the transfer-function formalism, which provides an analytic approximation to the dynamical evolution of gravitational waves since their production in the primordial universe up to the present. Then, we discuss the essential features of the NANOGrav signal, which a primordial spectrum of gravitational waves should reproduce in order to explain it. We also study observational constraints to primordial gravitational waves arising from interferometric observatories and Big Bang nucleosynthesis. We reach our climax in Figs. \ref{omegaT}–\ref{omegar}, exemplifying features that could allow a model of the early universe to account for the production of NANOGrav's signal while respecting those observational constraints. We conclude with a brief discussion of a specific model where those features are realized, in a forward-looking note of what could come next in studying primordial gravitational waves in light of NANOGrav's result.
\fancyhead[r]{\bfseries\thepage}
\fancyhead[l]{\bfseries \leftmark}

\chapter{\textsc{Gravitational waves}}\label{GWchapter}

\section{General relativity refresher}
\label{GRprimer}
In general relativity, gravitation comes from masses curving spacetime, such that the gravitational field is not an abstract force field as in classical electromagnetism, but the spacetime metric itself. The metric is a rank-2 tensor carrying information about the geometry of a mathematical object, like a manifold; more specifically, it has a central role in defining distances on that object. The infinitesimal distance element $ds$ between two points in a manifold is given by
\begin{equation}
\label{metric}
    ds^2=g_{\mu\nu}dx^\mu dx^\nu,
\end{equation}
where $g_{\mu\nu}$ are the components of the metric tensor and $dx^\mu$ are the components of the infinitesimal separation vector between the two points. In Eq. \ref{metric} and throughout this dissertation, we will be employing the Einstein summation convention, whereby the presence of repeated indices, one subscript and one superscript, implies a summation: $\sum_\alpha V^\alpha V_\alpha\equiv V^\alpha V_\alpha$. Moreover, as usual in this context, Greek indices label the four spatiotemporal coordinates of spacetime, while Latin indices refer to the three spatial ones. 

\begin{figure}
    \centering
    \includegraphics[width=0.5\linewidth]{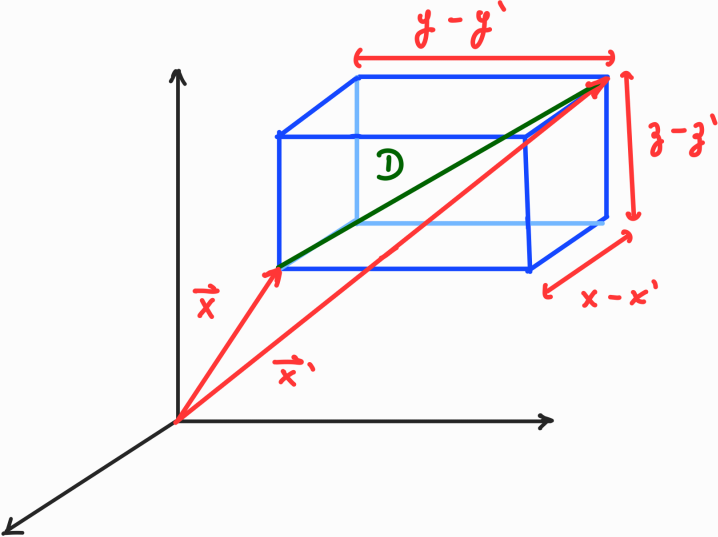}
    \caption{The distance $D$ between vectors $\Vec{x}$ and $\Vec{x}'$ in three-dimensional Euclidean space is one of the most intuitive cases to compute. Pictorially, it corresponds to the diagonal of a parallelpiped and, hence, the result $D^2=(x-x')^2+(y-y')^2+(z-z')^2$ may be derived from two sequential applications of the Pythagorean theorem.}
    \label{paralelepipedo}
\end{figure}

By high school, one will usually have learned a simple specific case of how to calculate the distance between two points from their separation as in Eq. \ref{metric}: the distance between two points in three-dimensional Euclidean space. In Cartesian coordinates, one can specify the position of two points with two vectors, say $\Vec{x}=(x,y,z)$ and $\Vec{x}'=(x',y',z')$. Then, the distance between $\Vec{x}$ and $\Vec{x}'$, which corresponds to the size of a parallelepiped's diagonal as illustrated in Fig. \ref{paralelepipedo}, is given by $\sqrt{(x-x')^2+(y-y')^2+(z-z')^2}$. This result, provable by applying the Pythagorean theorem twice, indeed matches the general prescription in Eq. \ref{metric}, given that the Euclidean metric is the identity matrix in Cartesian coordinates.

In general relativity, spacetime is a four-dimensional manifold. In the absence of masses, which source gravity, spacetime is flat, given that its curvature is the manifestation of gravity. Hence, adding a time $t$ coordinate on top of the $x$, $y$, and $z$ spatial coordinates, the distance element in empty space is commonly given by
\begin{equation}
    ds^2=\eta_{\mu\nu}dx^\mu dx^\nu=-dt^2+dx^2+dy^2+dz^2,
\end{equation}
where $\mathbf{\eta}=\mathrm{diag}(-1,1,1,1)$ and the sign difference between the temporal and spatial components reflects the hyperbolic geometry of spacetime. This sign difference is used to establish a cornerstone of relativity, light's propagation with a constant speed of $c=1$ in any reference frame. This is done by positing that light traverses $ds=0$ paths, so-called null or lightlike trajectories, so that $(dx^2+dy^2+dz^2)/dt^2=1$, i.e. the infinitesimal spatial displacement divided by the infinitesimal temporal displacement, which describes the speed, is 1 for light in any reference frame.

Physicist John Wheeler famously described the general-relativistic picture of gravity as ``spacetime tells matter how to move; matter tells spacetime how to curve.'' This idea is cast into a mathematical form in Einstein's equations:
\begin{equation}
    \label{Einstein}
    G_{\mu\nu}(g_{\mu\nu})=\frac{8\pi G}{c^4}T_{\mu\nu}.
\end{equation}
On the right-hand side, the energy-momentum tensor, $T_{\mu\nu}$, characterizes the content and motion of matter. On the left-hand side, the Einstein tensor $G_{\mu\nu}$, a function of the metric, relates to how spacetime is curving. There is a clear differential geometry algorithm leading from the metric to the Einstein tensor, whereby one calculates the Christoffel symbols $\Gamma^\lambda_{\mu\nu}$ from the metric and its derivatives, then the Riemann tensor $R^\lambda_{\;\mu\sigma\nu}$ from the Christoffel symbols and its derivatives, and then the Ricci tensor and the Ricci scalar from contractions of the Riemann tensor: $R_{\mu\nu}=R^\alpha_{\;\mu\alpha\nu}$ and $R=R^\beta_{\;\beta}$; the Einstein tensor is just $G_{\mu\nu}=R_{\mu\nu}-R g_{\mu\nu}/2$. 

\section{Gravitational-wave solution on a flat background}
\label{GWderivation}
Gravitational waves are oscillating perturbations that propagate on top of some background spacetime. Thinking of spacetime, with its malleable curvature, as the water in a pond, gravitational waves may be pictured as the ripples produced by a rock thrown into the pond. Our goal now is to apply the general-relativistic dynamics to distill their wavelike character, which will be apparent, by the end of this section, to any reader who has taken a first course on oscillations and waves at the undergraduate level. We will specialize to gravitational waves propagating on an empty, i.e. flat, background, which is not only the easiest case to work out but is also a useful model to study the propagation of gravitational waves emitted by astrophysical sources.

As small perturbations on top of a background, gravitational waves are indeed perturbative phenomena, just like those usually studied toward the end of undergraduate quantum mechanics. The analysis we are about to undertake is to consider first-order deviations from flat spacetime in the solution of Einstein's equations, i.e. we shall linearize these equations. Therefore, let us consider a metric of the form
\begin{equation}
    \label{linearizedg}
    g_{\mu\nu}=\eta_{\mu\nu}+h_{\mu\nu},
\end{equation}
where $|h_{\mu\nu}|\ll1\;\forall\;\mu,\;\nu$ everywhere. Given the invariance of general relativity under coordinate transformations, one might worry about this constraint on the numerical values of the components of the metric perturbation. Thus, it is important to clarify that the complete meaning of what we require of $h_{\mu\nu}$ is specifically that there exists a reference frame in which $|h_{\mu\nu}|\ll1$ holds in a sufficiently large region of spacetime \cite{Maggiore1}. With this, to linear order in $h_{\mu\nu}$, the Christoffel symbols are:
\begin{align}
    \Gamma_{\mu\nu}^\lambda&=\frac{1}{2}g^{\lambda\alpha}\left(\partial_\mu g_{\nu\alpha} + \partial_\nu g_{\mu\alpha} - \partial_\alpha g_{\mu\nu}\right)\\
    &=\frac{1}{2}\eta^{\lambda\alpha}\left(\partial_\mu h_{\nu\alpha} + \partial_\nu h_{\mu\alpha} - \partial_\alpha h_{\mu\nu}\right),
\end{align}
where, in going from the first to the second line, derivatives of the full metric become derivatives of the perturbation because the components of the flat background metric are all constant, and the inverse metric components out front become components of the inverse Minkowski metric since the terms coming from the perturbation are second-order and are, thus, discarded. Next, we determine the Riemann tensor components:
\begin{align}
    R^\lambda_{\;\sigma\mu\nu}&=\partial_\mu\Gamma^\lambda_{\nu\sigma}-\partial_\nu\Gamma^\lambda_{\mu\sigma}+\Gamma^\lambda_{\mu\alpha}\Gamma^\alpha_{\nu\sigma}-\Gamma^\lambda_{\nu\alpha}\Gamma^\alpha_{\mu\sigma} \\
    &=\frac{1}{2}\eta^{\lambda\alpha}(\partial_\mu\partial_\sigma h_{\alpha\nu}+\partial_\nu\partial_\alpha h_{\mu\sigma}-\partial_\mu\partial_\alpha h_{\sigma\nu}-\partial_\nu\partial_\sigma h_{\alpha\mu}).
\end{align}
We can lower the first index of the Riemann tensor using the Minkowski metric (given that, again, the expression is already first-order in the perturbation): $R_{\lambda\sigma\mu\nu}=\eta_{\lambda\beta}R^\beta_{\;\sigma\mu\nu}=(\partial_\mu\partial_\sigma h_{\lambda\nu}+\partial_\nu\partial_\lambda h_{\mu\sigma}-\partial_\mu\partial_\lambda h_{\sigma\nu}-\partial_\nu\partial_\sigma h_{\lambda\mu})/2$, where we have used that $\eta_{\beta\lambda}\eta^{\lambda\alpha}=\delta_\beta^{\;\alpha}$. Nearing the writing of the Einstein equation, we can then determine the Ricci tensor and scalar:
\begin{align}
    \label{linearricci}&R_{\sigma\nu}=\eta^{\mu\lambda}R_{\lambda\sigma\mu\nu}= \frac{1}{2}(\partial^\lambda\partial_\sigma h_{\lambda\nu}+\partial_\nu\partial^\mu h_{\mu\sigma}-\square h_{\sigma\nu}-\partial_\nu\partial_\sigma h)\\
    &R=\eta^{\nu\sigma}R_{\sigma\nu}=\partial^\lambda\partial^\nu h_{\lambda\nu}-\square h,
\end{align}
where $\square\equiv\partial_\alpha\partial^\alpha$ and $h\equiv h^\alpha_{\;\alpha}$. Finally, we can write the dynamical equation for the gravitational wave, the Einstein equation $R_{\sigma\nu}-R g_{\sigma\nu}/2=0$—with a $0$ on the right-hand side since we are specializing to propagation in a vacuum:
\begin{equation}
\label{lineareinstein}
    \partial^\lambda\partial_\sigma h_{\lambda\nu}+\partial_\nu\partial^\mu h_{\mu\sigma}-\square h_{\sigma\nu}-\partial_\nu\partial_\sigma h + \eta_{\sigma\nu}\square h -\eta_{\sigma\nu}\partial^\lambda\partial^\mu h_{\lambda\mu}=0.
\end{equation}

We may exploit the gauge freedom in general relativity to simplify this result, which, so far, is an equation governing an object with ten degrees of freedom—the ten independent components of a symmetric rank-2 tensor in four spacetime dimensions. Gauge freedom is the phenomenon whereby the predictions of a physical theory do not change when some function that is a part of the theory is changed in a specific way. For example, we usually learn in high school that, in classical electromagnetism, all that matters for determining measurable quantities—like the electric current and the power output of a circuit—is the electric potential difference between two points $\phi(\vec{x_b})-\phi(\vec{x_a})$. Therefore, if the electric potential is modified through a so-called gauge transformation as $\phi(\vec{x})\rightarrow\phi'(\vec{x})=\phi(\vec{x})+c$, where $c$ is a constant, the physical predictions extracted from electromagnetic theory using either $\phi(\vec{x})$ or $\phi'(\vec{x})$ will be the same. This means that the value of the electric potential alone is not directly measurable. One may hence interpret $\phi(\vec{x})$ as being a solely mathematical device within the theory's framework without representing anything physically real. 

General relativity's gauge invariance comes from another cornerstone of the theory: that performing a change of coordinates to a metric does not change the physical spacetime that it describes. In other words, the coordinate system is an arbitrary choice that should not impact the predictions one can extract from the theory. Consider the coordinate transformation $x'^\mu=x^\mu-\xi^\mu$. Then, since the metric is a rank-2 tensor, it will transform under a coordinate change through contraction with two Jacobians: $g'_{\mu\nu}(x')=g_{\alpha\beta}(x)\frac{\partial x^\alpha}{\partial x'^\mu}\frac{\partial x^\beta}{\partial x'^\nu}$. For the transformation in question, to linear order in small terms, we thus have
\begin{equation}
\label{metrictrans}
    h'_{\mu\nu}(x')=h_{\mu\nu}(x)+\partial_\mu\xi_\nu+\partial_\nu\xi_\mu. 
\end{equation}

We can separate $h_{ij}$ in parts with and without trace: $h_{ij}=(h_{ij}-\delta_{ij}h^k_{\;k}/3)+\delta_{ij}h^k_{\;k}/3$, where $\delta_{ij}$ are the components of an identity matrix, i.e. a Kronecker delta. It will prove to be useful to employ gauge freedom—which relates to the four dimensions of spacetime and, therefore, involves four degrees of freedom—to impose the so-called transverse gauge condition:
\begin{align}
    \label{trans1}&\partial^i(h_{ij}-\delta_{ij}h^k_{\;k}/3)=0;\\
    \label{trans2}&\partial^ih_{i0}=0.
\end{align}
Can we always choose a suitable $\xi^\mu$ to ensure these conditions will hold? Yes! Let us study Eq. \ref{trans1} first. Suppose we started with metric components $h_{ij}$ that did not satisfy this condition and used a change of coordinates to have $h'_{ij}$ do. Then, according to Eq. \ref{metrictrans}, we would be asserting that $\partial^i[h_{ij}+\partial_i\xi_j+\partial_j\xi_i-\delta_{ij}(h^k_{\;k}+2\partial_k\xi^k)/3]=0$. Applying the derivative $\partial^j$ to both sides of this equation and simplifying, we arrive at
\begin{equation}
\label{Poisson}
    \nabla^2\partial^i\xi_i=-\frac{3}{4}\partial^i\partial^j(h_{ij}-\delta_{ij}h^k_{\;k}/3),
\end{equation}
a Poisson equation for $\partial^i\xi_i$ whose source term is known. Let us proceed with an unproven claim: $\xi_i=\partial_i\beta$ with a suitably chosen $\beta$ can enforce the transverse gauge condition. Therefore, Eq. \ref{Poisson} is of the form $\nabla^2\nabla^2\beta\equiv\nabla^2\alpha=source$. Consequently, we will be able to determine the desired $\xi_i$ by solving two nested Poisson equations, problems for which we are guaranteed to find a solution imposing adequate boundary conditions! The proof that we can always impose Eq. \ref{trans2} is similar: we end up with a Poisson equation for $\xi_0$ by requiring $\partial_0\xi_i=0$, a choice we are free to make since the coordinate transformation brought about by $\xi^\mu$ is arbitrary and making $\xi_i$ time-independent does not spoil the the first part of the transverse gauge (Eq. \ref{trans1}), which was established through a differential equation for $\xi_i$ involving only its spatial derivatives (Eq. \ref{Poisson}).

Besides the gauge choice, further simplification of $h_{\mu\nu}$ comes from imposing on it conditions consistent with it being a solution of its equation of motion, the linearized Einstein equation in vacuum Eq. \ref{lineareinstein}. For example, writing out the $(0,0)$ component of this equation, we have:
\begin{equation}
\label{einstein00}
\begin{aligned}
    2(\partial_0\partial^ih_{i0}-\partial_0^2h_{00})+\partial_0^2h_{00}-\nabla^2h_{00}-\partial_0^2h+\partial_0^2h-\nabla^2h&\\+\partial_0^2h_{00}+\partial_i\partial_jh^{ij}+2\partial_0\partial_ih^{0i}&=0.
\end{aligned}
\end{equation}
After canceling out terms remembering that $h=h^\alpha_{\;\alpha}=h^k_{\;k}-h_{00}$, Eq. \ref{einstein00} simplifies to $\partial_i\partial_jh^{ij}-\nabla^2h^k_{\;k}=0$. Finally, using the transverse gauge condition in Eq. \ref{trans1}, we are left with $\nabla^2h^k_{\;k}=0$. A corollary of Liouville's theorem tells us that, if $u$ is a bounded harmonic function in $\mathbb{R}^n$ and $u(\mathbf x)\rightarrow0$ as $|\mathbf x|\rightarrow\infty$, then $u=0$. As a result, assuming that $h^k_{\;k}$ vanishes at infinity, the $(0,0)$ component of the equation of motion informs that $h^k_{\;k}=0$. 

\begin{figure}
    \centering
    \includegraphics[width=0.67\linewidth]{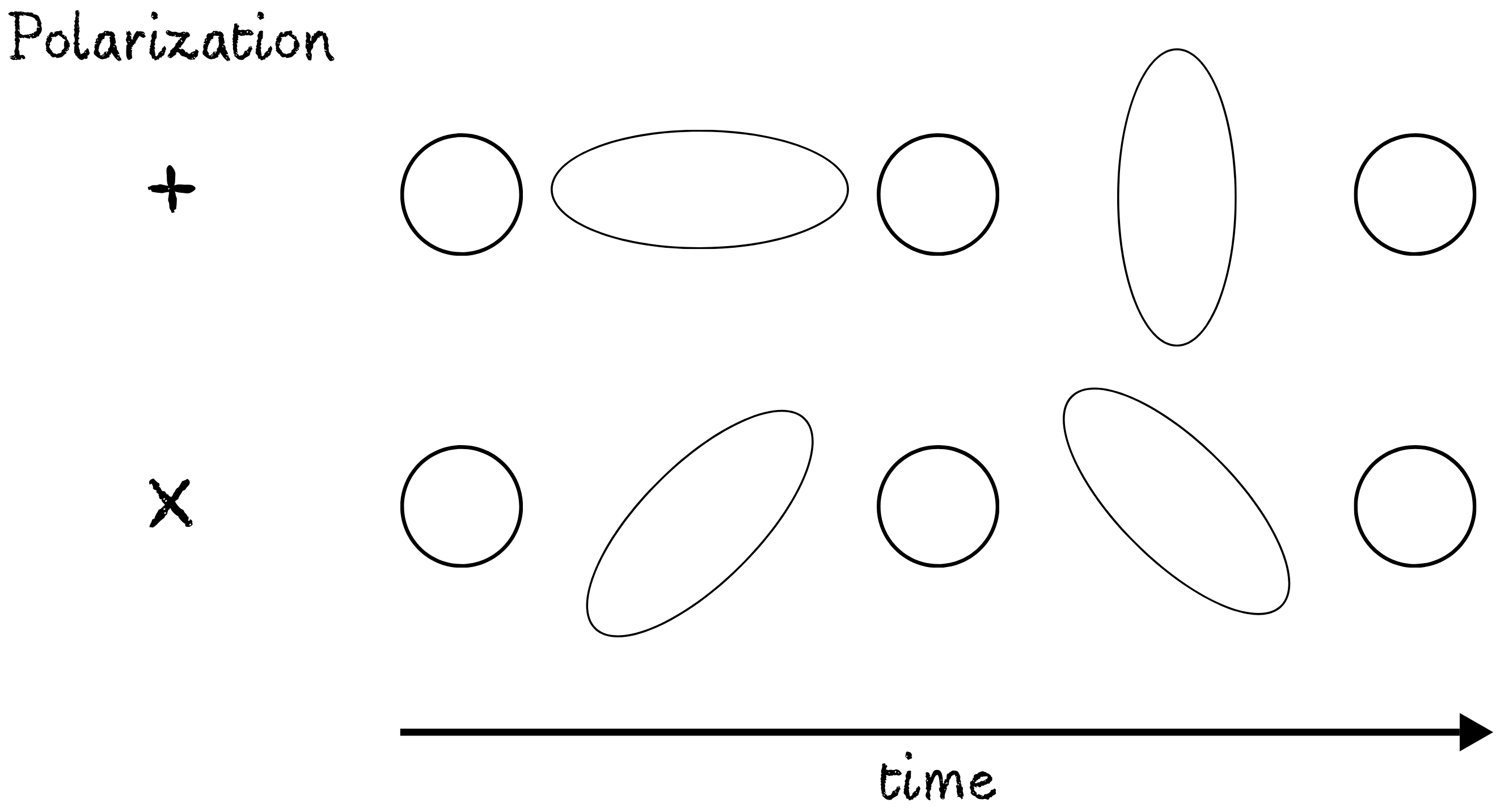}
    \caption{A qualitative depiction of the effect caused by gravitational waves with plus and cross polarizations when they pass through a region containing a ring of massive particles.}
    \label{pluscross}
\end{figure}

If one works out the $(0,i)$ and $(i,j)$ components of Eq. \ref{lineareinstein}, one will find out respectively that, with reasonable boundary conditions, $h_{0i}=0$ and $h_{00}=0$. The gauge conditions and the conditions derived from the equations of motion, specific to solving the Einstein equations in a vacuum, are collectively referred to as the transverse-traceless (TT) ``gauge''. While this nomenclature is widespread in the literature, one should be careful about its precise meaning since, as shown, only the transversality was imposed through a choice of gauge; the tracelessness, on the other hand, was merely consistent with the vacuum equations of motion in transverse gauge. In TT gauge, the wavelike character of small perturbations to a background flat spacetime is finally blatantly revealed, since Eq. \ref{lineareinstein} reduces to
\begin{equation}
\label{waveeq}
    \square h_{\mu\nu}^{TT}=0,
\end{equation}
homogeneous wave equations for each component of the metric perturbation, where the $TT$ superscript is added to remind us that this simplification of the equation of motion is gauge-specific. As one might recall from their first college-level introduction to waves, Eq. \ref{waveeq} is suitably solved by a plane wave,
\begin{equation}
\label{onewave}
    h_{\mu\nu}^{TT}(x)=C_{\mu\nu}e^{ik^\alpha x_\alpha},
\end{equation}
where $k_\alpha k^\alpha=0$—suggesting that gravitons in quantum field theory are massless since $k^\alpha$ is a particle's four-momentum—and the coefficients $C_{\mu\nu}$ are constants and, in this case, $C_{00}=C_{0i}=C^k_{\;k}=0$. Note that the transverse gauge condition Eq. \ref{trans1} implies that $\partial^jh^{TT}_{jk}=ik^jh_{jk}^{TT}=0$, justifying the gauge's name—the lines and columns of the symmetric matrix representing the spatial part of the TT metric perturbation (i.e. $h_{ij}^{TT}$) are orthogonal to the direction of propagation $\vec{k}/|\vec{k}|$. A common choice is to orient the z-axis along the gravitational-wave direction of propagation, such that the metric perturbation matrix becomes
\begin{equation}
\label{matrixeq}
\mathbf{h}=
\begin{bmatrix}
0 & 0 & 0 & 0\\
0 & h_+ & h_\times &0 \\
0 & h_\times & -h_+ & 0 \\
0 & 0 & 0 & 0
\end{bmatrix}e^{ik^\alpha x_\alpha}.
\end{equation}
As clearly expressed now in Eq. \ref{matrixeq}, the sequence of gauge choices and equation-of-motion constraints imposed onto $h_{\mu\nu}$ eliminated eight of its original ten degrees of freedom. The remaining two are interpreted as the polarizations of the gravitational wave, much like in the case of electromagnetic waves. The polarization are called ``plus'' ($+$) and ``cross" ($\times$) due to the effect of their passage through a region containing a ring of massive particles, which is illustrated in Fig. \ref{pluscross} and can be demonstrated using the geodesic deviation equation.

An interesting difference between electromagnetism and Einsteinian gravitation is that in the former, due to its linear equation of motion, electromagnetic waves cannot source other electromagnetic waves, while the opposite happens in the latter. To see this, we need to go further in our perturbative approach and consider second-order contributions:
\begin{equation}
    \begin{aligned}
    &g_{\mu\nu}=\eta_{\mu\nu}^{(0)}+h_{\mu\nu}^{(1)}+h_{\mu\nu}^{(2)}\\
    &R_{\mu\nu}=R_{\mu\nu}^{(0)}+R_{\mu\nu}^{(1)}+R_{\mu\nu}^{(2)},
    \end{aligned}
\end{equation}
where $R_{\mu\nu}^{(0)}$ is calculated from the background metric, $O\left(R_{\mu\nu}^{(1)}\right)=O\left(h_{\mu\nu}^{(1)}\right)$, and $O\left(R_{\mu\nu}^{(2)}\right)=O\left(h_{\mu\nu}^{(2)}\right)=O\left(h_{\mu\nu}^{(1)}\right)^2$. Since we are working with a flat background, $R_{\mu\nu}^{(0)}=0$. The first-order vacuum equation is just $R^{(1)}_{\mu\nu}[h^{(1)}]=0$, where $R^{(M)}_{\mu\nu}[h^{(N)}]$ is the $\mathrm{M^{th}}$-order part of the Ricci tensor calculated from $h_{\mu\nu}^{(N)}$. That is what we have already worked out in this section to solve for $h_{\mu\nu}^{(1)}$—the equivalence of $G_{\mu\nu}=R_{\mu\nu}-Rg_{\mu\nu}/2=0$ and $R_{\mu\nu}=0$ is easily seen by taking the trace of the former equation. Finally, $h_{\mu\nu}^{(2)}$ may be determined from the second-order vacuum Einstein equation $R^{(2)}_{\mu\nu}[h^{(1)}]+R^{(1)}_{\mu\nu}[h^{(2)}]=0$—the latter term is obtained by replacing $h_{\mu\nu}^{(1)}\rightarrow h_{\mu\nu}^{(2)}$ in Eq. \ref{linearricci} and the former is given by \cite{Carroll_2019}:
\begin{equation}
\label{ricci2}
\begin{aligned}
    R^{(2)}_{\mu\nu}[h]=&\frac{1}{2}\bigg[ h^{\rho\sigma}\partial_\mu\partial_\nu h_{\rho\sigma}+\frac{1}{2}(\partial_\mu h_{\rho\sigma})\partial_\nu h^{\rho\sigma}+(\partial^\sigma h^\rho_{\;\nu})\partial_\sigma h_{\rho\mu}-(\partial^\sigma h^\rho_{\;\nu})\partial_\rho h_{\sigma\mu}\\&-h^{\rho\sigma}\partial_\rho\partial_\mu h_{\nu\sigma}-h^{\rho\sigma}\partial_\rho\partial_\nu h_{\mu\sigma}+\partial_\sigma(h^{\rho\sigma}\partial_\rho h_{\mu\nu})-\frac{1}{2}(\partial_\rho h_{\mu\nu})\partial^{\rho}h\\&-\left(\partial_\sigma h^{\rho\sigma}-\frac{1}{2}\partial^{\rho}h\right)\partial_\mu h_{\nu\rho}-\left(\partial_\sigma h^{\rho\sigma}-\frac{1}{2}\partial^{\rho}h\right)\partial_\nu h_{\mu\rho}\bigg],
\end{aligned}
\end{equation}
where we have used $h$ as shorthand notation for $h^{(1)}$.

It is enlightening to bring the second-order vacuum equation to the $G^{(2)}_{\mu\nu}[h^{(1)}]+G^{(1)}_{\mu\nu}[h^{(2)}]=0$ form and suggestively rearrange it:
\begin{equation}
    R^{(1)}_{\mu\nu}[h^{(2)}]-\frac{1}{2}\eta^{\rho\sigma}R^{(1)}_{\rho\sigma}[h^{(2)}]\eta_{\mu\nu}=8\pi G\left\{\frac{-1}{8\pi G}\left[R^{(2)}_{\mu\nu}[h^{(1)}]-\frac{1}{2}\eta^{\rho\sigma}R^{(2)}_{\rho\sigma}[h^{(1)}]\eta_{\mu\nu}\right]\right\}\equiv8\pi Gt_{\mu\nu}.
\end{equation}
Note that there are no terms involving $h_{(1)}^{\rho\sigma}R^{(1)}_{\rho\sigma}[h^{(1)}]$ because of the first-order vacuum equation. There are several good reasons to interpret $t_{\mu\nu}$ as indeed the energy-momentum tensor associated with the gravitational field (for a longer list, see Section 7.6 in Ref. \cite{Weinberg:1972kfs}). Among them, there is the fact that it is quadratic in the field—in this case, the perturbation to the metric—as it commonly happens for conventional spacetime contents, like scalar and electromagnetic fields. Moreover, it is symmetric and locally conserved ($\partial_\mu t^{\mu\nu}=0$ exactly, i.e. $t^{\mu\nu}$ as defined obeys the linearized Bianchi identities). Nevertheless, $t^{\mu\nu}$ is not an actual tensor: as it turns out, it is not invariant under coordinate transformations. A gauge-invariant redefinition of $t^{\mu\nu}$ may be achieved by averaging this quantity as currently defined over several wavelengths—the idea is to capture enough information about the spacetime curvature in a small region to construct a gauge-invariant quantity. This works out to give Eq. 7.165 in Ref. \cite{Carroll_2019}. 

Since we are concerned with the propagation of gravitational waves away from the source, we may specialize to the TT gauge to derive a result that, although not general, will suffice for our purposes. To further simplify Eq. \ref{ricci2}, we can note that practical examples of gravitational waves involve bounded oscillations. So, consider, for instance, a bounded wavelike function $f(t-x)\equiv f(\tau)$ in two-dimensional spacetime. $\braket{\partial_\tau f}\equiv T^{-1}\int_0^T\mathrm d\tau\; \partial_\tau f(\tau)=[f(T)-f(0)]/T\rightarrow0$ in the limit $T\rightarrow\infty$. From this, using integration by parts, it follows that $\braket{X\partial_\tau Y}=-\braket{Y\partial_\tau X}$ if $X$ and $Y$ are bounded functions of $\tau$. We can use this result to send several terms in Eq. \ref{ricci2} to $0$ because of the transversality condition. After applying this, note, in order to set the third term in Eq. \ref{ricci2} to $0$, that, as previously derived, $\square h_{\mu\nu}^{TT}=0$. All that is left is, therefore,
\begin{equation*}
    \left\langle R^{(2)}_{\mu\nu}[h] \right\rangle = -\frac{1}{4}\left\langle\left(\partial_\mu h_{\rho\sigma}^{TT}\right)\left(\partial_\nu h^{\rho\sigma}_{TT}\right)\right\rangle.
\end{equation*}
Again using integration by parts with the averaging bracket and $\square h_{\mu\nu}^{TT}=0$, we learn that $\eta^{\rho\sigma}R^{(2)}_{\rho\sigma}[h^{(1)}]=0$, finally leading to the redefinition
\begin{equation}
\label{energyinGWs}
    t_{\mu\nu}\equiv \frac{1}{32\pi G}\left\langle\left(\partial_\mu h_{\rho\sigma}^{TT}\right)\left(\partial_\nu h^{\rho\sigma}_{TT}\right)\right\rangle.
\end{equation}

\section{Detecting gravitational waves}
This current year of 2025 marks the $\mathrm{10^{th}}$ anniversary of arguably the most emblematic milestone in the history of gravitational-wave science: the first detection of gravitational waves emitted by a localized source in the sky—in this case, the merger of a black hole binary—performed by the LIGO–Virgo collaboration (LIGO is the Laser Interferometer Gravitational-Wave Observatory) \cite{PhysRevLett.116.061102}. Published in 2016, it was among the great revolutions of this century in physics and astronomy more broadly, triggering a streamlined awarding of the Nobel Prize in Physics to the leaders of that endeavor in 2017.

On the physics side, it joined the hall of central pieces of empirical confirmation of general relativity, alongside, for example, the verification of the theory's predictions for the precession of the perihelion of Mercury and the bending of light by the Sun in eclipses. To extract from the signal measured by the LIGO and Virgo detectors information about the source—e.g. that it was a merging binary system with a total mass $M$, a position in the sky $(\theta,\phi)$, at a distance $D$, etc—, the team relies on sophisticated numerical solutions to Einstein's equations for the dynamics of a binary of compact objects—an umbrella term encompassing white dwarfs, neutron stars, and black holes, which are the densest objects found in the universe and, therefore, may be capable of producing gravitational waves that are loud enough to be detected at astronomical distances. Employing a technique for data processing called matched filtering (to be discussed in Section \ref{signal_analysis}), the detected gravitational-wave waveform is compared against numerous pre-saved ones, which were simulated for different input parameters, to determine the statistical significance of sources with different parameters having originated the detected signal. The precision reached by the LIGO–Virgo team in this process is remarkable and has rightly been celebrated again and again in the media in the past decade as confirmation of what Einstein had predicted a century before—that ripples in the fabric of spacetime can be generated and travel like waves.

On the astronomy side, gravitational waves inaugurate a brand new avenue for the observation of the universe. For most of human history, we relied exclusively on electromagnetic waves to observe the cosmos—first just through visible light and, more recently, accessing other regions of their frequency spectrum with telescopes. Being of a different nature from photons, gravitational waves are emitted by astronomical sources in totally different contexts, so that, with them, we may gain observational access to previously undetectable dark sources or learn complementary, previously inaccessible information about sources that have been studied only through their light. To emphasize the significance of this revolution, a metaphor has commonly been used: astronomy had had only vision, but, now, a new sense has awakened; it gained hearing. Target sources for gravitational-wave astronomy are numerous both at the astrophysical and cosmological scales, and include the already detected binaries of neutron stars and stellar-mass black holes, as well as systems containing white dwarfs or primordial, intermediate-mass, or supermassive black holes—these systems might be seen in early inspiral or merger in bound orbits as well as in parabolic or hyperbolic encounters, all of which are of interest to astronomers—, in addition to supernovae, cosmic strings, and early-universe processes like an electroweak phase transition, cosmic inflation, and reheating.

In this section, we shall start by going over the two currently most successful and well-known techniques for gravitational-wave detection: interferometry and pulsar timing. The former will be described qualitatively in Section \ref{LVK}, while for the latter, because it has been more relevant for the search of primordial gravitational waves—and, hence, for this dissertation—, we will also go over its basic functioning quantitatively in Section \ref{PTA}. Then, in Section \ref{signal_analysis}, we will expand on the general formalism employed in gravitational-wave data analysis.

\subsection{Interferometry}
\label{LVK}
Gravitational-wave interferometers, such as the operational LIGO \cite{aligo2015}, Virgo \cite{Acernese_2014}, and the Kamioka Gravitational Wave Detector (KAGRA) \cite{kagra2019}, as well as the planned Cosmic Explorer \cite{2019BAAS...51g..35R} and Einstein Telescope \cite{Punturo_2010}, rely on the interferometry technique, following the same essential structure used by Michelson and Morley in their historic failed attempt to detect the luminiferous aether. 

\begin{figure}
    \centering
    \includegraphics[width=0.75\linewidth]{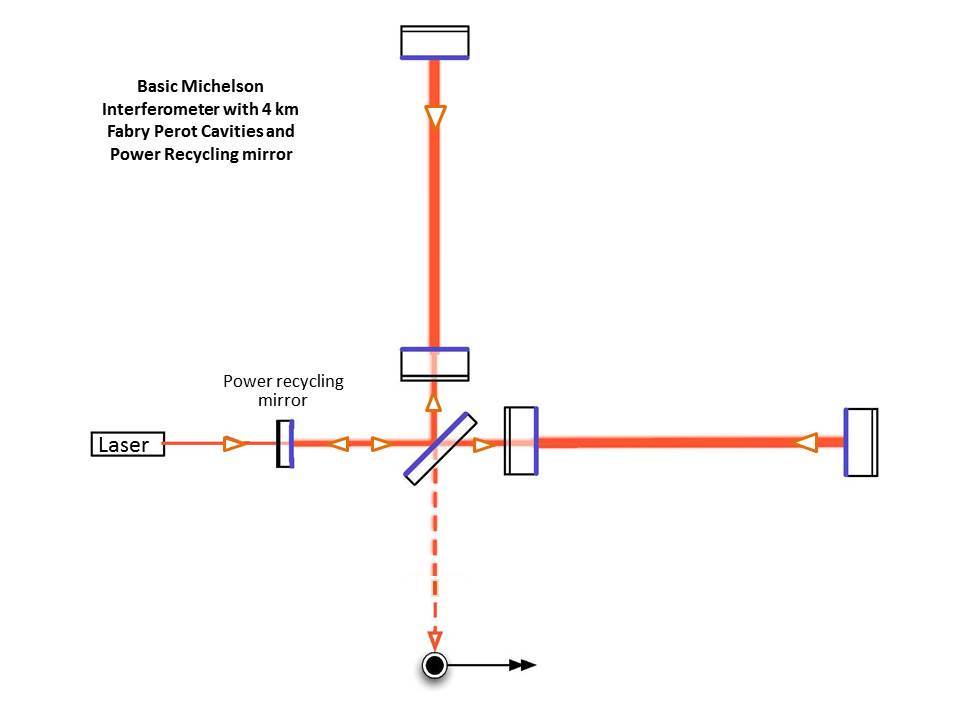}
    \caption{Cartoon illustration of the LIGO interferometer reproduced from Ref. \cite{ligo_interferometer}. Vertical and horizontal rectangles in the figure represent different mirrors, while the tilted rectangle represents the beam splitter. Arrows on the red lines indicate the path taken by the laser light. As noted and represented in the figure, the LIGO interferometer has sophisticated features that make it more than a basic Michelson-Morley interferometer. In the 4-km Fabry-Pérot cavities, the laser beams are reflected back and forth around three hundred times along the arm of the interferometer before being allowed to return to where they will interfere with one another, increasing photons' effective traveled distance and lasers' effective power within each arm. The power recycling mirror further enhances the lasers' power within the Fabry-Pérot cavities by ensuring that almost all light exiting them is redirected back, with only a small fraction being directed to the photodetector that will record the interference pattern.}
    \label{interfig}
\end{figure}

An interferometer's fundamental structure consists of two beams of light emitted in two perpendicular directions. As illustrated in Fig. \ref{interfig}, at a given distance from their emission point, those light beams are reflected back and rejoined at a common point, interfering with one another. Specific variations in time of the interference pattern created by the recombined light beams may serve as a proxy for the passage of a gravitational wave through the interferometer. This is because, given that their paths are perpendicular, a gravitational wave will stretch and contract the space through which each beam propagates differently—as can be seen in Fig. \ref{pluscross}. A change to the length of light's path brought about by the gravitational wave will delay or anticipate light's arrival at the point where it interferes with the other beam, an effect that can be understood as a change in the electromagnetic wave's phase. As a gravitational wave with specific orientation and polarization goes through the interferometer, the phase difference between the perpendicular light beams will change in a specific and predictable way, giving rise to specific and predictable interference patterns—which are what one will directly read off from the experiment and, given what has been just discussed, use as data to look for evidence of the passage of a gravitational wave.

\subsection{Pulsar timing}
\label{PTA}
\begin{figure}
    \centering
    \includegraphics[width=0.5\linewidth]{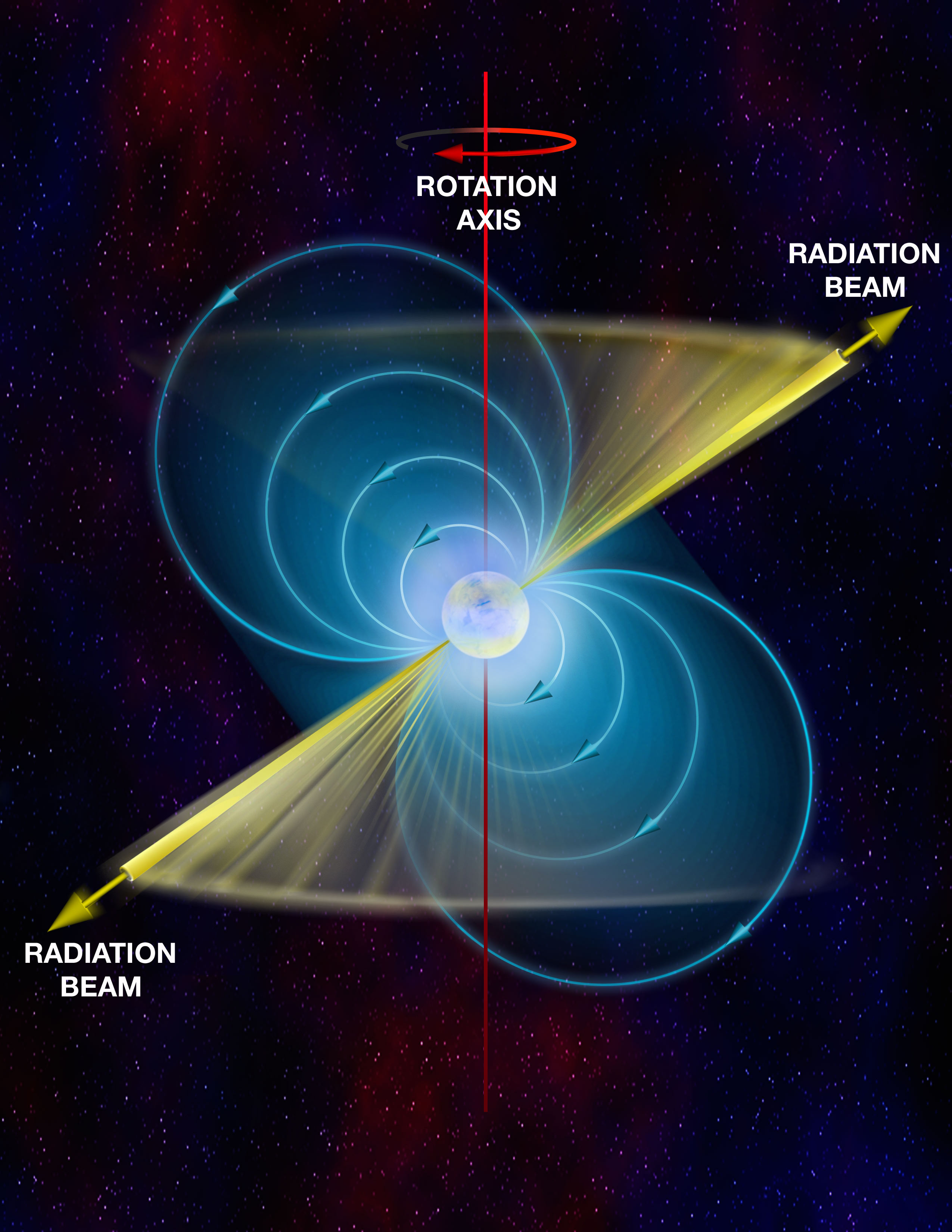}
    \caption{An artist's interpretation of a pulsar reproduced from Ref. \cite{nrao_parts_pulsar}. The white sphere in the center represents the neutron star, while the blue curves represent its magnetic field lines, and the yellow tubes represent its jets of electromagnetic radiation, whose rotation follows that of the star.}
    \label{pulsar}
\end{figure}

Pulsar timing arrays (PTAs), like NANOGrav \citep{NANOGrav:2023gor}, the European Pulsar Timing Array \cite{2016MNRAS.458.3341D, 2023EPTAdiscovery}, the Parkes Pulsar Timing Array \cite{2013PASA...30...17M, 2023ParkesPTA, 2023Parkes_PTADR}, the Chinese Pulsar Timing Array \cite{2023ChinesePTA}, and the MeerKAT Pulsar Timing Array \cite{2020PASA...37...28B}, rely on the precise timing of the periodic electromagnetic signals from pulsars to search for gravitational waves. Pulsars are magnetized neutron stars that rapidly rotate, with periods in the range from $10^{-3}$ to 10 s \cite{2016era..book.....C}. Pulsars steadily emit anisotropic radio jets, as illustrated in Fig. \ref{pulsar}. As they rotate, a radio jet might periodically enter and leave an astronomical observer's field of view like a cosmic lighthouse, giving rise to an observed intensity profile with periodic peaks—as if the source were pulsing. 

The rotation of millisecond pulsars—those with periods of a few ms—is particularly stable and well-measured: the change in their spin periods $\dot{P}$  is $\dot{P}\lesssim10^{-19}$ ss$^{-1}$ \cite{2008LRR....11....8L}. Therefore, if we were observing a pulsar close by, we could predict with high accuracy how much time should elapse from one brightness peak to the next. The fact that pulsars are at least kiloparsecs away from Earth complicates the problem: to predict the pulsar's radiant-flux profile, besides the star itself, we also need to model the propagation of photons from the star to our telescopes, taking into account effects like Shapiro time delay, parallax, and the effect of the dispersion measure (i.e. the integrated column density of free electrons between an observer and a light source) created by the intergalactic medium and solar wind. In this complication, however, also lies the enabler for gravitational-wave searches, since gravitational waves are one of the phenomena that may alter light's travel time from the pulsar to the Earth—and they do so in a predictable way. This is because if a photon, in its way from a pulsar to the Earth, passes through a region where a gravitational wave is traveling, the stretching and contracting of space provoked by the gravitational wave will alter the time spent by the photon to traverse that region. Therefore, deviations of a specific form in the time of the pulses in a pulsar's observed-flux time series may reveal that gravitational waves crossed the path of light from the pulsar to the Earth. Let us take a look at how exactly this happens, closely following Ref. \cite{Maggiore:2018sht} and largely reproducing Section III A of Ref. \cite{alves2024artificialprecisiontimingarray}.
 
Here, we will initially work in $c=1$ units and reinsert factors of $c$ at the end. We consider a coordinate system with its origin at the observer on Earth, in which a pulsar is at position $\Vec{r_s}=d_s\mathbf{\hat{x}}$ on the x-axis. Following the null (i.e. $ds^2=0$) trajectory $\Vec{x}(t)$ of a photon emitted at time $t_e$ from the star to the telescope, 
\begin{equation}
    dx^2=\frac{dt^2}{1+h^{TT}_{xx}(t,\Vec{x}(t))}.
\end{equation}
The photon is traveling from positive $x$ to the origin, so we get the solution moving in the $-x$ direction:
\begin{equation}
    dx=\frac{-dt}{\sqrt{1+h^{TT}_{xx}(t,\Vec{x}(t))}}\approx -\left[1-\frac{1}{2}h^{TT}_{xx}(t,\Vec{x}(t))\right]dt
\end{equation}
to first order in the perturbation $h_{xx}^{TT}$.

When dealing with a single pulsar, one can always choose the coordinate system so that the light source sits along one of the axes. However, to proceed to a result for a light source on an arbitrary direction $\mathbf{\hat p}$, which will be useful when we deal with multiple beacons, we replace $h_{xx}^{TT}$ with $\hat p^i \hat p^j h^{TT}_{ij}$ \cite{Maggiore:2018sht}.

Next, we can integrate from $t_e$ to the observation time $t_o$ so that
\begin{equation}
    d_s=t_o-t_e-\frac{\hat p^i \hat p^j}{2}\int_{t_e}^{t_o}\mathrm{d}t' \;h^{TT}_{ij}(t',\Vec{x}(t')).
\end{equation}
We are concerned with first-order perturbations and the integrand $h^{TT}_{ij}$ is already a first-order term. Therefore, within the integral, we can consider the unperturbed trajectory of the photon,
\begin{subequations}
    \begin{align}
        t_o&=t_e+d_s \\
        \Vec{x_0}(t)&=(t_e-t+d_s)\mathbf{\hat{p}},
    \end{align}
\end{subequations}
such that
\begin{equation}
\label{timing1}
    t_o=t_e+d_s+\frac{\hat p^i \hat p^j}{2}\int_{t_e}^{t_e+d_s}\mathrm{d}t' \;h^{TT}_{ij}(t',(t_e+d_s-t')\mathbf{\hat{p}}).
\end{equation}

The fundamental idea in pulsar timing is to detect gravitational waves through deviations in the time elapsed between observed pulses. Therefore, consider the immediate next pulse, emitted by the satellite at time $t_e+\tau$, where $\tau$ is the rotational period of the pulsar. The second pulse will be observed at $t_o'$. The same derivation that led to Eq. \ref{timing1} gives, by replacing $t_o$ and $t_e$ with $t_o'$ and $t_e+\tau$,
\begin{equation}
\label{timing2}
    t_o'=t_e+\tau+d_s+\frac{\hat p^i \hat p^j}{2}\int_{t_e+\tau}^{t_e+\tau+d_s}\mathrm{d}t' \;h^{TT}_{ij}(t',(t_e+\tau+d_s-t')\mathbf{\hat{p}}).
\end{equation}
Changing the integration variable in Eq. \ref{timing2} as $t'\rightarrow t'-\tau$ and subtracting Eq. \ref{timing1} from it, it becomes clear that the effect of the GW is a pulsing period shift as perceived by the observer: 
\begin{subequations}
\label{deltatau}
\begin{align}
    &t_o'-t_o=\tau+\Delta\tau \\
    &\Delta\tau=\frac{\hat p^i \hat p^j}{2}\int_{t_e}^{t_e+d_s}\mathrm{d}t' \; \left[h^{TT}_{ij}(t'+\tau,\Vec{x_0}(t'))-h^{TT}_{ij}(t',\Vec{x_0}(t'))\right].
\end{align}
\end{subequations}

For the usual PTA, the typical source period $\tau_{p}$ is of order 1 ms while the target GW frequency $\omega_{gw,p}$ is of order $0.1\mathrm{yr^{-1}}$, such that $\omega_{gw,p}\tau_{p}\ll1$. In this regime, we may Taylor expand $h^{TT}_{ij}(t'+\tau,\Vec{x_0}(t'))$ around $t'$ in powers of $\tau$ and discard second-order and higher terms:
\begin{equation}
    h^{TT}_{ij}(t'+\tau,\Vec{x_0}(t'))= h_{ij}^{TT}(t',\Vec{x_0}(t'))+\frac{\partial}{\partial t'}h_{ij}^{TT}(t',\Vec{x_0}(t'))\tau
\end{equation}
so that Eq. \ref{deltatau} becomes
\begin{equation}
\label{approx-integral}
    \frac{\Delta\tau}{\tau}=\frac{\hat p^i \hat p^j}{2}\int_{t_e}^{t_e+d_s}\mathrm{d}t' \; \left[\frac{\partial}{\partial t'}h_{ij}^{TT}(t',\Vec{x})\right]_{\Vec{x}=\Vec{x_0}(t')}.
\end{equation}

To integrate Eq. \ref{approx-integral}, we now specialize to a monochromatic GW—which, depending on the observational timescale, may be a good approximation for several astrophysical sources—propagating along direction $\mathbf{\hat n}$ in TT gauge: 
\begin{equation}
    h_{ij}^{TT}(t,\Vec{x})=\mathcal{A}_{ij}(\mathbf{\hat{n}})\cos{\left[\omega_{gw}(t-\mathbf{\hat{n}}\cdot \Vec{x})\right]},
\end{equation}
With this, Eq. \ref{approx-integral} becomes:
\begin{equation}
\label{pre-template}
    \frac{\Delta\tau}{\tau}=\frac{\hat p^i \hat p^j\mathcal{A}_{ij}}{2(1+\mathbf{\hat p}\cdot\mathbf{\hat n})}\{\cos[\omega_{gw}t_o]-\cos[\omega_{gw}(t_e-d_s\mathbf{\hat p}\cdot\mathbf{\hat n})]\}.
\end{equation}

PTA signal processing can be done with either the so-called redshift, expressed in Eq. \ref{pre-template}, or with its integral, the timing residual $R(t)$:
\begin{equation}   
    \label{template}
    R(t)  = \int_0^t   \mathrm{d} t' \; \frac{\Delta \tau}{\tau}(t') = \frac{\hat p^i \hat p^j\mathcal{A}_{ij}}  {2(1+\mathbf{\hat p}\cdot\mathbf{\hat n})} \frac{1}{\omega_{gw}}  \Big\{\sin{[\omega_{gw}t]} - \sin{\Big[\omega_{gw}\Big(t - \frac{d_s}{c}\Big(1 + \mathbf{\hat p}\cdot\mathbf{\hat n}\Big)\Big) \Big]} \Big\}, 
\end{equation} 
where we have now reinserted factors of $c$ and set $t_o=t'$ and $t_e=t'-d_s/c$, and the constant term from integration was dropped—for PTA data analysis, one will ultimately be interested in the frequency integral of $R(t)$'s Fourier transform, and that constant term will contribute a delta function centered at $f=0$ to the integrand, which will be outside of the range of integration. Without going that deep into the math, it is still possible to develop some good intuition for why the constant term that should be in Eq. \ref{template} can be dropped. In signal processing in general—and surely in gravitational-wave signal processing specifically, as we shall see in Section \ref{signal_analysis}—, it is commonly useful to take a set of data points collected in a sequence of temporal instants—hence forming a time series—and take its Fourier transform to study how oscillatory modes of different frequencies contribute to compose it. Now, a detector can only resolve a mode of vibration if there has been enough time for its oscillatory behavior to unfold, meaning that the duration of the observation $T$ has to contain at least one period of the mode's oscillation, which is $1/f$ for a mode with frequency $f$. This requirement can be understood as a lower bound for the mode frequencies $f_{obs}$ that can be resolved with a detector: $f_{obs}\geq1/T$. Since the frequency-domain representation of a signal that is constant in the time domain is a monochromatic wave with frequency $0$, a PTA would need an infinite amount of time to resolve the constant term's contribution to the Fourier transform of the signal $R(t)$, such that it will be invisible in real observations with a finite duration.

\subsection{General gravitational-wave signal analysis}
\label{signal_analysis}
Now, with a more concrete idea of how an experiment can be concocted for the detection of gravitational waves in mind, let us take a step back and consider the general mathematical formalism underlying gravitational-wave data analysis. Plots describing the emission of gravitational waves by different sources and the sensitivity of different detectors abound in the astrophysics and cosmology literature. Different quantities, like the spectral amplitude, the characteristic strain, and the energy density, are commonly used to convey the same basic information—the intensity of gravitational waves produced or detectable—in different ways. The goal of this section is to mathematically spell out the precise meaning of those different quantities, in preparation for our later goal of quantitatively describing gravitational waves produced by cosmic inflation and observational constraints on them imposed by existing results from PTAs and interferometers. 

The physical quantity one is interested in measuring, $h_{ij}^{TT}(t)$, is a tensor, while the output of a detector is a time series, a scalar function $h(t)$. Surely, the output signal $h(t)$ is a function of what causes it, $h_{ij}^{TT}(t)$. This relationship will be commonly linear: $h(t)=D^{ij}h_{ij}^{TT}(t)$, where the so-called detector tensor $D^{ij}$ is a constant that depends on the detector's geometry.

Of course, the gravitational-wave signal of interest will not be the only input affecting the data collected by the detector. For example, if a person starts jumping right next to LIGO's laser, the vibrations caused on the ground will be transmitted to the laser, affecting the phase of the emitted light and, hence, the interference pattern in which gravitational-wave signals are searched for—this is an extreme and silly example that would completely hinder LIGO's search for binary-compact-object-merger gravitational waves; these are so faint that, being sensitive to them, the detector is also sensitive to tectonics, wind outside the interferometer, and even trucks passing in nearby roads and people talking in LIGO's facilities during working hours. Whatever processes affecting what is read off from the detector that are not the physical phenomena one is interested in observing—gravitational waves in this case—are understood as noise. Then, the full time-series data stream $d(t)$ recorded by the detector may be understood as a superposition of the signal of interest $h(t)$ and noise $n(t)$:
\begin{equation}
\label{dhn}
    d(t)=h(t)+n(t).
\end{equation}

An important descriptor of the noise's behavior is its autocorrelation function $\mathcal{R}(\tau;t)\equiv\braket{n(t+\tau)n(t)}$, where $\braket{\cdot}$ denotes an ensemble average. $\mathcal{R}$ describes the correlation between the element at time $t$ of the noise time series with the one at time $t+\tau$, after a time interval $\tau$ has elapsed. In general, it is reasonable to expect that, if $\tau$ is very large, the statistical connection between $n(t)$ and $n(t+\tau)$ will be very weak. More specifically, it is reasonable to expect that $\mathcal{R}$ will decay sufficiently fast when $|\tau|\rightarrow\infty$ for its Fourier transform to be well-defined:
\begin{equation}
\label{defPSD}
    \frac{1}{2}S_n(f)\equiv\int_{-\infty}^\infty \mathrm{d}\tau \;\mathcal{R}(\tau;t)e^{2\pi i f\tau}.
\end{equation}
$S_n(f)$ is called the (one-sided) noise power spectral density (PSD), and we are about to see why it deserves its own name. Before that, let us note in passing that its square root $\sqrt{S_n(f)}$, called the spectral amplitude or amplitude spectral density, is one common representation of a detector's sensitivity as a function of frequency. $n(t)$ is dimensionless, so $\mathcal{R}$ is dimensionless and the spectral amplitude has units of $\mathrm{Hz}^{-1/2}$.

The ensemble average in the definition of $\mathcal{R}$ is an average over realizations of the system. This poses a problem since only the single, physical realization of the detector is available in reality. However, we can assume for simplicity that the noise is stochastic and stationary—as it turns out, this will often be the case in reality. That means that the noise's statistical properties do not change with time, so that the noise autocorrelation function is $t$-independent. Consequently, $\mathcal{R}$ may be computed from an average over time: given that the noise's statistical properties do not change in time, subsets of the full time series with the same duration may function as the different realizations of the system needed for the ensemble average. Another important implication of assuming stationary noise is the invariance of $\mathcal{R}$ under time translations, which implies $\mathcal{R}(-\tau)=\braket{n(t-\tau)n(t)}=\braket{n(t-\tau+\tau)n(t+\tau)}=\mathcal{R}(\tau)$. As a result, $S_n(f)=S_n(-f)$. Together with the reality of $\mathcal{R}(\tau)$, from which it follows that $S_n(-f)=S_n^*(f)$, that also implies the reality of $S_n(f)$. Still exploring that former result, we see from inverting Eq. \ref{defPSD} that
\begin{equation}
\label{r0}
    R(0)=\braket{n(t)^2}=\braket{n(0)^2}=\frac{1}{2}\int_{-\infty}^\infty\mathrm{d}f\;S_n(f) =\int_0^\infty \mathrm{d}f \; S_n(f),
\end{equation}
where the second equality came from the assumption that the noise is stationary. This tells us that the integral of the (one-sided) noise PSD over all physical (positive) frequencies gives the average of the square of the noise, which is a constant in time under our assumption of stationarity. Eq. \ref{defPSD} is the more general and rigorous definition of the noise PSD $S_n(f)$. We can derive another, more restricted definition, which is nonetheless commonly found in the literature, from Eq. \ref{r0}, valid when the Fourier transform of $n(t)$ is well-defined—which need not be the case, since the noise need not decay as $|t|\rightarrow\infty$:
\begin{align}
\label{PSDrestrictedintegral}
    &\braket{n(0)^2}=\int_{-\infty}^\infty\mathrm{d}f\int_{-\infty}^\infty\mathrm{d}f'\braket{\tilde n(f')\tilde n^*(f)}=\frac{1}{2}\int_{-\infty}^\infty\mathrm{d}f\;S_n(f) \\
    \label{PSDrestricted}
    &\therefore\braket{\tilde n(f')\tilde n^*(f)}=\frac{1}{2}S_n(f)\delta(f-f').
\end{align}
This tells us that Fourier modes of stationary noise are uncorrelated, and that the average of the squares of the modes is proportional to the noise PSD.

An important challenge in gravitational-wave astronomy is that, in spite of the unprecedented levels of isolation from noise achieved in detectors like LIGO, Virgo, and KAGRA, the signal of interest is so faint that in practice $|h(t)|\ll|n(t)|$. But the detection of gravitational waves has been successful due to an ingenious data processing technique, matched filtering, which enables the extraction of a low signal from loud noise if the signal $h(t)$ is known beforehand. This works by convolving the signal with a real filter function $K(t)$ of Fourier transform $\tilde K(f)$,
\begin{equation}
    \hat d=\int^\infty_{-\infty}\mathrm{d}t\;d(t)K(t),
\end{equation}
choosing the filter function so as to maximize the signal-to-noise ratio $\mathcal{S/N}$. $\mathcal{S}$ is defined as the expectation value of $\hat d$ when the signal is present, while $\mathcal{N}$ is defined as the root-mean-square value of $\hat d$ when the signal is absent ($h(t)=0$):
\begin{equation}
\label{signal}
    \mathcal{S}=\int^\infty_{-\infty}\mathrm{d}t\;K(t)\braket{d(t)}=\int^\infty_{-\infty}\mathrm{d}t\;K^*(t)h(t)=\int^\infty_{-\infty}\mathrm{d}f\;\tilde K^*(f)\tilde h(f) 
\end{equation}
\begin{equation}
    \label{noise}
    \begin{aligned}
    \mathcal{N}^2&=\braket{n(t)^2}=\int^\infty_{-\infty}\int^\infty_{-\infty}\mathrm{d}t\mathrm{d}t'\;K(t)K(t')\braket{n(t)n(t')}\\
    &=\int^\infty_{-\infty}\int^\infty_{-\infty}\mathrm{d}t\mathrm{d}t'\;K(t)K^*(t')\int^\infty_{-\infty}\int^\infty_{-\infty}\mathrm{d}f\mathrm{d}f'\braket{\hat n^*(f)\hat n(f')}e^{2\pi i(ft-f't')} \\
    &=\frac{1}{2}\int_{-\infty}^\infty\mathrm{d}f\;S_n(f)\tilde K(f)\tilde K^*(f),
    \end{aligned}
\end{equation}
where we have used that $\braket{n(t)}\equiv0$ (a possible definition since the noise is stationary), $K(t)\in \mathbb{R}$, $\int_{-\infty}^\infty\mathrm{d}t\; \exp{(2\pi i (f-f')t)}=\delta(f-f')$, and Eq. \ref{PSDrestricted}. The expression for $\mathcal{S}^2/\mathcal{N}^2$ may be cast into a simple form by defining the following inner product \cite{Finn_1992}:
\begin{align}
    &(\tilde A(f)|\tilde B(f))\equiv 2\: \mathrm{Re}\left[\int_{-\infty}^\infty\mathrm{d}f\;\frac{\tilde A^*(f)\tilde B(f)}{S_n(f)}\right]\\
    &\therefore \frac{\mathcal{S}^2}{\mathcal{N}^2}=\frac{(\tilde K(f)S_n(f)|\tilde h(f))^2}{(\tilde K(f)S_n(f)|\tilde K(f)S_n(f))}.
\end{align}
According to the Cauchy-Schwarz inequality, a general property of inner products, $|(\tilde A|\tilde B)|^2\leq(\tilde A|\tilde A)(\tilde B|\tilde B)$, with equality happening if and only if the entries $\tilde A$ and $\tilde B$ are the same up to a constant scalar factor $a$, $\tilde A=a\tilde B$. Consequently, $\mathcal{S}^2/\mathcal{N}^2\leq2(\tilde h(f)|\tilde h(f))$. In order to maximize the signal-to-noise ratio, we therefore need $\tilde K(f)S_n(f)\propto \tilde h(f)$. So, we have found that the filter function that will optimally extract the signal from the noise is $\tilde K(f)= a\: \tilde h(f)/S_n(f)$ ($a$ is arbitrary in this case since it will cancel out in the expression for $\mathcal{S}/\mathcal{N}$). The signal-to-noise ratio then simplifies to:
\begin{equation}
\label{SNR1}
    \frac{\mathcal{S}^2}{\mathcal{N}^2}=(\tilde h(f)|\tilde h(f))=2\:\mathrm{Re}\int_{-\infty}^\infty\mathrm{d}f\;\frac{|\tilde h(f)|^2}{S_n(f)}=4\int_{0}^\infty\mathrm{d}f\;\frac{|\tilde h(f)|^2}{S_n(f)},
\end{equation}
where the last equality comes from the already established reality and evenness of $S_n(f)$ and by the relationship $|\tilde h(f)|^2=\tilde h(f)\tilde h^*(f)=\tilde h^*(-f)\tilde h(-f)$ implied by the reality of $h(t)$. Eq. \ref{SNR1} may be recast into  an enlightening form by defining the characteristic strain or characteristic amplitude $h_c$ and the noise amplitude $h_n$:
\begin{align}
    \label{characteristics}
    &h_c(f)\equiv2f|\tilde h(f)|;\;h_n(f)\equiv\sqrt{fS_n(f)} \\
    \label{SNRintegral}
    &\therefore\frac{\mathcal{S}^2}{\mathcal{N}^2}=\int_{-\infty}^\infty\mathrm{d}(\ln f)\left[\frac{h_c(f)}{h_n(f)}\right]^2.
\end{align}
This way, the area between the source and detector curves on a log-log scale is related to the signal-to-noise ratio by Eq. \ref{SNRintegral}, allowing the reader to develop a qualitative idea of a source's detectability by simply eyeballing how far on top of the $h_n(f)$ curve the $h_c(f)$ curve sits. Moreover, inverting the relationship for $h_n(f)$ in Eq. \ref{characteristics}, we get that the spectral amplitude is $\sqrt{S_n(f)}=h_n(f)f^{-1/2}$. We may then define an analogous quantity describing not the noise but the signal of interest as $\sqrt{S_h(f)}\equiv h_c(f)f^{-1/2}$ \cite{Moore_2014}. 

Besides discussing the detectability of gravitational waves in terms of strain, it is also common to do so in terms of their energy density $\rho_{gw}$ using Eq. \ref{energyinGWs}—as elaborated in Section \ref{introcosmo}, $\rho_{gw}$ comes from the $00$-component of $t_{\mu\nu}$. As a starting point to that end, let us write the generalization of the solution in Eq. \ref{onewave} to the wave equation by superposing linearly independent plane waves:
\begin{equation}
\label{hijexpansion}
\begin{aligned}
    h_{ij}^{TT}(x)&=\int\frac{\mathrm d^3k}{(2\pi)^3}[\mathcal A_{ij}(\vec k)e^{ik_\mu x^\mu}+\mathcal A^*_{ij}(\vec k)e^{-ik_\mu x^\mu}]\\
    &=\int_0^\infty\mathrm df \;f^2\int_{S^2}\mathrm d\hat n\;[\mathcal A_{ij}(f,\hat n)e^{-2\pi if(t-\hat n\cdot\vec x)}+\mathcal A^*_{ij}(f,\hat n)e^{2\pi if(t-\hat n\cdot\vec x)}],
\end{aligned}
\end{equation}
where again $\mathcal A^k_{\;k}=k^j\mathcal{A}_{jk}=0$ due to the choice of transverse-traceless gauge and $k^0=|\vec k|=\omega=2\pi f$ such that $k^\nu k_\nu=0$, ensuring compliance with the equation of motion; $\hat n\equiv\vec k/|\vec k|$. The transversality condition implies that perturbations to the geometry of space caused by the gravitational wave only take place in the plane whose normal is $\hat n$, i.e. components of $h_{ij}^{TT}$ in the direction of $\hat n$ are $0$. We can omit these null components and replace $h_{ij}^{TT}$ with $h_{ab}^{TT}$, where $a,b\in\{1,2\}$ refer to two directions that span that transverse plane. Since this choice of directions in the transverse plane is arbitrary, there may exist a polarization angle $\psi$ between one's choice of axes and the axes defined by a gravitational wave's $+$ and $\times$ polarizations. We can introduce the polarization tensors $e_{ab}^A(\hat n)$, $A\in\{+,\times\}$, to write $h_{ab}^{TT}$ in a general form for any polarization angle:
\begin{equation}
    e_{ab}^+(\hat n)=\hat u_a\hat u_b-\mathrm{\hat v}_a\mathrm{\hat v}_b;\; e_{ab}^\times(\hat n)=\hat u_a\mathrm{\hat v}_b+\mathrm{\hat v}_a\hat u_b,
\end{equation}
where $\hat u$, $\mathrm{\hat v}$, and $\hat n$ form an orthonormal set. One can easily verify that this definition is normalized as $e_{ab}^A(\hat n)e_{A'}^{ab}(\hat n)=2\delta^A_{A'}$. Moreover, this definition reproduces Eq. \ref{matrixeq} for $\hat n=\hat z$, $\hat u=\hat x$ and $\mathrm{\hat v}=\hat y$, because then $e^{+}=\begin{bmatrix}
    1&0\\0&-1
\end{bmatrix}$ and $e^{\times}=\begin{bmatrix}
    0 & 1\\1& 0
\end{bmatrix}$. Hence, for a general polarization angle, we may define the Fourier amplitudes of the gravitational wave components as \cite{Maggiore1}:
\begin{equation}
    f^2\mathcal A_{ab}(f,\hat n)\equiv\sum_{A=+,\times}\tilde h_A(f,\hat n)e^A_{ab}(\hat n).
\end{equation}
It is useful to define $\tilde h$ for negative frequencies as $\tilde h_A(-f,\hat n)=\tilde h^*_A(f,\hat n)$ to render Eq.~\ref{hijexpansion} in a Fourier-transform-like form:
\begin{equation}
    h_{ab}^{TT}(t,\vec x)=\int_{-\infty}^\infty\mathrm df\int_{S^2}\mathrm d\hat n \sum_{A=+,\times}\tilde h_A(f,\hat n)e_{ab}^A(\hat n)e^{-2\pi if(t-\hat n\cdot \vec x)}.
\end{equation}
This form of the plane-wave expansion of transverse-traceless gravitational waves, which is very common in the literature, will now come in handy to evaluate $\rho_{gw}=t_{00}$ from Eq.~\ref{energyinGWs}:
\begin{equation}
\begin{aligned}
    &32\pi G\rho_{gw}=\left\langle\dot h^{ab}_{TT}\dot h^{TT}_{ab}\right\rangle=\left\langle\dot h_{TT}^{ab}\left(\dot h^{TT}_{ab}\right)^*\right\rangle=\\
    &\int_{-\infty}^\infty\mathrm df\int_{-\infty}^\infty\mathrm df'\int_{S^2}\mathrm d\hat n\int_{S^2}\mathrm d\hat n' \sum_{A,A'}\braket{\tilde h_A(f,\hat n)\tilde h^*_{A'}(f',\hat n')}e^{ab}_A(\hat n)e_{ab}^{A'}(\hat n')4\pi^2ff'e^{2\pi i[t(f'-f)-\vec x\cdot(\hat n'-\hat n)]},
\end{aligned}
\end{equation}
where the second equality comes from the reality of $h_{ab}^{TT}(t,\vec x)$. In analogy with Eq. \ref{PSDrestrictedintegral}, we would like to have 
\begin{equation*}
   \int_{-\infty}^\infty\mathrm df\int_{-\infty}^\infty\mathrm df'\int_{S^2}\mathrm d\hat n\int_{S^2}\mathrm d\hat n' \sum_{A,A'} \braket{\tilde h_A(f,\hat n)\tilde h^*_{A'}(f',\hat n')}=\int_{0}^\infty\mathrm df\;S_h(f).
\end{equation*}
Hence, for a collection of superposed stochastic gravitational waves—often called a stochastic background—that is stationary, unpolarized, isotropic, and Gaussian \cite{mingarelli2025understandingomegamathrmgwfgravitationalwave,flannagansgwb}:
\begin{equation}
\label{correlationAnf}
    \braket{\tilde h_A(f,\hat n)\tilde h^*_{A'}(f',\hat n')}=\frac{\delta_{AA'}}{2}\frac{\delta(\hat n-\hat n')}{4\pi}\frac{\delta(f-f')}{2}S_h(f).
\end{equation}
The last two terms involving the frequency are analogous to our result in Eq. \ref{PSDrestricted} for the noise. The term involving $A$ expresses the fact that waves with different polarizations are uncorrelated, and similarly for the term involving the direction of propagation $\hat n$. Noting that $\sum_A e^A_{ab}e^{ab}_A=4$, we can finally write
\begin{equation}
\label{rhogw}
    \rho_{gw}=\int_0^\infty\mathrm df\;\frac{\pi f^2}{4G}S_h(f),
\end{equation}
where it has been used, like before, that $S_h(f)$ is even. The integrand in Eq. \ref{rhogw} is the energy per unit volume per unit frequency in gravitational waves, often called the spectral energy density $S_E(f)$. As we shall see in Section \ref{cosmochapter}, there exists a critical value for the total energy density contained in the universe that will make the current constant-time slice of the universe exactly flat, which is $\rho_c=3H_0/8\pi G$, where the Hubble constant $H_0$ describes the current rate of expansion of the universe. It is common in cosmology to form ratios between $\rho_c$ and the energy density $\rho_s$ of a certain species $s$ (e.g. photons or dark energy) as $\Omega_s\equiv\rho_s/\rho_c$. For gravitational waves, however, it is more common to define 
\begin{equation}
\label{omegagw}
    \Omega_{gw}(f)\equiv\frac{1}{\rho_c}\frac{d\rho_{gw}}{d(\ln f)}=\frac{fS_E(f)}{\rho_c}=\frac{2\pi^2f^2h_c(f)^2}{3H_0},
\end{equation}
where the last equality, which is recurrent in the literature and important for our later purposes, comes from the definition of $S_h$ in the paragraph immediately after Eq. \ref{SNRintegral}.

\section{Gravitational-wave background from a distribution of binaries}
\label{2/3sec}
The focus of this dissertation is the study of observables related to gravitational waves produced by early-universe processes—for us, more specifically, cosmic inflation. We will be mainly interested in models that can match the $\Omega_{gw}(f)$ for a stochastic gravitational-wave background for which NANOGrav has been accumulating statistical evidence. Although a primordial origin has been among the systematically discussed proposals to explain the origin of the NANOGrav signal, the community has also been entertaining—perhaps as its top candidate, at least for some considerable time—the superposition of gravitational waves from binaries of compact objects from all sky as responsible for the background PTAs are starting to see. Therefore, before starting our discussion of cosmology, let us derive some relevant results regarding a gravitational-wave background from binary systems.

The formalism for gravitational-wave generation from the second time derivative of the source's mass quadrupole moment is rich and interesting, but, to obtain the results we ultimately want, we may take a much simpler route, as long as we accept without deriving one well-known result: that the gravitational-wave frequency is twice the orbital frequency of the circular binary that produces it. Consider a binary of point masses $m_1$ and $m_2$ respectively at positions $\vec r_1$ and $\vec r_2$ in circular orbits—bound binaries in the universe generally have small eccentricity, so this will serve as a good approximation. A well-known result in Newtonian mechanics is that each of the two masses will describe circular trajectories around their center of mass $\vec r_{CM}=(m_1\vec r_1+m_2\vec r_2)/M$, where $M\equiv m_1+m_2$ is the total mass. Conveniently setting $\vec r_{CM}$ as the origin of the coordinate system and defining $\vec r\equiv\vec r_1-\vec r_2$, we have
\begin{equation}\label{kepler1}\vec r_1=\frac{m_2}{M}\vec r.\end{equation}
In this Newtonian approximation we are carrying out, we may then write $m_1$'s centripetal acceleration:
\begin{equation}
    \label{kepler2}\frac{Gm_2}{r^2}=\frac{v_1^2}{r_1}=\omega^2 r_1,
\end{equation}
where we wrote $m_1$'s angular frequency $\omega$ without a subscript since, as it turns out, $\omega_1=\omega_2$ in this setting. Plugging Eq. \ref{kepler1} into Eq. \ref{kepler2}, we obtain Kepler's third law:
\begin{equation}
\label{kepler}
    \omega^2=\frac{GM}{r^3}.
\end{equation}
Making use of Eqs. \ref{kepler1}–\ref{kepler}, the total energy of the binary is:
\begin{equation}
\label{energybinary}
    E_{b}=\frac{1}{2}m_1v_1^2+\frac{1}{2}m_2v_2^2-\frac{Gm_1m_2}{r}=-\frac{1}{2}\frac{Gm_1m_2}{r}.
\end{equation}
It is common to define the chirp mass $\mathcal M\equiv (m_1m_2)^{3/5}/M^{1/5}$ and employ Eq. \ref{kepler} to rewrite Eq. \ref{energybinary} as
\begin{equation}
    E=-\left(\frac{G^2\mathcal M^5\omega_{gw}^2}{32}\right)^\frac{1}{3}=-\left(\frac{\pi^2G^2\mathcal M^5f_{gw}^2}{8}\right)^\frac{1}{3},
\end{equation}
where we have made use of $\omega_{gw}=2\omega$.

In a vacuum and, consequently, in the absence of friction and similar dissipative effects, Newtonian physics would predict that $E$ would remain constant, and the system would be stationary with a constant orbital separation. However, as we have discussed toward the end of Section \ref{GWderivation}, gravitational waves have an associated energy density. Consequently, as the binary radiates and its energy is dissipated in the form of gravitational waves, its orbital separation decreases (see Eq. \ref{energybinary}) and its orbital frequency increases (see Eq. \ref{kepler}). It is therefore the case that the energy lost by the binary as it evolves from frequency $f_s$ to frequency $f_s+df_s$ is precisely the energy carried by the gravitational waves thus produced:
\begin{equation}
\label{dEgwdf}
    \frac{dE_{gw}}{df_s}df_s=\frac{\pi}{3}\left(\frac{G^2\mathcal M^5}{\pi f_s}\right)^\frac{1}{3}df_s.
\end{equation}
We have started using the subscript $s$ for the frequency to indicate we are dealing with frequencies in the frame of the source. As we will discuss in Section \ref{cosmochapter}, the universe is expanding, such that all points in it are getting farther away from all other points. Effectively, then, astronomical objects that follow the expansion of the universe, moving with the so-called Hubble flow, develop a receding relative velocity with respect to one another. And, if the source of a wave is moving away, then the wave's frequency is Doppler shifted down—it is, using astronomical terms, redshifted (the reference to red comes from it being the color of visible light with the lowest frequency). The frequency $f$ observed by an astronomical observer far away from the source is usually parametrized as $f=f_s/(1+z)$, where $z\geq0$ is the source's redshift.

With Eq. \ref{dEgwdf} at hand, we can now reproduce the result due to Ref. \cite{Phinney:2001di} for the spectrum of stochastic gravitational waves produced by the binaries distributed across the cosmos. Consider that radiating binaries occur with a number density per unit redshift $N(z)$. As we will discuss in Section \ref{cosmochapter}, our universe is isotropic and homogeneous. As a consequence, the $\rho_{gw}$ that reaches observers on Earth must equal the sum of energy densities emitted at each redshift:
\begin{equation}
\label{rhophinney}
    \rho_{gw}=\int_0^\infty\mathrm dz\int_0^\infty\mathrm df_s\; N(z) \frac{dE_{gw}}{df_s}\frac{1}{1+z},
\end{equation}
where the factor of $(1+z)^{-1}$ accounts for the redshifting of the gravitons' energies. Then, we may equate Eqs. \ref{rhophinney} and \ref{rhogw} and use Eq. \ref{omegagw} to change variables from $S_h$ to $\Omega_{gw}$:
\begin{equation}
\label{lucasmartins}
    \int_0^\infty\mathrm dz\int_0^\infty\mathrm df_s\; N(z) \frac{dE_{gw}}{df_s}\frac{1}{1+z}=\int_0^\infty\frac{\mathrm d f}{f}\rho_c\Omega_{gw}(f).
\end{equation}
We may insert a factor of $f_s/f_s$ in the integrand of the left-hand side of Eq. \ref{lucasmartins} and use the fact that $f_s^{-1}\mathrm df_s=f^{-1}\mathrm df$ to conclude that
\begin{equation}
\label{omegagwphinney}
    \Omega_{gw}(f)=\frac{1}{\rho_c}\int_0^\infty\mathrm dz\;\frac{N(z)}{1+z} \left.\frac{dE_{gw}}{df_s}f_s\right|_{f_s=f(1+z)}.
\end{equation}
Plugging Eq. \ref{dEgwdf} into Eq. \ref{omegagwphinney}, the result is:
\begin{equation}
\label{phinney}
    \Omega_{gw}(f)=\frac{8\pi^{5/3}}{9H_0^2}(G\mathcal M)^{5/3}f^{2/3}\int_0^\infty\mathrm d z\;\frac{N(z)}{(1+z)^{1/3}}.
\end{equation}
The key piece of information we have gained by deriving Eq. \ref{phinney} is that $\Omega_{gw}(f)\propto f^{2/3}$ for the stochastic gravitational-wave background produced by a homogeneous and isotropic distribution of binaries. This power-law spectral index of $2/3$ is very commonly cited in the literature when studying this source's contribution to the stochastic background of gravitational waves.
\chapter{\textsc{Inflationary cosmology}}
\label{cosmochapter}
\section{Introduction to cosmology}
\label{introcosmo}
Cosmology is the branch of science concerned with the study of the universe considered as a system. Hence, it involves the study of the origin, evolution, composition, and shape of everything that exists considered as a whole—or, perhaps, of the region we can in principle observe, the observable universe, and its content considered as a whole (it is completely reasonable to consider the possibility that what we can in principle access empirically is but a fraction of the entirety of the cosmos). One of the cornerstones of modern physical cosmology is the cosmological principle, the observation that, at large scales, the universe is homogeneous and isotropic. Let us take a closer look at what this means. 

\begin{figure}
    \centering
    \includegraphics[width=0.95\linewidth]{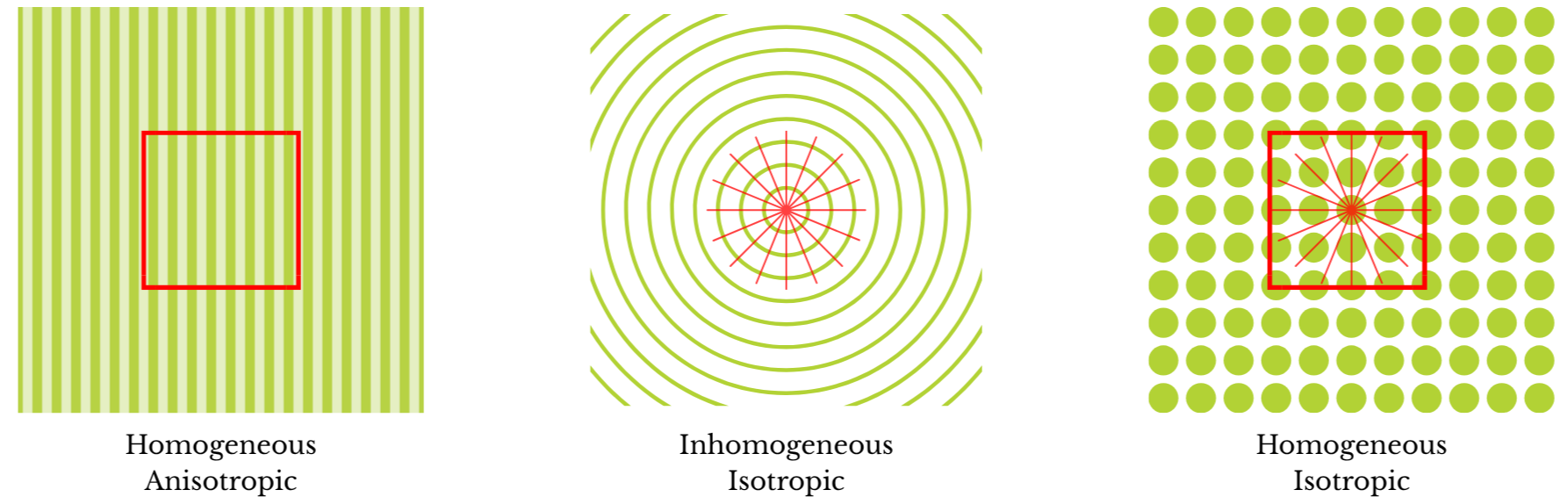}
    \caption{Two-dimensional patterns in green, with drawings in red that help convey points about the observer's viewpoint. The pattern on the left is homogeneous when considering scales larger than the stripe width, as the one demarcated by the red square, for otherwise there is variation between regions within light and dark stripes. The vertical direction is preferential, so it is not isotropic. The pattern in the middle is isotropic around its center since, as highlighted by the radial lines in red, the alternation of green and white along any direction extending outward from the center is exactly equal. It is, however, inhomogeneous: imagine dragging a square like the red one in the left panel around this figure and easily convince yourself that the pattern bounded by it changes around the figure. The pattern on the right is both homogeneous and isotropic in light of what has already been discussed for the first two panels.}
    \label{homoiso}
\end{figure}

Homogeneity characterizes those things that do not have preferred regions along their extension, that is, things whose attributes do not change from place to place. Isotropy is the property of being direction-independent, such that attributes of something isotropic will be the same regardless of the direction along which one looks. Nothing is better than a drawing to explain the subtle difference between these concepts, and this is done in Fig. \ref{homoiso}, which depicts in green different two-dimensional patterns that are homogeneous, isotropic, or both.

One might be confused by the statement that the universe we live in is homogeneous and isotropic. The Earth, a highly complex, inhomogeneous, and anisotropic system, is surrounded by large regions of near vacuum and has neighbors, like the moon, the sun, Mars, and Venus, that constitute abrupt and huge departures from that near vacuum, being also very different from one another in volume, mass, composition, etc. As clarified in the caption of Fig. \ref{homoiso}, homogeneity and isotropy may be scale-dependent notions. This is certainly the case for the cosmological principle: not only is the universe highly inhomogeneous and anisotropic at solar-system-sized scales, it remains so even at the galaxy-cluster-sized scales; the cosmological principle is a good description of the universe at the scale of superclusters, of hundreds of megaparsecs ($1\mathrm{\;Mpc}\approx3.1\times10^{19}\mathrm{\;km}$; the aphelion of Neptune, the outermost solar-system planet, is of $\sim1.5\times10^{-10}\mathrm{\; Mpc}$).

In general relativity, the set of possible spacetime metrics is considerably restricted by the cosmological principle, which requires that, for our universe's spacetime, there must exist a slicing of it in a series of constant-time-$t$ isotropic and homogeneous spatial hypersurfaces—this slicing constitutes the so-called comoving reference frame. Cross terms of the form $dtdx^i$ and $dx^idx^j$, $i\neq j$, need to be absent due to isotropy: 
\begin{equation}
    \label{protoFRW}
    ds^2=-A(t,\vec x)dt^2+B_{ij}(t,\vec x)dx^idx^j,
\end{equation}
with $B_{ij}=0$ if $i\neq j$. To see this, suppose, for example, the line element were $ds^2=-dt^2+dtdy+dx^2+dy^2+dz^2$. In this case, the spacetime interval traversed by a particle moving along the $y$ direction would be different from that traversed by a particle moving along the $-y$ direction, in clear violation of isotropy. 

The spatial components of the four-velocity of comoving observers $u^i$ have to be zero, otherwise a preferred direction would be picked, also in violation of isotropy. From their normalization $u^\mu u_\mu=-1$, it follows that $u^0=A^{-1/2}$. We can then construct their acceleration from $a^\mu=u^\nu\nabla_\nu u^\mu=du^\mu/d\tau+\Gamma^\mu_{\nu\sigma} u^\nu u^\sigma$. From the constraints already in place on the four-velocity and the metric, we know that $a^i=\Gamma^i_{00}/A=(1/2A)B^{ij}\partial A/\partial x^j$. Again, to prevent the appearance of a three-vector field picking a preferred direction in space, it must be true that $a^i=0$, implying that $A(t,\vec x)=A(t)$. Hence, we can redefine $t$ by absorbing $A(t)$ into it through $dt\equiv\sqrt{A(t')} dt'$, meaning that the proper time measured by the comoving observer's clock—which is $\vec x$-independent—will be used as the temporal coordinate. 

It must also be true that the coefficient of $dx^idx^j$ is separable in functions of $t$ and $\vec x$, $B(t,\vec x)\equiv a(t)^2 \gamma_{ij}(\vec x)$, since otherwise the time-dependent behavior would stretch/contract different regions or directions of space differently, such that an initially homogeneous and isotropic solution would be driven away from homogeneity and isotropy by time evolution. Since we are working with an isotropic space, proceeding in spherical coordinates $r$, $\theta$, and $\varphi$ will allow for extra simplification of the result's form: $\gamma_{ij}(\vec x)dx^idx^j\equiv\zeta(r) dr^2+\psi (r)^2d\theta^2+\chi(r)^2\sin^2\theta d\varphi^2$. If $\psi\neq\chi$, the length of meridians would differ from that of the equator on the same spherical shell, rendering constant-$r$ surfaces oblate. Therefore, a truly spherical slicing of space, where isotropy will manifest itself as invariance of physical quantities on constant-$r$ surfaces, will have $\psi=\chi$. We could work with a general $\psi(r)$ without breaking isotropy, but it is convenient to perform the redefinitions $\psi(r)\rightarrow r$ and $\zeta(r)/\psi'(r)^2\rightarrow\zeta(r)$, so as to have $r$ be the aeral radius. Finally, to determine $\zeta$, we invoke the homogeneity requirement by imposing that the Ricci curvature scalar of spatial hypersurfaces, $^3R$, is constant. Applying the process summarized in Section \ref{GRprimer} to go from the spatial metric to $^3R$, we arrive at $^3R=2\{1-d[r/\zeta(r)]/dr\}/r^2\stackrel{!}{\equiv}\text{constant}\equiv6k$. This is easily integrated to $\zeta(r)=[1-kr^2+C/r]^{-1}$, where C is an integration constant that is set to $0$ to avoid a singularity at the origin. (Note that the singularity at $r=k^{-1/2}$ for $k>0$ is a coordinate singularity, removable e.g. by the coordinate transformation $r=k^{-1/2}\sin\beta$.) Our final result is the famous Friedmann-Lemaître-Robertson-Walker (FLRW, often FRW) line element describing a homogeneous and isotropic spacetime like our universe: 
\begin{equation}
\label{frwmetric}
ds^2=-dt^2+a(t)^2\left[\frac{dr^2}{1-kr^2}+r^2\left(d\theta^2+\sin^2\theta d\varphi^2\right)\right].
\end{equation}
$k=0$ describes a flat universe, while $k<0$ describes a constant-negative-curvature universe, like a hyperboloid, and $k>0$, a constant-positive-curvature universe (a 2-sphere is an example of a two-dimensional geometry with constant positive curvature). The former two geometries are open and have an infinite volume, while the latter is closed and has a finite volume. $a(t)$ is usually called the scale factor; it is the factor scaling all spatial distances as time passes: the physical/proper distance between any two points $\vec x$ and $\vec y$ at time $t_2$ will be scaled by a factor of $a(t_2)/a(t_1)$ when compared to that at time $t_1$.

\begin{figure}
    \centering
    \includegraphics[width=0.75\linewidth]{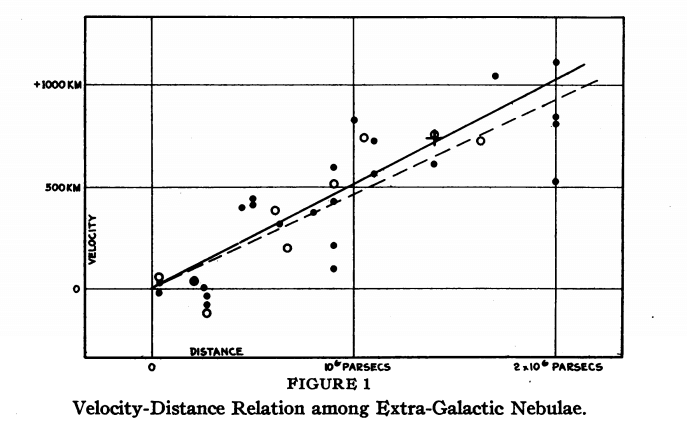}
    \caption{Radial velocity of extragalactic nebulae as a function of their distance to the Earth, reproduced from Edwin Hubble's 1929 paper \cite{hubble}.}
    \label{hubblep}
\end{figure}

Almost 100 years ago, in 1929, Edwin Hubble published his seminal finding that extragalactic nebulae were moving faster away from the Earth the farther they were from it, as shown in Fig. \ref{hubblep}. This was an early landmark in the history of the idea of the expansion of the universe, which physicists would later interpret as a growing scale factor. To get a rough idea of why Hubble's result can be interpreted as $\dot a>0$, one can reason as follows: if all spatial distances are stretched by the same factor as $a$ increases, over a timescale in which $a$ triples, a point that is initially at a distance $d$ from us will get farther by $2d$, while another one initially at a distance $2d$ will get farther by $4d$. Therefore, to traverse longer distances in the same time, farther points have to be moving faster than closer ones.

Eq. \ref{frwmetric} is the starting point to derive general-relativistic equations of motion for the cosmos. As introduced in Section \ref{GRprimer}, general relativity involves two main ingredients: the geometry of spacetime and its content. Having the necessary information about the former from Eq. \ref{frwmetric}, we now need to discuss the latter, represented by the stress-energy tensor components $T^{\mu\nu}$. Although there exists a formal definition of $T^{\mu\nu}$ in the Lagrangian formulation of general relativity as a function of the variation of the action with respect to metric components, the physical meaning of $T^{\mu\nu}$ is the flux of the $\mu$-component of four-momentum across a constant-$x^\nu$ hypersurface: $T^{\mu\nu}=d^3p^\mu/\prod_{\sigma\neq\nu}dx^\sigma$ \cite{Carroll_2019}. Therefore, considering from now on the rest frame of the fluid to avoid the need to account for bulk motion, $T^{00}=d^3p^0/dx^1dx^2dx^3\equiv dE/dV$ is the energy density and $T^{ii}=d^2p^i/dtdA^i_\perp\equiv dF^i/dA^i_\perp$ is the pressure along the $x^i$-direction ($dA^i_\perp$ stands for the area element perpendicular to the $x^i$-direction). Moreover, for example, for $i\neq j$, $T^{ij}=d^2p^i/dtdA_\perp^j\equiv dF^i/dA^j_\perp$ is a shearing stress, which, like pressure, is a force per unit area. The difference between them lies in the fact that pressure from one fluid element onto another is applied outward or inward, whereas shearing stress is applied along some direction in their surface of contact, like the effect of viscosity. Consequently, for an isotropic fluid, there can be no shearing stresses since they would establish preferred directions within the fluid, while there can be pressure so long as it has the same value along every direction: the fluid element will push outward or inward with the same intensity $P$ in every direction, so that $T^{ij}=P\delta^{ij}$. Similar reasoning leads to the conclusion that $T^{0i}=\rho \delta^{0i}$.

Now that all ingredients are at our disposal, let us cook the Einstein equations for the FLRW universe. Using Mathematica to work out nonzero components of $\Gamma^\alpha_{\mu\nu}=g^{\alpha\beta}(\partial_\mu g_{\nu\beta}+\partial_\nu g_{\mu\beta}-\partial_\beta g_{\mu\nu})/2$, we find
\begin{align*}
    &\Gamma^0_{rr}=\frac{a\dot a}{1-kr^2}, \;\Gamma^0_{\theta \theta}=r^2 a \dot a,\; \Gamma^0_{\varphi\varphi}=r^2 a \dot a \sin^2\theta, \\
    &\Gamma^r_{0r}=\Gamma^r_{r0}=H, \; \Gamma^r_{rr}=\frac{kr}{1-kr^2},\; \Gamma^r_{\theta\theta}=kr^3-r,\;\Gamma^r_{\varphi\varphi}=(kr^3-r)\sin^2\theta,\\
    &\Gamma^\theta_{0\theta}=\Gamma^\theta_{\theta0}=H,\;\Gamma^\theta_{r\theta}=\Gamma^\theta_{\theta r}=\frac{1}{r},\;\Gamma^\theta_{\varphi\varphi}=-\cos\theta\sin\theta,\\
    &\Gamma^\varphi_{0\varphi}=\Gamma^\varphi_{\varphi0}=H,\; \Gamma^\varphi_{r\varphi}=\Gamma^\varphi_{\varphi r}=\frac{1}{r}, \; \Gamma^\varphi_{\theta\varphi}=\Gamma^\varphi_{\varphi\theta}=\cot \theta,
\end{align*}
where $H(t)\equiv\dot a/a$ is the Hubble parameter. Mathematica can help us keep pushing through the calculations to now obtain the nonzero components of the Ricci tensor $R_{\nu\sigma}=R^\mu_{\;\nu\mu\sigma}=\partial_\mu\Gamma^\mu_{\nu\sigma}-\partial_\sigma\Gamma^\mu_{\nu\mu}+\Gamma^\mu_{\rho\mu}\Gamma^\rho_{\nu\sigma}-\Gamma^\mu_{\rho\sigma}\Gamma^\rho_{\nu\mu}$: 
\begin{align*}
    &R_{00}=-3\frac{\ddot a}{a}, \;R_{rr}=\frac{2(k+\dot a^2)+a\ddot a}{1-kr^2},\\
    &R_{\theta\theta}\sin^2\theta=R_{\varphi\varphi}=r^2[2(k+\dot a^2)+a\ddot a]\sin^2\theta.
\end{align*}
The Ricci tensor is then $R=g^{\nu\mu}R_{\mu\nu}=6(k+\dot a^2+a\ddot a)/a^2$ and, finally, the Einstein tensor $G_{\mu\nu}=R_{\mu\nu}-Rg_{\mu\nu}/2$ is
\begin{align*}
    &G_{00}=\frac{3(k+\dot a^2)}{a^2},\;G_{rr}=\frac{k+\dot a^2+2a\ddot a}{kr^2-1},\\
    &G_{\theta\theta}\sin^2\theta=G_{\varphi \varphi}=-r^2(k+\dot a^2+2a\ddot a)\sin^2\theta.
\end{align*}
From the $00$-component of the Einstein equation, $G_{00}=8\pi GT^{00}$, we obtain one of the most recurrent and insightful equations in cosmology, Friedmann's equation: 
\begin{equation}
    \label{friedmann}
    H^2=\frac{8\pi G}{3}\rho-\frac{k}{a^2}.
\end{equation}
The physical reasoning that led to the form $\mathrm{diag}(\rho,P,P,P)$ for the stress-energy tensor was frame-specific and implicitly assumed Minkowski spacetime. The covariant generalization of that result, which describes the so-called perfect fluid, is \cite{Carroll_2019}
\begin{equation}\label{perfectfluid}T^{\mu\nu}=(P+\rho)u^\mu u^\nu+Pg^{\mu\nu}.\end{equation} Note that, in the rest frame of the fluid, where spatial components of its four-velocity are 0, $T^{ij}=Pg^{ij}$. Besides, it turned out that $G_{ij} = -g_{ij}(k+\dot a^2+2a\ddot a)/a^2$. As a result, $H^2+2\ddot a/a=-8\pi GP-k/a^2$; we subtract from this Eq. \ref{friedmann} to obtain the better known acceleration equation:
\begin{equation}
\label{acc}
\frac{\ddot a}{a}=-\frac{4\pi G}{3}(\rho+3P).
\end{equation}
So far, we have derived two equations for the three unknowns $a$, $P$, and $\rho$. The usual route to a solution is to find an equation of state, which informs $P$ as a function of $\rho$. For matter components of cosmological interest, such as gases of non-relativistic or relativistic particles, the equation of state will usually take on the simple form
\begin{equation}
P=w\rho.
\end{equation}
The outward, positive, constant pressure of an isotropic gas of particles comes from the particles' momenta, which may give rise to a force during collisions. Consequently, in the non-relativistic limit, in which the average speed of particles is 0, the pressure is also 0, such that $w_m=0$.

We may succinctly apply the tools of classical statistical mechanics to obtain the equation of state for another case of cosmological interest, a gas of volume $V$ of $N$ ultrarelativistic particles at temperature $T$, of energy $E=\sum_{j=1}^N\sqrt{m_j^2+|\vec p_j|^2}\approx\sum_{j=1}^N|\vec p_j|$. The canonical partition function in this case is: 
\begin{equation*}
    Z=\frac{1}{h^{3N}N!}\int \prod_{i=1}^N\mathrm d^{3}q_i\;\mathrm d^{3}p_i \; e^{-\beta p_i}=\frac{(4\pi V)^N}{h^{3N}N!}\left(\int_0^\infty \mathrm dp\;p^2 e^{-\beta p}\right)^N=\frac{1}{N!}\left[\frac{8\pi V}{(h\beta)^{3}}\right]^N,
\end{equation*} 
where $\beta\equiv1/(k_BT)$. Defining $z\equiv e^{\mu\beta}$ from the chemical potential $\mu$, we can compute the pressure and the energy using the grand canonical partition function $\Xi$ \cite{salinas}:
\begin{align*}
&PV=\beta^{-1}\ln\Xi=\beta^{-1}\ln\sum_{N=0}^\infty z^NZ=\frac{8\pi Vz}{h^3\beta^4}\\ &U=-\frac{\partial}{\partial\beta}\ln\Xi=3\frac{8\pi V z}{h^3\beta^4},
\end{align*}
where we have used $\sum_{n=0}^\infty x^n/n!=e^x$. Clearly, $U=3PV\therefore P=\rho/3$, i.e. $w_r=1/3$ for an ultrarelativistic species.

We may learn more about the dynamics of the universe by studying the energy-momentum conservation equations $\nabla_\mu T^{\mu\nu}=\partial_\mu T^{\mu\nu}+\Gamma^\mu_{\mu\alpha}T^{\alpha\nu}+\Gamma^\nu_{\alpha\beta}T^{\alpha\beta}=0$, which hold generally in general relativity—the ``covariant divergence'' of the Einstein tensor vanishes identically. Specifically, the $0^{th}$ component for the perfect fluid in its rest frame in an FLRW spacetime will read 
\begin{equation}
\label{conservationcosmo}
\begin{aligned}
    0&=\partial_0T^{00}+\Gamma^i_{0i}T^{00}+\Gamma^0_{ii}T^{ii}=\dot\rho+3H\rho+Hg_{ii}Pg^{ii}\\
    &=\dot\rho+3H\rho(1+w).
\end{aligned}
\end{equation}
If we specialize to the case where $w$ is a constant, which is a good model of the universe during several stages of cosmic history, this equation can be easily integrated to yield
\begin{equation}
    \label{rho(t)}
    \rho(t)=\rho(t_0)\left[\frac{a(t)}{a(t_0)}\right]^{-3(1+w)}.
\end{equation}
Hence, pressureless dust is diluted by the expansion of the universe as one would intuitively think: with a constant energy, $\rho_m\propto1/\mathrm{volume}\propto a^{-3}$. Meanwhile, the dilution of a relativistic gas is more sensitive to changes in the scale factor: $\rho_r\propto a^{-4}$. With the information discussed so far in this section, one can now understand a central part of the story told by modern cosmology: that the universe started in a hot and dense phase where most of its content's energy was in the form of relativistic matter (often called the radiation-dominated phase) and, subsequently, its composition was dominated by non-relativistic matter (often called the matter-dominated phase). It is simple: nowadays, relativistic and non-relativistic species are both present in the universe. Given that the history of the universe has involved its monotonic expansion as far as we can tell, in the far past, $a\rightarrow0\therefore\rho_r\gg\rho_m$. Afterward, as the increase of the scale factor diluted the relativistic component more intensely than pressureless dust, it became the case that $\rho_r\ll\rho_m$.

\section{Cosmic inflation: what role does it play in our account of the history of the universe?}
\label{introinflation}
Throughout the $\mathrm{20^{th}}$-century, the use of physics to study the universe as a whole went from mostly speculative theoretical efforts by Einstein, Friedmann, Lemaître, and many others to an increasingly data-informed endeavor. In this context, the Big Bang model arose as a powerful paradigm: the spatially flat, isotropic, and homogeneous universe was very hot and dense in the far past and has been expanding according to the Friedmann equation ever since. First, it went through a radiation-dominated and, then, through a matter-dominated phase—the discovery that we are currently in a third, dark-energy-dominated phase only came about in the 1990s, a decade later than the story we are about to tell about inflation. By the late 1970s, the Big Bang model was very well established, but puzzles remained. Cosmic inflation was first proposed to solve these puzzles \cite{Guth:1980zm,Guth:1979bh}, and its success grew beyond that, arriving in the 2020s as the most studied and established model of the very early universe, the standard used in textbooks. Still, our account of the very early universe is far from complete. Inflation is not a single model, but rather a broad class of models—like hybrid inflation \cite{Linde_1994} and G-inflation \cite{Kobayashi_2010} to name two—, many of which presently fit the available data well (see Ref. \cite{2016A&A...594A..20P}, especially Section 6 and its much reproduced Fig. 12). Moreover, there exists an exciting scholarly conversation at the foundational level (see e.g. \cite{Ijjas_2013, SMEENK2014122, carroll2018falsifiabilitynormalsciencemultiverse}) regarding the predictability (or lack thereof) of inflation in the context of eternal inflation leading to a multiverse \cite{Guth_2007}. Although undoubtedly an exciting topic, we shall not dive any deeper into it—and those references shall remain as strong reading recommendations to the reader. Besides the numerous inflationary models, there also exist alternative models of the early universe such as bouncing and cyclic cosmologies \cite{Khoury_2001,Steinhardt_2002,Ijjas_2019, penroseccc} and string gas cosmology \cite{BRANDENBERGER1989391}. Therefore, before working out the dynamics of the simplest inflationary models, which is central to this dissertation, we shall describe some of the problems with the Big Bang model for which cosmic inflation was first proposed as a solution, in order to understand the essential features a theory of the very early universe should have to be compelling, be it some inflationary model or something else.

\subsection{The flatness problem}
\label{flatnessproblem}

If $\rho=\rho_c\equiv3H^2/8\pi G$ in Friedmann's equation (\ref{friedmann}), it will be the case that $k=0$, i.e. the universe's spatial slices will be flat. We can track the deviation of the universe's energy density $\rho$ from the critical density $\rho_c$ that makes it flat by forming the density parameter $\Omega(t)\equiv\rho(t)/\rho_c(t)$, so that $\Omega=1$ corresponds to a flat universe. The Friedmann equation may thus be rewritten as
\begin{equation}
\label{friedmannomega}
    1=\Omega(t)-\frac{k}{a^2H^2}=\Omega_0-\frac{k}{H_0^2},
\end{equation}
where it is customary (not just in the context of this discussion but usually in cosmology) to use the subscript $0$ to indicate quantities that refer to the present epoch and to set $a_0\equiv1$ (the value of $a(t)$ itself is not measurable; only the ratio $a(t_2)/a(t_1)$ between the value of the scale factor at two different times has physical significance). Since $\Omega_0$ and $H_0$ are observables, we may use Eq. \ref{friedmannomega} to eliminate $k$ from Friedmann's equation: $k=H_0^2(\Omega_0-1)$. Then:
\begin{equation*}
    \frac{H^2}{H^2_0}=\frac{\rho(t)}{\rho_{c,0}}-\frac{\Omega_0-1}{a^2}.
\end{equation*}
We may use Eq. \ref{rho(t)} to write $\rho(t)=\sum_i\rho_{i,0}/a(t)^{3(1+w_i)}$, where the sum over $i$ is a sum over different constituents of the universe's energy content, e.g. pressureless dust and an ultrarelativistic gas. By defining $\Omega_{k}(t)\equiv1-\Omega(t)$, we bring Friedmann's equation to an insightful form:
\begin{equation}
\label{H(a)}
    H(t)^2=H_0^2\left[\frac{\Omega_{k,0}}{a(t)^2}+\frac{\Omega_{m,0}}{a(t)^3}+\frac{\Omega_{r,0}}{a(t)^4}+\frac{\Omega_{1,0}}{a(t)^{3(1+w_1)}}+\ldots\right],
\end{equation}
where $\Omega_k$ looks like a density parameter for spatial curvature, which is diluted as $\propto a^{-2}$ as the universe expands. Note, moreover, that $\Omega_k$ is a measure of how much the universe's density deviates from the critical density, i.e. how far the density is from the exact value that renders the universe spatially flat. 

Suppose the universe's only constituents are pressureless dust and the ultrarelativistic gas—which is a good approximation for the early universe and was the paradigm for its full history in the late 1970s when inflation came about, decades before the discovery of dark energy in the late 1990s. Then, Eqs. \ref{friedmannomega} and \ref{H(a)} can be combined to yield
\begin{equation}
\label{planura}
    \Omega_{k}(t)=\frac{\Omega_{k,0}}{\Omega_{k,0}+\Omega_{m,0}\: a(t)^{-1}+\Omega_{r,0}\:a(t)^{-2}}.
\end{equation}
Therefore, since the $a$-dependent terms in the denominator were much larger in the early universe than today, if today $|\Omega_{k,0}|$ is small, then $|\Omega_k|$ must have been much smaller in the past: $|\Omega_k(t_{\mathrm{early \:universe}})|\ll|\Omega_{k,0}|\ll1$. And, as it turns out, today $|\Omega_{k,0}|$ is indeed very small. State-of-the-art observational constraints from the cosmic microwave background combined with gravitational lensing and baryon acoustic oscillations indicate $|\Omega_{k,0}|\lesssim0.005$ \cite{Planck:2018vyg}—i.e. the universe we live in is geometrically flat or very close to it. For example, as further addressed in Section \ref{bbn}, Big Bang nucleosynthesis, when the universe's low-mass chemical elements were formed from nuclear reactions in the radiation-dominated epoch, is arguably the earliest moment in the universe's history whose physics details we are confident about, with $a(t_{BBN})\approx3.6\times10^{-9}$. Consequently, given that $\Omega_{m,0}$ is $O(1)$ and $\Omega_{r,0}$ is $O(10^{-4})$, $|\Omega_k(t_{BBN})|\lesssim10^{-15}$. The deviation of the density parameter $\Omega$ from $1$ could have happened only after the $\mathrm{15^{th}}$ decimal place during Big Bang nucleosynthesis. This is impressive! If nothing else is added to the theory of the history of the universe, then $\rho$ will have been extremely close to $\rho_c$ for the entirety of the cosmos' history, with spectacular many-decimal-places agreement in the early universe.

Physicists tend to feel more comfortable about their theories when relatively broad ranges of possible values for their input parameters could work to explain how nature behaves; otherwise, they will often feel bothered about the presence of a fine-tuning problem, whereby the exact values of many decimal places of a dimensionless parameter in the theory need to be specified as an ``axiom'', a starting point of the theory. Of course, the truth about nature regarding some finely-tuned parameter might end up being that the parameter's value is a brute fact, with no explanation needed, period, end of story. Indeed, one could posit $\Omega(t_{\mathrm{Big Bang}})=1.0000\ldots$ to however many decimal places one would like as an initial condition for the Big Bang model and move on. Nonetheless, physicists will feel much more tranquil if, for a theory initially containing a fine-tuning problem, they can find a dynamical mechanism to explain why the value of the parameter in question evolved to a very specific number. And we often succeed! Note that, if one works with a universe dominated by a single component with equation-of-state parameter $w$ instead of radiation and non-relativistic matter, Eq. \ref{planura} reads
\begin{equation}
\label{omegak(t)}
    \Omega_{k}(t)=\frac{\Omega_{k,0}}{\Omega_{k,0}+\Omega_0a(t)^{-(1+3w)}}.
\end{equation}
It follows that an epoch in which $w<-1/3$ drives $|\Omega_k|$ close to, and not away from, 0. Note, in light of the acceleration equation (\ref{acc}), that this constraint on $w$ corresponds to $\ddot a>0$. Therefore, if, in the very early universe, there was a sufficiently long phase of sufficiently accelerated expansion, an arbitrarily high initial value of $|\Omega_k|$ (i.e. a universe that started out arbitrarily away from flatness) would have been sufficiently lowered (i.e. the universe would have been brought sufficiently close to flatness) by the start of the radiation-dominated phase to match observational constraints. We henceforth regard such an accelerated phase preceding radiation domination as a feature of a good theory of the very early universe\footnote{Although using the names ``early universe'' and ``very early universe'' might make it sound like we are committed to the existence of a beginning, a first temporal slice of the universe, all those should be taken to mean, at least up to this point where we are not committed to a more specific model, is that the ``early universe'' refers to the earliest moments of the radiation-dominated epoch whose physics is constrained by observation in detail—i.e. roughly around Big Bang nucleosynthesis—, and the ``very early universe'' is whatever came before.}.

More recently, it has been pointed out that $|\Omega_k|\ll1$ in the early universe is only surprising if one were to assume that initial conditions for the universe were picked from a uniform distribution for the value of $\Omega_k$. One could posit that, in the absence of knowledge about what the initial condition for $\Omega_k$—call it $\Omega_k(t_i)$—should have been, attributing equal probabilities to every possibility would respect the a priori notion that we do not know that some specific value of $\Omega_k$ should have been preferred over any other. This reasoning, however, is flawed: to know that the probability distribution of $\Omega_k(t_i)$ was uniform would be to know \textit{something} about it; to truthfully be following the epistemic status of \textit{not knowing anything} about $\Omega_k(t_i)$, one should answer ``I don't know'' if asked what the probability distribution for $\Omega_k(t_i)$ would have been. Suppose one, on the other hand, considers a probability measure over trajectories in phase space for a general-relativistic description of the universe \cite{Gibbons:1986xk}. In this case, the flatness problem dissolves, for one finds out that the probability that an FLRW spacetime is nearly spatially flat is close to 1 \cite{Carroll:2014uoa}. This more careful treatment of fine-tuning does not do away, nevertheless, with the horizon problem, which we shall discuss next.

\subsection{The horizon problem}
\label{thehorizonproblem}

As just discussed, solving the equations of motion with $k=0$ is a good description of our universe, so we shall often proceed in that way, as will be the case in this subsection. If such a universe has a single component with equation-of-state parameter $w$, Friedmann's equation (\ref{friedmann}) can be easily integrated, again choosing $a(t_0)=1$, to 
\begin{equation}
\label{a(t)}
    a(t)=\left(\frac{t}{t_0}\right)^{\frac{2}{3(1+w)}}
\end{equation}
in light of Eq. \ref{rho(t)}. This solution can help us study a remarkable feature of FLRW spacetimes: they may contain particle horizons, finite causally connected regions. The intuitive idea is simple: if, for $t\rightarrow0$, spatial distances also approach 0 for some values of $w$, including $w_m=0$ and $w_r=1/3$, then the universe started from a point\footnote{Or from an infinitesimal volume, if one would rather dispense with the complex discussion, unnecessary for ours, regarding whether it makes sense for a spatially three-dimensional universe to instantaneously pop into existence out of a zero-dimensional initial condition.}. $t=0$ hence sets an initial time for which particles could possibly have started following some trajectory in the universe, so that only a finite amount of time is available for any particle to be following some path in the universe. As a consequence, our intuition will rightly tell us (at least for some cases of interest, although counterintuitive exceptions of interest also exist—more on this later) that any causal contact in the universe, which can happen at most as fast as the speed of light, can only happen within a maximum distance: light propagating with a finite speed for a finite time can only move a finite distance away. 

Let us take a look at how that unfolds for the flat radiation-dominated early universe. Consider a photon ($ds=0$) traversing a radial path from time $t_i$ up to time $t$. The comoving distance $d_C$ thus traversed will be 
\begin{equation}
    d_C=c\int_{t_i}^t\frac{\mathrm dt}{a(t)}=c\int_{t_i}^t\frac{\mathrm dt}{\sqrt{t/t_0}}=2c\sqrt{t_0}\left(\sqrt t-\sqrt{t_i}\right),
\end{equation}
where we have inserted factors of the speed of light $c=1$ for clarity. Therefore, the physical/proper distance traversed will be $d_H=a(t)d_C=2c\left( t-\sqrt{tt_i}\right)$. For $t_i=0$\footnote{The reader might be wondering why we introduced what now might seem as an unnecessary complicating step of calling the initial time $t_i$ and then setting it to $0$ instead of just integrating from $0$ from the beginning. This was a hint toward the fact that, for certain values of $w$, $d_C(t_i=0)$ is undefined, and $\lim_{t_i\rightarrow 0}d_C=\infty$, so that a horizon is absent. Moreover, there exist cases, e.g. $w=-1$ as we will see soon, in which $a(t)\neq0\;\forall \;t$ and, therefore, integration can be performed from $t\rightarrow-\infty$.}, $d_H=2ct=c/H(t)$: this is the size of the particle horizon in a radiation-dominated flat universe, meaning that causal contact can only have happened between points separated by a proper distance of at most $c/H(t)$ by time $t$. In the static Minkowski space, we would of course have obtained $d_H=ct$; the extra factor of $2$ in our result is attributed to the fact that distances traversed by the photon at earlier times will, by later times, have been enlarged.

Up to $\sim$400,000 yr after Big Bang nucleosynthesis (i.e. redshift $z\sim1100$), the universe's content was opaque to photons, such that light coming from earlier times was constantly scattered and hence could not freely propagate and be observed by our telescopes in the present. We call the last scattering surface (LSS) the spatial hypersurface with $t=t_{LSS}$ at which the universe's content transitioned from being opaque to photons to not being, such that light emitted at $t>t_{LSS}$ has usually been able to reach our observatories. The light we detect all over the sky coming from the LSS, which has reached the present redshifted\footnote{The expansion of spatial distances in the universe Doppler shifts waves just like a moving source in Newtonian physics.} to microwave frequencies, is the famous cosmic microwave background (CMB). Its temperature has been measured by missions such as COBE, WMAP, and Planck to be highly homogeneous, with typical variations at the order of 1 part in $10^5$. A strong pillar for the cosmological principle, this observation is also the empirical starting point for the horizon problem.

With the nontrivialities associated with working with a non-Euclidean spacetime geometry, there exist several different measures of distance in cosmology, each of which is carefully defined. Consider an arc in the flat FLRW sky at $t=t_e$ connecting points with the same $r$ and $\varphi$ coordinates, observed angular size $\Delta \theta$, and proper size $l=\int ds=a(t_e)r\Delta\theta$. Then, the angular diameter distance is defined as $d_A\equiv l/\Delta\theta$, which is akin to asserting $radius=(arc\;length)/(angular\;size)$ for a circle. The angular diameter distance from the Earth to the last scattering surface is $d_A(t_{LSS})\approx 10$ Mpc, while $d_H(t_{LSS})\approx0.3$ Mpc. This means that, as viewed from the Earth, the angular size of a particle horizon within the LSS is $\Delta \theta=d_H/d_A\sim0.01\;\mathrm{rad}\sim1\;\mathrm{deg}$. Given that a full spherical shell has $\sim$43,000 $\mathrm{deg^2}$, we have just learned that what we see today as the last scattering surface consisted of many thousands of causally disconnected patches in the cosmos. How could a collection thousands of such regions—whose contents, let it be stressed, could not have had any sort of causal influence on one another—be so homogeneous as to have a constant temperature up to $O(10^{-5})$ relative deviations? This is the horizon problem! One way out—inflation's way out, as we will shortly see—is that the story of a universe that starts radiation-dominated and transitions to matter domination is incomplete, such that, early on, there was causal contact between the LSS patches that brought them to thermal equilibrium. If such a fix to the Big Bang model could not be found, we would have to posit a clumsy initial condition for the universe: that those thousands of causally disconnected patches happened to start out with the same homogeneous temperature, up to tiny fluctuations—the inelegance of such a component for a theory of the universe can be understood as a fine-tuning of initial conditions, as discussed for the flatness problem in Section \ref{flatnessproblem}.

\subsection{The monopole problem}
\label{themonopoleproblem}
The flatness and horizon problems are more general, model-independent issues plaguing the original Big Bang model, whereas the monopole problem is importantly dependent on the underlying particle theory governing the very early universe. Not only that, the specific particle physics framework used to first formulate the monopole problem, the Grand Unified Theory (GUT) of Georgi and Glashow, has now been ruled out by experiments. However, it was working on the monopole problem that directly led Alan Guth to discover cosmic inflation as a solution to the flatness and horizon problems, besides the monopole one itself. Given its historical importance, we shall then provide a brief qualitative description of the monopole problem, inspired by that in Ref. \cite{Guth:1997wk}, leaving out most of the particle physics details and focusing on its connection with the horizon problem. 

The monopole problem's monopoles are magnetic ones. The inexistence of magnetic monopoles is imparted into the minds of young physicists early on through the homogeneous Maxwell equation $\vec\nabla\cdot\vec B=0$. Would Maxwell's equations not be more elegant if there were a source there, in parallel with its electric counterpart $\vec\nabla\cdot\vec E=\rho_e$? Well, indeed one can make Maxwell's mathematical formalism consistent with magnetic monopoles \cite{Tong2018GaugeTheory}. But, as it turns out, we have never observed magnetic monopoles in nature, leading to electromagnetic theory being constructed the way it was. The monopole problem in cosmology refers to specific particle physics models predicting the overproduction of magnetic monopoles in the very early universe, in such a way that their abundance in the present universe would be so large that their repeated detection should have been inevitable.

\begin{figure}
    \centering
    \includegraphics[width=\linewidth]{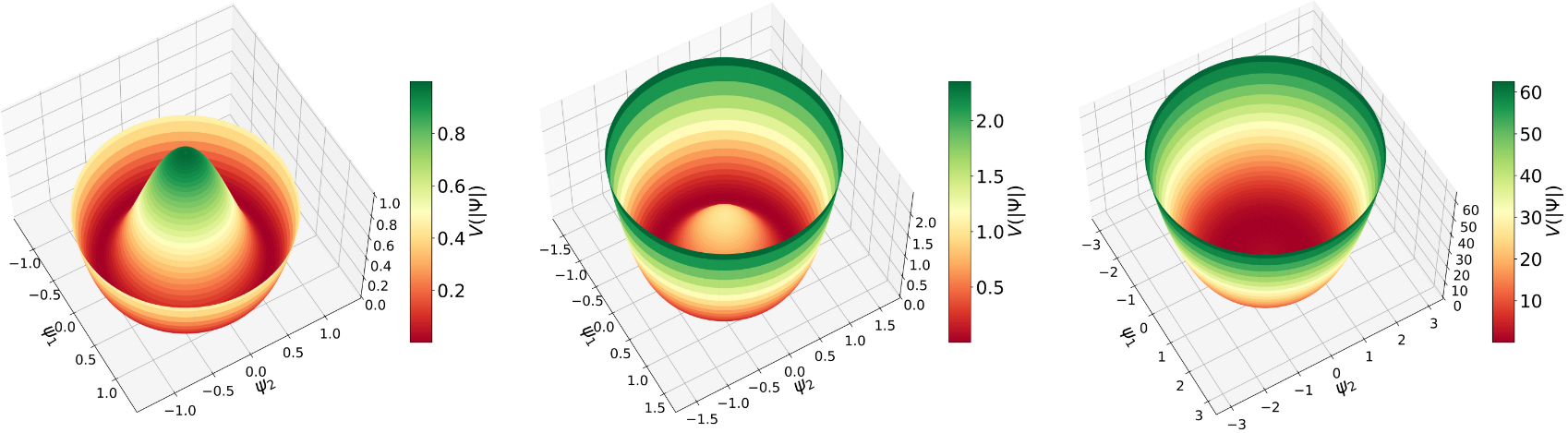}
    \caption{The potential energy of a set of two scalar fields $V(|\vec\Psi|)$ from Eq. \ref{mexico} for $v=1$ as a function of field values, plotted from $|\vec\Psi|=0$ up to maximum values of $|\vec\Psi|$ that increase from the left to the mid and from the mid to the right plots.}
    \label{SSB}
\end{figure}

The simplest particle physics toy model in which there are magnetic monopoles contains three Higgs fields $\phi_1(x)$, $\phi_2(x)$, and $\phi_3(x)$, which are scalar quantum fields. We may summarize the configuration of fields at each point in spacetime by forming a vector $\vec\Phi(x)=(\phi_1(x),\phi_2(x),\phi_3(x))$. The potential energy carried by the Higgs fields is a function of $|\vec\Phi|^2=\sum_i |\phi_i|^2$. Higgs fields are known for their Mexican-hat type potentials, as exemplified for a two-field system $\vec\Psi=(\psi_1,\psi_2)$ by
\begin{equation}
\label{mexico}
    V(|\vec\Psi|)=(\psi_1^2+\psi^2_2-v)^2.
\end{equation}
A potential like this one is known to give rise to spontaneous symmetry breaking (SSB). As illustrated in Fig. \ref{SSB}, considering energy scales $\sim v$, there is a prominent peak at the center of the potential energy profile. The system's spontaneous tendency to minimize its energy will be concretized, therefore, by a vacuum that has $\braket{|\vec\Psi|}=v$, for which reason $v$ is called the vacuum expectation value (vev) of the field. However, for energy scales $\gg v$, the central local peak in the potential is negligible; the potential energy plot is effectively a paraboloid, and $\braket{|\vec\Psi|}=0$ due to symmetry. The transition from high to low energies thus induces a symmetry-breaking phase transition in the system, whereby $\vec\Psi$ goes from averaging to $(0,0)$ to averaging to any $(\psi_1,\psi_2)\;\mathrm{such\; that}\;\psi_1^2+\psi_2^2=v^2$.

On top of that, the theory informs that a gradient in the direction of $\vec\Phi$ also contributes to the total energy. Consequently, in the vacuum after SSB, the field on average takes on $\vec\Phi(x)=v\vec n$, where $\vec n$ is a constant unit-norm vector field, and configurations where the direction of the field vector $\vec\Phi(x)/|\vec\Phi(x)|$ varies in space are more energetic than the vacuum even if $|\vec\Phi(x)|=v$ everywhere. In particular, as it turns out, a configuration of the field whereby $\vec\Phi(x)$ is $\vec 0$ at the origin and points radially outward elsewhere, with magnitude increasing from $0$ and asymptoting to $v$, is a magnetic monopole. Comparing the vacuum and magnetic-monopole configurations of the fields—the former, homogeneous, and the latter, inhomogeneous—, we can make the rough assertion that magnetic monopoles are associated with systems of Higgs fields with a larger ``degree of chaos''.

Now, consider the situation of this three-field system in the hot very early universe. At earlier times, with a very high thermal energy available, the Higgs had not undergone SSB and averaged to $\vec 0$. Once the universe cooled sufficiently, the Higgs' expected magnitude jumped to the vev, with the direction of $\vec\Phi$ being established randomly at each point. Subsequently, the field vectors in neighboring points would spontaneously align so as to decrease the energy. Yet, there is a limit to the alignment that could have taken place, set by the particle horizon we have just studied in the previous subsection: with no causal contact, there could not have been a correlation between the directions to which the Higgs field vectors in two points outside of one another's horizon aligned. This statement—which has been put forth vaguely but can certainly be made clearer by including all the relevant math—provides a lower limit to the degree of chaos in the Higgs fields, which, in turn, translates into a lower limit to magnetic monopole production. The monopole problem is the fact that, even at the level of this lower bound, which—after assuming some particle theory to describe what the monopoles consist of—only takes into account the impossibility of causal contact across large enough regions, magnetic monopoles are still overproduced in the early universe and should have an appreciable abundance in the present, as was shown by John Preskill \cite{Preskill:1979zi}.

In these terms, it becomes clear that a solution to the horizon problem might automatically solve the monopole problem. If the horizon distance during the Higgs phase transition was much larger than assumed in the original Big Bang model, then the degree of alignment in the fields could have been much higher, suppressing the formation of monopoles.

\section{Inflationary dynamics}
As previously stated, nowadays there exists a wide range of models under the cosmic inflation umbrella. Let us then study a very simple inflationary model to develop some general ideas of inflation and see dynamical solutions to the problems discussed in Section \ref{introinflation}. Consider a scalar field $\phi$ with Lagrangian\footnote{Keeping in mind that we are using the $(-,+,+,+)$ metric signature.}
\begin{equation}
\label{inflatonlag}
    \mathcal L^{(M)}=-\frac{1}{2}g^{\mu\nu}\partial_\mu\phi\partial_\nu\phi-V(\phi).
\end{equation}
For inflation to be a successful theory, it will need the assumption that the energy content of the very early universe is dominated by $\phi$ during the inflationary phase; this field is thus called the inflaton, the driver of inflation. Then, we can write the action describing this system as
\begin{equation}
    S=\int\mathrm d^4x\sqrt{-g}\left(\mathcal L^{(M)}+\frac{R}{16\pi G}\right)=S^{(M)}+S^{(G)},
\end{equation}
where $R$ is as before the Ricci scalar and the factor $\sqrt{-g}\equiv\sqrt{-\det \mathbf g}$ is included in the integration measure to ensure its covariance, since $\mathrm d^4x$ is not a tensor but rather a tensor density \cite{Carroll_2019}. The term $\mathcal L^{(G)}\equiv R/16\pi G$, the Einstein-Hilbert Lagrangian, comes from the Lagrangian formulation of general relativity: when the least action principle is applied for variations in the metric $g_{\mu\nu}\rightarrow g_{\mu\nu}+\delta g_{\mu\nu}$ together with the definition 
\begin{equation}
T_{\mu\nu}\equiv\frac{-2}{\sqrt{-g}}\frac{\delta S^{(M)}}{\delta g^{\mu\nu}},
\end{equation} 
the resulting Euler-Lagrange equation of motion is Einstein's equation. To learn more about the dynamics of $\phi$, we may instead apply the least action principle for variations in the inflaton field $\phi\rightarrow \phi+\delta\phi$:
\begin{equation*}
    \delta S=\int\mathrm d^4x \sqrt{-g}\left[\frac{\partial \mathcal L^{(M)}}{\partial\phi}\delta\phi+\frac{\partial \mathcal{L}^{(M)}}{\partial (\partial_\mu\phi)}\delta(\partial_\mu\phi)\right]=0.
\end{equation*}
We may integrate the second term by parts, using the fact that the boundary term is 0 because $\delta\phi=0$ at the boundary, as usual in the calculus of variations. The resulting Euler-Lagrange equation is:
\begin{equation*}
    \frac{\partial\mathcal L^{(M)}}{\partial\phi}-\frac{1}{\sqrt{-g}}\partial_\mu\left(\sqrt{-g}\frac{\partial\mathcal L^{(M)}}{\partial(\partial_\mu\phi)}\right)=-\frac{\partial V}{\partial\phi}+\frac{1}{\sqrt{-g}}\partial_\mu\left(\sqrt{-g}\partial^\mu\phi\right)=0
\end{equation*}
for the Lagrangian \ref{inflatonlag}. Note that the operator acting on $\phi$ in the second term is the covariant generalization of the d'Alambertian, $\square\phi=\nabla^\mu\nabla_\mu\phi=(1/\sqrt{-g})\partial_\mu(\sqrt{-g}\partial^\mu\phi)$. 

It is simplest to proceed assuming homogeneous and isotropic initial conditions for the inflaton field in the region that inflates. As we are about to find out, over the inflationary epoch, the scale factor grows by many orders of magnitude. It follows that the initial inflaton-filled region that got inflated to produce the current observable universe is relatively very small. Therefore, a priori, one may feel comfortable assuming homogeneous and isotropic initial conditions for the field, not in an arbitrarily large universe, but rather in a tiny patch of it. Whether, in practice, this initial condition would involve a bothersome amount of fine-tuning is a cumbersome question we shall ignore. It is worth noting, nonetheless, that progress is actively being made on studying the robustness of the inflationary mechanism to inhomogeneous initial conditions, as summarized in Sections 3.3 and 3.4 of Ref. \cite{Flauger:2022hie}. In any case, we shall work with the assumption of homogeneous and isotropic initial conditions, absolutely usual throughout the historical development of inflationary theory. In this context, the inflaton field can only be a function of time and not of spatial position, i.e. $\phi=\phi(t)$. Then, given that $\sqrt{-g}=a(t)^3r^2\sin\theta/\sqrt{1-kr^2}$, one can easily compute the d'Alambertian above to finally arrive at the equation of motion for a scalar field in an FLRW spacetime:
\begin{equation}
    \label{inflODE}
    \ddot\phi+3H\dot\phi+\partial_\phi V=0.
\end{equation}
The specific form of $V$ is a modeling choice we shall make later. Hence, what remains to be done aiming at a solution for $\phi(t)$ is to determine $H(t)$ from Einstein's equations. We have already derived the relevant Einstein equations, Eqs. \ref{friedmann} and \ref{acc}, and the missing ingredient to use them is the specific form of $T_{\mu\nu}$ for the matter Lagrangian we are studying. Applying the matrix identity $\delta g=g g^{\mu\nu}\delta g_{\mu\nu}$ and carrying out differentiation, the result is
\begin{equation*}
    T_{\mu\nu}=\frac{-2}{\sqrt{-g}}\frac{\delta S^{(M)}}{\delta g^{\mu\nu}}=\partial_\mu\phi\partial_\nu\phi-g_{\mu\nu}\left[\frac{1}{2}g^{\lambda\sigma}\partial_\lambda\phi\partial_\sigma\phi+V(\phi)\right].
\end{equation*}
Again considering the rest frame of the perfect fluid filling the FLRW universe and keeping in mind that $\partial_i\phi=0$, this reduces to 
\begin{align}
    \label{rhoinfl}&\rho=\frac{\dot\phi^2}{2}+V(\phi)\\
    \label{pinfl}&P=\frac{\dot\phi^2}{2}-V(\phi).
\end{align}

\begin{figure}
    \centering
    \includegraphics[width=0.425\linewidth]{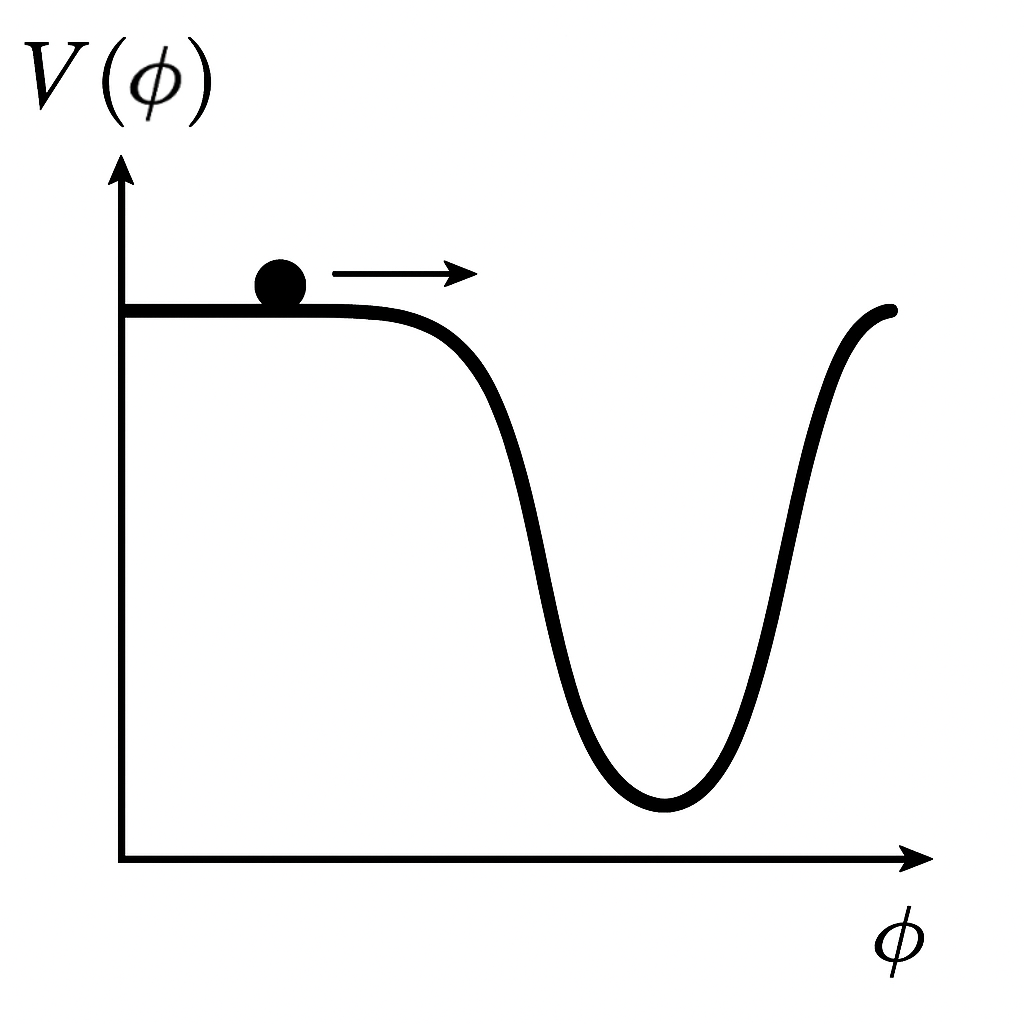}
    \caption{On top of a curve of potential energy $V$ as a function of field value $\phi$, a ball slowly moves in a typical region of slow roll—a plateau of the $V(\phi)$ curve for which $V(\phi)\gg \dot\phi^2$—, as a metaphor for the evolution of $\phi(t)$ and $V(\phi(t))$. Were we examining a particle at position $x$ under the influence of a gravitational potential $V(x)$ of the same shape, the behavior of rolling slowly toward positions where $V$ is smaller, which is metaphorical for the scalar field as it takes place in field space, would be the particle's concrete behavior in physical space.}
    \label{slowroll}
\end{figure}

After this modest Lagrangian mechanics workout, we are now armed with Eqs. \ref{inflODE}–\ref{pinfl}, which, supplemented by Eqs. \ref{friedmann} and \ref{acc}, provide a description of the matter and spacetime dynamics of a scalar-field-filled FLRW world. What specific choices turn this more general setting into an inflating universe? A simple, broad, and common class of inflationary models arises in slow roll, which refers to when the contribution of $\dot\phi$ to the inflaton's four-momentum is negligible, i.e. $\dot\phi^2\ll V(\phi)$. The system's time evolution may be summarized by an ordered pair parametrized by $t$, $(\phi(t),V[\phi(t)])$. The name ``slow roll'' comes from picturing this ordered pair as a ball that sits on top of the $V(\phi)$ curve in the $\phi V$-plane, whose motion, i.e. whose \textit{rolling} on top of the potential energy curve, is slow since $\dot\phi$ is small. This is illustrated in Fig. \ref{slowroll}.

During slow roll, therefore, $\rho\approx V(\phi)\approx-P$, i.e. $w\approx-1$. This is remarkable and has many implications. First, note, from Eq. \ref{rho(t)}, that $\rho(t)\approx\rho(t_0)[a(t)/a(t_0)]^0=\rho(t_0)$, i.e. the energy density remains constant as the scale factor increases. Surprising! Instead of being diluted by the expansion of the universe, which is intuitively expected and does happen for pressureless dust and an ultrarelativistic gas, $\rho$ for the slow-roll scalar field stays constant even as the volume of space increases, such that the expansion of the universe leads to an increase in the total energy in the inflaton field. 

It is worth noting in passing that this $w=-1$ component and its odd behavior are in fact what we call dark energy. The fact that a $w=-1$ component is not only present in the universe but currently dominates its total energy density, with $\Omega_{\Lambda,0}\approx0.6889$ \cite{Planck:2018vyg}, is the most recent addition to the standard cosmological model, $\Lambda$CDM (where CDM stands for cold dark matter), with decisive observational evidence being found in 1998 \cite{SupernovaSearchTeam:1998fmf,SupernovaCosmologyProject:1998vns}. The dark energy is usually denoted by $\Lambda$ because of the cosmological constant: general relativity is consistent with the addition of the term $\Lambda g_{\mu\nu}$, where $\Lambda$ is a constant, to the left-hand side of Einstein's equations (\ref{Einstein}). As it turns out, the dynamics of a $\Lambda=0$ spacetime filled by a $w=-1$ fluid is equivalent to that of an empty $\Lambda>0$ spacetime, called de Sitter space. The nature of dark energy is one of the big open questions in cosmology: we do not know if it is a cosmological constant or some other form of energy with $w=-1$ and, if the latter, we do not have a good particle physics theory to describe such a thing.

As discussed in Section \ref{flatnessproblem}, the acceleration equation (\ref{acc}) implies that $w=-1<-1/3$ drives a phase of accelerated expansion of the universe, which will solve the flatness problem if this phase lasts long enough. With a scale-factor-independent $\rho$, then, the second term in Friedmann's equation (\ref{friedmann}), which is $\propto a^{-2}$, will rapidly become negligible compared to the first. Consequently, $H$ will tend to behave as a constant as long as slow roll persists. Since $H=\dot a/a$, we find that the scale factor thus expands exponentially during inflation: \begin{equation*}
    a(t)\propto e^{Ht}.
\end{equation*} 

The inflationary phase in the very early universe cannot last forever; rather, it has to eventually end and be followed by the radiation-dominated phase, for which we have strong evidence—it has got to be there. Reheating is the process whereby the inflaton, after the slow-roll regime comes to an end, decays into a hot soup of standard model particles, such that the enormous energy it gained during inflation is deposited into the current matter content of the universe. For now, let us not discuss reheating and focus on the end-of-slow-roll aspect of the end of inflation. To establish some rough quantitative measure of whether slow-roll is taking place or not, it is usual to form so-called slow-roll parameters:
\begin{align}
    \label{epsilon}&\epsilon\equiv - \frac{\dot H}{H^2} \\
    &\eta\equiv\epsilon-\frac{\ddot\phi}{H\dot\phi},
\end{align}
which are perturbative as long as slow roll happens, since their numerators contain time derivatives of quantities in their denominators. We can take a time derivative of Friedmann's equation (\ref{friedmann}) to obtain $2H\dot H\approx8\pi G(\partial_\phi V)\dot\phi/3$ (with the end of inflation under consideration, it should certainly be the case that inflationary expansion has driven $k/a^2$ very close to 0 compared to the other term in the right-hand side). With $|\ddot\phi|\ll|3H\dot\phi|$ during slow roll, Eq. \ref{inflODE} gives $\dot\phi\approx-(\partial_\phi V)/3 H$. As a result, $\dot H\approx-(\partial_\phi V)^2/6V$—which, note, is consistent with our previous statement that $H$ is approximately constant during inflation: given that slow roll happens on a plateau of the potential energy curve, $\partial_\phi V\approx0$. These results allow for a determination of $\epsilon$ as a function of the potential and its derivative; one can proceed similarly for $\eta$ by starting with the time derivative of the field's equation of motion (\ref{inflODE}) and employing slow-roll approximations; the results are:
\begin{align}
    &\epsilon\approx\frac{(\partial_\phi V)^2}{16\pi G V^2}\\
    &\eta\approx\frac{\partial_\phi^2 V}{8\pi G V}.
\end{align}

Since $\epsilon$ and $\eta$ are of perturbative order during slow roll, we may assert that slow roll has ended when they reach order 1. In fact, when $\epsilon$ is exactly 1, the acceleration of the scale factor ends identically. This is easy to see: $\dot H=\frac{d}{dt}(\dot a/a)=- H^2+\ddot a/a=\dot H+\ddot a/a$, where the last equality holds for $\epsilon=1$.

Let us now take a look at how inflation can solve the flatness and horizon problems—and, as discussed in Section \ref{themonopoleproblem}, the monopole problem as well, since solving the horizon problem basically meant solving the monopole problem as a result in the historical context in which inflation appeared. According to Eq. \ref{omegak(t)}, the relative change in $|\Omega_k|$ from the beginning to the end of the inflationary epoch is
\begin{equation}
    \label{inflationflatness}
    \frac{|\Omega_k(t_{end})|}{|\Omega_k(t_i)|}=\frac{a(t_{end})^{-2}}{a(t_i)^{-2}}.
\end{equation} 
From $t_{end}$ up to $t_{BBN}$ in the radiation-dominated era, it is
\begin{equation}
    \label{BBNflatness}
    \frac{|\Omega_k(t_{BBN})|}{|\Omega_k(t_{end})|}=\frac{a(t_{BBN})^{2}}{a(t_{end})^{2}}=\frac{t_{BBN}}{t_{end}},
\end{equation}
where the last equality follows from Eq. \ref{a(t)}. To remove from the Big Bang model the flatness problem, which is a fine-tuning problem, it should suffice to have $|\Omega_k(t_i)|=O(1)$. In this case, according to Eq. \ref{inflationflatness}, $|\Omega_k(t_{end})|=e^{-2H(t_{end}-t_i)}\equiv e^{-2N}$, where the number of $e$-foldings $N$ is the number of times $a$ was scaled by a factor of $e$ across some time interval. As a result, $|\Omega_k(t_{BBN})|=e^{-2N}t_{BBN}/t_{end}\lesssim10^{-15}$, where the inequality comes from the discussion in Section \ref{flatnessproblem}. For a typical slow-roll-inflation value of $t_{end}$, a few tens of $e$-foldings, often $N\sim 60$, will solve the flatness problem.

As put forth in Section \ref{thehorizonproblem}, for a radiation-dominated universe, the proper size of the horizon is $d_H=2ct$. Consider, as a reasonable example, that the universe started radiation-dominated, underwent cosmic inflation, and, then, was reheated back to radiation domination. In this case, if inflation starts at time $t_i$ and ends at time $t_{end}$, the horizon size at the end of inflation will be:
\begin{equation*}
    d_H(t_{end})=a(t_{end})c\int_0^{t_{end}}\frac{\mathrm dt}{a(t)}=e^Nc\left(2t_i+a(t_i)\int_{t_i}^{t_{end}}\frac{\mathrm dt}{a(t_i)\exp[H(t_i)(t-t_i)]}\right)\approx3e^Nct_i,
\end{equation*} 
where $a(t_{end})=e^Na(t_i)$, $t_i\sim1/H(t_i)$, and $t_{end}\gg t_i$. A typical value considered in this context is $t_i=10^{-36}$ s, for which $d_H(t_i)\approx6\times10^{-28}$ m and $d_H(t_{end})\approx10$ m for 60 $e$-foldings! Then, if we evolve $d_H$ up to $t_{LSS}$,
\begin{equation}
\label{horizonsolved}
    d_H(t_{LSS})\approx\frac{a(t_{LSS})}{a(t_{end})}3e^Nct_i+2ct_{LSS}\approx130\mathrm{\;Mpc},
\end{equation}
where the first approximation follows from $t_{LSS}\gg t_i$ and from the fact that, although the LSS is in the matter-dominated universe, $d_H$'s order-of-magnitude behavior does not change if we approximate the full post-inflationary period as radiation-dominated—and $2ct_{LSS}$ is the subdominant term in that expression anyway. In Eq. \ref{horizonsolved}, $a(t_{LSS})=(1+z_{LSS})^{-1}\approx1/1100$ (where the first equality here comes from the definition of redshift), $t_{LSS}\approx3.4\times10^5$ yr, and $a(t_{end})=2\times10^{-27}$ for inflation happening at the typical scale associated with $t_i=10^{-36}$ s. Recalling that $d_A(t_{LSS})\approx10$ Mpc, we see that this amount of inflation is sufficient to solve the horizon problem: now, rather than thousands of causally disconnected regions on the LSS sky, we have that even antipodal points on the LSS sky are within the same particle horizon and, thus, causally connected.
\chapter{\textsc{Inflationary gravitational waves}}
\label{IGWchapter}
\section{Tensorial metric perturbations during inflation}
Now that we have firmly established the concepts of gravitational waves and inflation, we are ready to combine them into our topic of interest, inflationary gravitational waves. As we have seen in Section \ref{GWchapter}, gravitational waves are small tensorial perturbations to some background metric—Minkowski was the case studied there. Now, we shall perturb the FLRW metric describing the expanding homogeneous and isotropic universe, which was worked out in Section \ref{cosmochapter}. This will not be an exercise that is interesting merely mathematically. Rather, this is an essential step in describing the universe we live in, which is homogeneous and isotropic in a coarse-grained sense on large scales; on smaller scales, the evolution of deviations away from homogeneity and isotropy is, among other things, precisely the history of the formation of structure in the universe—galaxies, stars, planets...—, hence composing one of the richest and most relevant topics for study within cosmology. 

Gravitational waves, the tensorial modes, are not the most general class of perturbations to a metric: there are also scalar and vectorial degrees of freedom. Although our topic of interest is indeed gravitational waves, it is worth constructing the most general perturbed FLRW metric to then comment on the role of the scalar and vectorial perturbations for our universe. Thereafter, we may specialize to the dynamics of gravitational waves neglecting other perturbations thanks to the so-called decomposition theorem, according to which, in general relativity, scalar, vectorial, and tensorial perturbations evolve independently at first order in perturbation theory \cite{Dodelson:2020bqr}. This means that, if the initial conditions contain, for example, scalar perturbations to the metric, these will not, at linear order, induce vectorial or tensorial ones as it evolves.

Akin to the process in Section \ref{GWderivation}, we may use $h_{\mu\nu}$ to denote small perturbations in the line element \ref{frwmetric}, specializing to the spatially flat case, which provides a very good description of our universe, and adopting Cartesian coordinates:
\begin{equation*}
    ds^2=[-1+h_{00}(t,\vec x)]dt^2+2a(t)h_{0i}(t,\vec x)dtdx^i+a(t)^2[\delta_{ij}+h_{ij}(t,\vec x)]dx^idx^j,
\end{equation*}
where $\delta_{ij}$ are the components of a Kronecker delta. We have been referring to scalars, vectors, and tensors having in mind representations of the SO(3) group, that is, considering objects' behaviors under rotations in three-dimensional space. Therefore, $h_{00}$ is a scalar, $h_{0i}$ is a vector, and $h_{ij}$ is a (rank-2) tensor. 

To start, we may adopt a conventional relabeling: $h_{00}(t,\vec x)\equiv-2A(t,\vec x)$. Next, we may apply a well-known result studied in undergraduate physics: vector fields—at least sufficiently well-behaved ones, which will generally be the topic of interest in physics—may in general be written as a sum of a curl-free and a divergence-free part. Recalling that the curl of the gradient of a scalar function is identically 0, we may write $\vec{h_0}$ as the (co-)vector whose components are $h_{0i}$ and hence define $\vec{h_0}\equiv-\vec\nabla B-\vec S$, where $\vec\nabla\cdot\vec S=0$. In index notation, this is $h_{0i}=-\partial_i B-S_i$, with $\partial_i S^i=0$. Finally, we want the most general decomposition of $h_{ij}$, which must be symmetric since the full metric tensor has to be symmetric by definition. Scalars can be used to this end in two ways. First, one can simply attach a scalar function of space and time to a Kronecker delta: $h_{ij}(t,\vec x)\supset 2 D(t,\vec x)\delta_{ij}$. Moreover, since derivatives commute, the second derivative of a scalar field can also contribute a symmetric term: $h_{ij}(t,\vec x)\supset -2 \partial_{i} \partial_{j} E(t,\vec x)$. We can also form a symmetrized first derivative of vector fields: $h_{ij}(t,\vec x)\supset \partial_iF_j(t,\vec x)+\partial_j F_i(t,\vec x)$. $\vec F$ should be divergenceless, because its term that has a divergence is already captured by the scalar degree of freedom $E$ (decomposing $F_i$ as done for $h_{0i}$, $\partial_j F_i=\partial_j(\partial_i \psi + \xi_i)$, with $\partial_i\xi^i=0$; $\partial_j\partial_i \psi$ may be absorbed into $\partial_j\partial_i E$, so that the only independent degree of freedom in $F_i$ will be the divergence-free term). The $4\times4$ symmetric metric tensor has 10 independent components. We have so far specified 8 of them: $A$, $B$, $\vec S$, $D$, $E$, and $\vec F$—note that, being both transverse, $\vec S$ and $\vec F$ account for two and not three degrees of freedom. The remaining two are the gravitational-wave, the tensorial degrees of freedom, $h_{ij}^{TT}$, as worked out in detail in Section \ref{GWderivation}. As a result, our decomposition of perturbations to the FLRW line element into scalar, vector, and tensor modes reads
\begin{equation}
\label{perturbedFRW}
\begin{aligned}
    ds^2=&[-1+A(t,\vec x)]dt^2+2a(t)[-\partial_iB(t,\vec x)-S_i(t,\vec x)]dtdx^i\\&+a(t)^2\{[1+2D(t,\vec x)]\delta_{ij}-2 \partial_i \partial_j E(t,\vec x)+\partial_iF_j(t,\vec x)+\partial_j F_i(t,\vec x)+h_{ij}^{TT}(t,\vec x)\}dx^idx^j.
    \end{aligned}
\end{equation}

As we will soon see, tensorial perturbations can be a robust prediction of models of the very early universe. Those primordial gravitational waves have not been detected yet, but the search for them is an active and growing endeavor in astronomy and the ultimate topic of interest of this dissertation. As it turns out, the vectorial ones, whose production is not appreciably predicted by most cosmological models, can be shown to decay as $a^{-2}$ or faster in an FLRW spacetime, being mostly erased by the expansion of the universe. Therefore, emphasis on them in the literature is sparse. Scalar perturbations, on the other hand, were, for a long time, the leading topic of interest in the study of the inhomogeneities of the universe. This is due to their connection to energy density perturbations: they source and are sourced by them. Density perturbations are protagonists of the story told by modern cosmology: billions of years ago, density fluctuations were very small—in the CMB time slice, for example, typical fluctuations were of order $10^{-5}-10^{-4}$. With time, overdense regions, toward which there is a stronger gravitational pull—and, by the way, this process can be modeled well with Newtonian gravitation for most of the evolution that has taken place in our universe—, started to attract more matter and consequently become increasingly overdense, leaving initially underdense regions increasingly more so. The clumps of matter that are thus formed evolve—through complicated routes, where other forces besides gravity start to play a dominant role—to be the structure in the universe: planets, stars, galaxies, galaxy clusters, asteroids, people, physics textbooks...

\section{Gravitational-wave dynamics in the expanding universe}
\label{gwdynamicsfrw}
As previously stated, the decomposition theorem in general relativity guarantees the independent evolution of scalar, vector, and tensor perturbations at linear order. Therefore, we may neglect scalar and vector degrees of freedom from now on, even if they were present, to study the dynamics of the tensor perturbations in the FLRW spacetime, i.e. cosmological gravitational waves. The process to be undertaken, therefore, is to set $A=B=S_i=D=E=F_i=0$ in the metric from Eq. \ref{perturbedFRW} and calculate from it the Christoffel symbols, then the Ricci tensor, the Ricci scalar, and, finally, the Einstein tensor. We have shown this process in more detail for gravitational waves on a flat background in Section \ref{GWderivation} and for an unperturbed FLRW spacetime in Section \ref{introcosmo}. Since the algorithm is precisely the same, and since the problem of interest now is precisely a combination of the two problems whose solution we have already shown, we shall omit more steps than before. If interested in more details, the reader is referred to Section 6.4 of Ref. \cite{Dodelson:2020bqr}. The spatial part of the Ricci tensor is then:
\begin{equation}
\label{ricciGWFRW}
    R_{ij}=g_{ij}\left(\frac{\ddot a}{a}+2H^2\right)+\frac{3}{2}a^2 H\partial_th_{ij}^{TT}+\frac{1}{2}a^2\partial^2_th_{ij}^{TT}-\frac{1}{2}\nabla^2h_{ij}^{TT}.
\end{equation}
where $g_{ij}$ refers to components of the full metric. It is often convenient to work in Fourier space, where $\partial_j\rightarrow ik_j$. We could compute the Ricci scalar from the Ricci tensor, but it is easy to see that it should vanish. Being a scalar, the factors of $h_{ij}^{TT}$ in it have to be contracted with other rank-1 or rank-2 tensors, but it turns out that all possible contractions are 0: contracting two factors of $h_{ij}^{TT}$ leads to a second-order term, discarded for our first-order analysis; contraction with $k^j$ is 0 because $h_{ij}^{TT}$ is transverse; contraction with $g^{jk}\propto\delta^{jk}$ is 0 because $h_{ij}^{TT}$ is traceless. As a result, using the notation of Ref. \cite{Dodelson:2020bqr} whereby $\delta \mathcal X$ refers to the first-order terms contained in $\mathcal X$, the first-order part of the Einstein equations read
\begin{equation}
\label{einsteinGWFRW}
    \delta \tilde G^{i}_{\;j}=\delta \tilde R^i_{\;j}=\delta^{ik}\left(\frac{3}{2} H\partial_t\tilde h_{kj}^{TT}+\frac{1}{2}\partial^2_t\tilde h_{kj}^{TT}+\frac{1}{2a^2}k^2\tilde h_{kj}^{TT}\right),
\end{equation}
where we have used that, to zeroth order, $g^{ik}=\delta^{ik}/a^2$, and tildes indicate a Fourier transform, i.e. position-dependence in Eq. \ref{ricciGWFRW} has become wave-number-dependence in Eq. \ref{einsteinGWFRW}. As in Eq. \ref{matrixeq}, we may, without loss of generality, specialize to a gravitational wave propagating in the $z$-direction. Then, dropping the tildes to make the notation cleaner,
\begin{equation}
\label{quase}
    \delta R^1_{\;1}-\delta R^2_{\;2}=2\delta R^1_{\;1}=3 H\partial_t h_++\partial^2_t h_++\frac{1}{a^2}k^2 h_+=\delta G^1_{\;1}-\delta G^2_{\;2}.
\end{equation}
It is common in cosmology to perform a coordinate transformation from time $t$ to conformal time $\eta$ such that the FLRW line element has an overall factor of $a^2$: 
\begin{equation*}
ds^2=a(\eta)^2\left[-d\eta^2+\frac{dr^2}{1-kr^2}+r^2d\Omega_2^2\right].
\end{equation*} 
$d\Omega_2^2$ is the line element on the 2-sphere, a common abbreviation for $d\theta^2+\sin^2\theta d\varphi^2$. Clearly, the coordinates are related by $d\eta=dt/a(t)$, such that the conformal time at time $t$ may be obtained from an integral of $1/a(t)$. The name ``conformal time'' comes from the fact that, switching to this measure of time, the scale factor can be removed from the line element by a conformal transformation, whereby, in general, $ds\rightarrow ds'=ds/C(x)$, where the conformal factor $C(x)$ is some function of space and time. Eq. \ref{quase} is commonly rewritten with $t\rightarrow \eta$:
\begin{equation}
\label{deltag}
    \delta G^1_{\;1}-\delta G^2_{\;2}=\frac{1}{a^2}\left(h_+''+2\frac{a'}{a}h'_++k^2h_+\right),
\end{equation}
where primes indicate derivatives with respect to $\eta$. 

One can show that the left-hand side of Eq. \ref{deltag} is zero by the Einstein equation in the absence of anisotropic stress \cite{Dodelson:2020bqr}. We may verify this for the case of inflation, in which, according to Eqs. \ref{perfectfluid} and \ref{pinfl}, $T^{i}_{\;j}=g^{ik}g_{kj}P=\delta^{i}_{\;j}[\dot\phi^2/2-V(\phi)]$ at the background level. First-order perturbations may be introduced to the inflaton matter field as $\phi=\bar\phi(t)+\delta\phi(t,\vec x)$, where the background value $\bar\phi$ is homogeneous and thus does not depend on $\vec x$. Then, $\dot\phi^2=\dot{\bar\phi}^2+2\dot{\bar\phi}\delta\dot\phi+O(\delta\dot\phi^2)$ and $V(\phi)=V(\bar\phi)+[\partial_\phi V(\phi)]\rvert_{\bar\phi}\:\delta \phi+O(\delta\phi^2)$. Collecting first-order terms, $\delta T^i_{\; j}=\delta^i_{\; j}\{\dot{\bar\phi}\delta\dot\phi-[\partial_\phi V(\phi)]\rvert_{\bar\phi}\:\delta \phi\}$, so that, clearly, $\delta T^1_{\;1}-\delta T^2_{\;2}=0=\delta G^1_{\;1}-\delta G^2_{\;2}$. As it turns out, $h_\times$ obeys the same equation, $h_\times''+2a'h'_\times/a+k^2h_\times=0$. Henceforth, we will use $h$ as an umbrella that encapsulates both, i.e. $h\in\{h_+,h_\times\}$.

We may perform a change of variables $\mathfrak h\equiv ah/\sqrt{16\pi G}$, where $1/\sqrt{16\pi G}$ is just a suitable normalization factor \cite{Dodelson:2020bqr}. After computing $\mathfrak h'$ and $\mathfrak h''$, substituting into Eq. \ref{deltag}, and simplifying, one finds
\begin{equation}
\label{mathfrakeom}
    \mathfrak h''+\left(k^2-\frac{a''}{a}\right)\mathfrak h=0.
\end{equation}
This second-order differential equation with no first-derivative terms looks like a harmonic oscillator with a variable frequency, $\omega_k(\eta)^2=k^2-a''(\eta)/a(\eta)$. We shall treat this harmonic oscillator quantum-mechanically, quantizing the tensorial perturbations to the metric for a simple theory of gravitons. Indeed, the canonical story told by contemporary cosmology is that it was quantum fluctuations in the fields—the inflaton and the metric—that originated tensorial and scalar perturbations and, therefore, primordial gravitational waves and density perturbations. Recalling the usual procedure from quantum mechanics and quantum field theory, we promote the variable $\mathfrak h$ to the operator $\hat{\mathfrak h}$ and write the quantum harmonic oscillator as a linear combination of creation and annihilation operators:
\begin{equation}
    \hat{\mathfrak h}(\vec k,\eta)=v(k,\eta)\hat a_{\vec k}+v^*(k,\eta)\hat a_{\vec k}^\dagger.
\end{equation}
The space spanned by the eigenstates $\ket {n,\vec k}$ of the number operator $N_{\vec k}\equiv \hat a_{\vec k}^{\dagger}\hat a_{\vec k}$ is called the Fock space. $\ket {n,\vec k}$ is interpreted as a state containing $n$ particles of wavenumber $\vec k$. The names annihilation and creation operators for $\hat a$ and $\hat a^\dagger$ respectively come from the fact that $\hat a\ket n\propto \ket{n-1}$ and $\hat a^\dagger\ket n\propto\ket{n+1}$, such that the action of the former is to ``destroy'' one particle, while the latter's is to ``create'' one. Hence, $\hat a\ket 0\equiv 0$. These operators satisfy the commutation relation \begin{equation}\label{aadagger}[\hat a_{\vec k},\hat a_{\vec{k'}}^\dagger]=(2\pi)^3 \delta(\vec{k}-\vec{k'}),\end{equation} reflecting the independent dynamics of modes with different wavenumbers. Clearly, the vacuum expectation value of $\hat{\mathfrak h}$ is 0:
\begin{equation*}
    \bra 0\hat{\mathfrak h}\ket 0= \bra 0\left(v\hat a_{\vec k}+v^*\hat a_{\vec k}^\dagger\right)\ket 0=v\bra 0\hat a_{\vec k}\ket 0+v^*\bra 0\hat a_{\vec k}^\dagger\ket 0=0,
\end{equation*}
given that $\bra 0 \hat a^\dagger=\left(\hat a\ket 0\right)^\dagger=0$. However, there exist quantum oscillations around this zero average, quantified by a nonzero variance and, consequently, by a nonzero expectation value of $\hat{\mathfrak h}^\dagger\hat{\mathfrak h}$:
\begin{equation*}
\begin{aligned}
    \bra0\hat{\mathfrak h}^\dagger (\vec k,\eta)\hat{\mathfrak h}(\vec{k'},\eta) \ket 0 =& \bra 0\left(v^*\hat a_{\vec k}^\dagger+v\hat a_{\vec k}\right)\left(v\hat a_{\vec k'}+v^*\hat a_{\vec k'}^\dagger\right)\ket 0\\
    =&|v|^2\bra0\hat a_{\vec k}\hat a_{\vec k'}^\dagger\ket0=|v|^2\bra0\left([\hat a_{\vec k},\hat a_{\vec k'}^\dagger]+\hat a_{\vec k'}^\dagger\hat a_{\vec k}\right)\ket0 \\
    =&|v|^2(2\pi)^3\delta(\vec k-\vec{k'}).
    \end{aligned}
\end{equation*}
Reverting back to $\hat h$, we end up with
\begin{equation}
\label{<h^2>}
    \bra0\hat h^\dagger(\vec k,\eta)\hat h(\vec{k'},\eta)\ket0=\frac{16\pi G}{a^2}|v(k,\eta)|^2(2\pi)^3\delta(\vec k-\vec{k'}).
\end{equation}

Let us henceforth drop hats on $h$ to simplify our notation. In analogous fashion to how we defined the noise PSD from the Fourier transform of the noise's autocorrelation function $\braket{n(t+\tau)n(t)}$ in Eq. \ref{defPSD}, we may define some quantities from the Fourier transform of $\braket{(h_{ij}^{TT})^\dagger(\tau,\vec x) \:h^{ij}_{TT}(\tau,\vec x)}$:
\begin{equation}
\label{powerspectrumworkout}
\begin{aligned}\braket{(h_{ij}^{TT})^\dagger(\eta,\vec x)\: h^{ij}_{TT}(\eta,\vec x)}=&\int\frac{\mathrm d^3k\;\mathrm d^3k'}{(2\pi)^6}\braket{(h_{ij}^{TT})^\dagger(\eta,\vec k)\: h^{ij}_{TT}(\eta,\vec{k'})}\:e^{i\vec x\cdot(\vec{k'}-\vec k)}\\
=&4\int\frac{\mathrm d^3k}{(2\pi)^3}\frac{16\pi G}{a^2}|v(k,\eta)|^2 \\
\equiv& 4\int\frac{\mathrm d^3k}{(2\pi)^3} P_h(k,\eta)\equiv\int\frac{\mathrm d^3k}{(2\pi)^3}P_T(k,\eta)\\
\equiv&\int\mathrm d(\ln k)\Delta_h^2(k,\eta),
\end{aligned}
\end{equation}
where the second equality follows from the fact that, once Eq. \ref{<h^2>} pertains to both $h_+$ and $h_\times$, $\braket{(h_{ij}^{TT})^\dagger(\eta,\vec k)\: h^{ij}_{TT}(\eta,\vec{k})}=2\braket{ h_+^\dagger h_+}+2\braket{ h_\times^\dagger h_\times}=4\braket{h^\dagger h}$. In sum, the definitions we have presented are:
\begin{align}
    \label{Ph}P_h(k,\eta)\equiv&\frac{16\pi G}{a^2}|v(k,\eta)|^2\\
    \label{PT}P_T(k,\eta)\equiv&4P_h(k,\eta)\\
    \label{Deltah}\Delta_h^2(k,\eta)\equiv&\frac{k^3}{2\pi^2}P_T(k,\eta).
\end{align}
$P_h$ is the power spectrum for a single polarization, while $P_T$ is the total power spectrum of tensor perturbations; $\Delta_h^2$ is often referred to as the dimensionless power spectrum.

Since the creation and annihilation operators are time-independent and linearly independent, the function $v(k,\eta)$ must also obey a differential equation as the one for $\mathfrak h$ in Eq. \ref{mathfrakeom}. To get a solution for it valid during inflation, we first need to compute $a''/a$. Given that $H$ is nearly constant during inflation, 
\begin{equation}
\label{eta(a)nao}
    \eta(t_f)-\eta(t_i)=\int_{t_i}^{t_f}\frac{\mathrm dt}{a(t)}=\int_{a_i}^{a_f}\frac{\mathrm da}{Ha^2}\approx\frac{1}{aH}
\end{equation}
since $a_f\gg a_i$ for $t_f$ considerably later than $t_i$ during inflation. Eq. \ref{eta(a)nao} informs how much conformal time elapses from an earlier to a later time during inflation. To define the value of $\eta$ for all times, we need to enforce some initial condition, which may be chosen arbitrarily. It is intuitive and usual to set $\left.\eta\right|_{\mathrm{end\:of\:inflation}}=0$, such that, during inflation,
\begin{equation}
    \label{eta(a)}
    \eta\approx-\frac{1}{aH}.
\end{equation}
Consequently, $a'=1/H\eta^2$ and
\begin{equation*}
    \frac{a''}{a}=-\frac{1}{a}\frac{2}{H\eta^3}=\frac{2}{\eta^2}.
\end{equation*}
The dynamics of tensor perturbations during inflation will then be completely determined by solving the differential equation
\begin{equation}
\label{vdyn}
    v''+\left(k^2-\frac{2}{\eta^2}\right)v=0.
\end{equation}
One can verify that the solution to this problem is \cite{Dodelson:2020bqr}
\begin{equation}
\label{vsol}
    v(k,\eta)=\frac{e^{-ik\eta}}{\sqrt{2k}}\left(1-\frac{i}{k\eta}\right).
\end{equation}
The $1/\sqrt{2k}$ normalization factor ensures that the commutator of $\hat a$ and $\hat a^\dagger$ in Eq. \ref{aadagger} corresponds to the canonical commutation relation for the field and its conjugate momentum, $i\delta(\vec x-\vec{x'})$. 

Note, in light of Eq. \ref{eta(a)}, that, in the far past, $k\gg1/|\eta|\rightarrow0$, so that Eq. \ref{vdyn} reduces to a simple harmonic oscillator's equation of motion with a constant frequency; correspondingly, the solution for $v$ in Eq. \ref{vsol} becomes a complex exponential, the well-known oscillatory solution to that simpler problem. 

In the opposite limit, after many e-foldings of inflation, $k\ll1/|\eta|$ and 
\begin{equation}
\label{vlim}
    v\approx\frac{-ie^{-ik\eta}}{ k\sqrt{2k}\eta}\therefore|v|^2\approx\frac{1}{2k^3\eta^2}.
\end{equation}
Therefore,
\begin{equation}
\label{frozenP}
    P_T(k,\eta)\propto\braket{\hat h^\dagger\hat h}\propto\frac{|v|^2}{a^2}\propto\frac{1}{k^3\eta^2a^2}\propto\frac{1}{k^3}\propto P_T(k),
\end{equation}
where the fourth relation follows from Eq. \ref{eta(a)}. The variance in tensorial fluctuations has, then, become time-independent, it has frozen, after sufficient inflation has taken place to make $k\ll1/|\eta|$. And we can neatly make sense of the physical picture behind this behavior. In the sense described in Section \ref{thehorizonproblem}, inflation's de Sitter universe does not have a particle horizon:
\begin{equation*}
    a(t_0)\int_{-\infty}^{t_0}\frac{\mathrm dt'}{a(t')}=\int_{-\infty}^{t_0}\frac{\mathrm dt'}{e^{H(t'-t_0)}}=\left[\frac{-1}{H}e^{-H(t'-t_0)}\right]_{t'\rightarrow-\infty}^{t'=t_0}=\infty,
\end{equation*}
where again we have used the convention $a(t_0)=1$. This means that regions that are arbitrarily far away at a present time $t_0$ had been in causal contact at some point in a sufficiently far past. Nevertheless, de Sitter does have an event horizon:
\begin{equation*}
    a(t_0)\int^{\infty}_{t_0}\frac{\mathrm dt'}{e^{H(t'-t_0)}}=\left[\frac{-1}{H}e^{-H(t'-t_0)}\right]^{t'\rightarrow\infty}_{t'=t_0}=\frac{1}{H},
\end{equation*}
meaning that some causal influence originating from the present time $t_0$ can only reach a physical distance of at most $1/H$, even if it is given an infinite amount of time to propagate.

Waves of any sort, including the inflationary gravitational waves under consideration, are local phenomena, requiring time for causal contact to be established between neighboring points and for this influence to propagate across many points so that its global oscillatory pattern may be formed. If, therefore, a wave's wavelength is much larger than the universe's event horizon, that causal contact required for the oscillation to take place will not be possible. Note that the late-time requirement we imposed to obtain Eq. \ref{frozenP}, $k\ll1/|\eta|$, coincides precisely with that scenario in light of Eq. \ref{eta(a)}: $k/a\ll H$\footnote{Note that, ever since we brought our discussion to wavenumber space, we have been dealing with comoving wavenumbers, and $k_{physical}=k_{comoving}/a$ because $k$ has dimensions of inverse length. It is worth emphasizing that $1/H$ is the physical, and not the comoving, horizon size.}. Therefore, indeed, after inflation has unfolded for long enough—and whether it has been long enough is $k$-dependent—, gravitational waves will have had their wavelengths stretched to super-horizon sizes, so that their oscillations will stop and their global profile will freeze. They will remain frozen until after the end of inflation; during some subsequent era, with the decrease of $H$, these gravitational waves will re-enter the horizon and their dynamical behavior will resume. 

The value to which $v$ and, thus, $P_T$ will freeze for $k\lesssim aH$ may be determined by plugging Eq. \ref{vlim} into Eq. \ref{Ph} using Eqs. \ref{PT}–\ref{eta(a)}:
\begin{equation}
\label{idealPT}
    P_{T,dS}(k)=\frac{64\pi G}{a^2}\frac{1}{2k^3\eta^2}=32\pi G H^2k^{-3}\therefore\Delta_{h,dS}^{2}=\frac{16 GH^2}{\pi},
\end{equation}
where we have used the subscript $dS$ to label quantities referring to exact de Sitter. As was explicitly indicated, much of the discussion we have just presented is approximate because $H$ is only approximately constant during inflation, such that spacetime is only approximately de Sitter. Let us consider small deviations in the $k$-dependence of $\Delta_{h}^2$ arising from a slowly-varying $H$ via a power-law parametrization with spectral index $n_T$:
\begin{equation}
    \label{powerlawdelta}\Delta_h^2(k)=\left.\frac{16GH^2}{\pi}\right|_{H=k/a}=\Delta_h^2(k_*)\left(\frac{k}{k_*}\right)^{n_T},
\end{equation}
Then,
\begin{equation*}
    n_T=\frac{d\ln\Delta_h^2}{d\ln k}=\left.2\frac{k}{H}\frac{dH}{dk}\right|_{H=k/a}=\left.2\frac{k}{H}\frac{dH}{d\eta}\frac{d\eta}{dk}\right|_{H=k/a},
\end{equation*}
where the first equality here comes from the second equality in Eq. \ref{powerlawdelta} and vice-versa. By Eq. \ref{epsilon}, $H'=a\dot H=-aH^2\epsilon$. By Eq. \ref{eta(a)}, 
\begin{equation*}
    \left.\frac{d\eta}{dk}\right|_{H=k/a}=\left.\frac{d}{dk}\left(-\frac{1}{aH}\right)\right|_{H=k/a}=-\frac{d}{dk}\frac{1}{k}=\frac{1}{k^2}.
\end{equation*}
Finally, then,
\begin{equation}
\label{slowrollnt}
    n_T=-2\epsilon \left.\frac{aH}{k}\right|_{H=k/a}=-2\epsilon.
\end{equation}
This result is central to our interests. During slow roll, $H^2\sim V(\phi)$ (see Eqs. \ref{friedmann} and \ref{rhoinfl}). Having in mind, then, that slow roll unfolds as schematized in Fig. \ref{slowroll}, it is easy to remember that $V$ slowly decreases during inflation. Consequently, so does $H$, and $\epsilon$, which has a minus sign in its definition and serves as a proxy for $\dot H$, is small and positive for a slowly decreasing $H$. As will be presented in Figs. \ref{omegaT}–\ref{omegar}, we are interested in primordial gravitational-wave spectra with $|n_T|\sim O(1)$ and $n_T>0$ to try to explain the signal observed by the NANOGrav pulsar timing array. Given that $|n_T|\ll1$ and $n_T<0$ were derived from general features of slow-roll inflation, a model of the early universe capable of producing a signal like the one observed by NANOGrav might require new ingredients in addition to or in place of a phase of slow-roll inflation. Nevertheless, it is not easy to get rid of a phase of slow-roll inflation altogether: although the fact that tensorial perturbations have not been measured leaves leeway for a broad range of theoretical models with different gravitational-wave spectra—as long as they do not violate observational constraints—, scalar modes, related to density perturbations, have been measured with high precision in the temperature anisotropies of the CMB, and slow-roll inflation is very successful in predicting scalar power spectra that fit the data exquisitely.

\section{Energy of primordial gravitational waves and the transfer function}
As discussed following Eq. \ref{frozenP}, after a tensor perturbation mode of wavenumber $\vec k$ is stretched to wavelengths larger than the cosmological horizon, its evolution remains time-independent until it re-enters the cosmological horizon during some later stage in the universe's history. Therefore, it is common to write the current gravitational-wave amplitude $h(\eta,\vec k)$—where, as in Section \ref{gwdynamicsfrw}, $h\in\{h_+,h_\times\}$—as a time-independent primordial part seeded during inflation and a time-dependent transfer function encapsulating the amplitude's evolution after horizon re-entry:
\begin{equation}
\label{transfer}
    h(\eta,\vec k)=h_{prim}(\vec k)\:T_h(k,\eta).
\end{equation}
$T_h$ is normalized such that $T_h\rightarrow 1$ as $k\rightarrow 0$, reflecting the fact that arbitrarily large-wavelength modes have always stayed outside the cosmological horizon and, thus, never picked up time dependence.

Before studying the form of the transfer function, let us determine the energy density of gravitational waves in terms of the variables in Eq. \ref{transfer}. Recall from Eq. \ref{energyinGWs} that
\begin{equation}
    \rho_{gw}=t_{00}=\frac{1}{32\pi G}\braket{\dot h^{TT}_{ij}\dot h_{TT}^{ij}}=\frac{1}{32\pi Ga^2}\left\langle\frac{d h^{TT}_{ij}}{d\eta} \frac{d h_{TT}^{ij}}{d\eta}\right\rangle.
\end{equation}
Then, in light of Eqs. \ref{<h^2>}–\ref{Deltah}, 
\begin{equation}
    \rho_{gw}=\frac{1}{32\pi Ga^2}\int\mathrm d(\ln k)\Delta_{h,prim}^2(k)[T_h'(k,\eta)]^2.
\end{equation}
Note that $\Delta^2_h\propto T_h^2$ because $\Delta_h^2\sim h^2$. We can also write the gravitational-wave density parameter following Eq. \ref{omegagw}:
\begin{equation}
\label{ultom}
    \Omega_{gw}=\frac{1}{\rho_c}\frac{d\rho_{gw}}{d\ln k}=\frac{\Delta_{h,prim}^2(k)}{12a(\eta)^2H(\eta)^2}[T_h'(k,\eta)]^2
\end{equation}
As discussed in Ref. \cite{Watanabe:2006qe}, 
since $T_h$ is usually given by linear combinations of Bessel-type functions—as we are about to touch upon—, for modes deep inside the horizon—i.e. $k\gg aH$, which are the modes of interest to study phenomena that are relevant to astronomy, such as gravitational waves observable by PTAs—, $[T_h'(k,\eta)]^2\approx k^2[T_h(k,\eta)]^2$. This approximation is often used in the literature.

Solving Eq. \ref{deltag} numerically is often manageable, and one could choose to proceed this way. Nevertheless, analytic approximations are frequently used in the literature, and our next step is to go over the construction of such an approximation.

During the matter-dominated phase, Eq. \ref{a(t)} informs that $a\propto t^{2/3}$, so that $\eta=\int\mathrm d t\;a(t)^{-1}\propto t^{1/3}$ and $a\propto \eta^2$. This means that, in Eq. \ref{deltag}, $2a'/a=4/\eta$. This differential equation can be shown to have as a solution
\begin{equation}
    h(k,\eta)=h_{in}(k) \frac{3j_1(k\eta)}{k\eta},
\end{equation}
where $h_{in}(k)$ is the initial condition and $j_1(x)=x^{-2}(\sin x-x\cos x)$ is a spherical Bessel function \cite{Maggiore:2018sht}. A factor of 
\begin{equation}
\label{bessel}
    \left(\frac{3j_1(k\eta)}{k\eta}\right)^2
\end{equation}
is hence included in the transfer function. The effect of the later, recent end of matter domination and transition to dark energy domination can be approximately accounted for by including a damping factor of 
\begin{equation}
\label{turner}
    \Omega_{m}^2
\end{equation}
in $T_h$ \cite{Turner:1993vb}. Note that this would be just $1$ had we not evolved toward dark energy domination. Next, we consider the final factors in the transfer function coming from earlier phases of the universe's evolution.

From statistical mechanics, we know that the energy density $\rho$ of some species of mass $m$ may be written as 
\begin{align}
    \label{rhostatmech}&\rho = \frac{g}{(2\pi\hbar )^3}\int\mathrm d^3p\;E(\vec p)f(\vec p), \\
    &f(\vec p)=\frac{1}{e^{E(\vec p)-\mu}\pm 1},\\ &E(\vec p)=\sqrt{|\vec p|^2+m^2},
\end{align}
where $g$ counts the number of internal degrees of freedom of this species (e.g. the number of possible spin values), $\mu$ is the chemical potential, and the plus sign in $f(\vec p)$ describes the statistics of fermions, while the minus sign, that of bosons. In the $T\gg m,\mu$ regime, Eq. \ref{rhostatmech} can be integrated to \cite{Kolb:1990vq}
\begin{align}
    \rho_{boson}&=\frac{\pi^2}{30}gT^4\\
    \rho_{fermion}&=\frac{7}{8}\rho_{boson}.
\end{align}
For a bath of relativistic particles in which photons have temperature $T$, this is often summarized as 
\begin{align}
   \label{rhorel} \rho &= \frac{\pi^2}{30}g_*T^4,\\
    \label{gstar}g_*&=\sum_{b=bosons}g_b\frac{T_b^4}{T^4}+\frac{7}{8}\sum_{f=fermions}g_f\frac{T_f^4}{T^4}.
\end{align}
$g_*$ is understood as an effective number of relativistic degrees of freedom. Eq. \ref{rhorel} can be used to compute an entropy density $s\equiv S/V$ for the universe. First, since entropy is extensive in the homogeneous and istropic universe, $s=\partial S/\partial V$. Then, if we recall the first law of thermodynamics, i.e. conservation of energy, whereby the heat exchanged is the same as the change in the internal energy added to the work:
\begin{equation}
\begin{aligned}
\label{termo1}
    &dQ=TdS=dU+dW=\partial_T UdT+(\rho+P)dV\\
    \therefore\;&dS=(\partial_T S)dT+(\partial_V S)dV=\frac{1}{T}\partial_T UdT+\frac{\rho+P}{T}dV.
\end{aligned}
\end{equation}
From there, we can read off $\partial_T S$ and $\partial_V S$. In particular, this informs that the entropy density is $s=(\rho+P)/
T$. In radiation domination, in line with Eq. \ref{rhorel}, this is
\begin{align}
    \label{srel}s &= \frac{2\pi^2}{45}g_{*s}T^3,\\
    \label{gstars}g_{*s}&=\sum_{b=bosons}g_b\frac{T_b^3}{T^3}+\frac{7}{8}\sum_{f=fermions}g_f\frac{T_f^3}{T^3}.
\end{align}
Note that, if all species are in thermal equilibrium—which was the case in the thermal bath of the early radiation-dominated universe—, $g_*=g_{*s}$. 

It is noteworthy that, for the entropy $S$ to be well-defined mathematically, it is necessarily true that $\partial^2S/\partial T\partial V=\partial^2S/\partial V\partial T$, so that, in light of Eq. \ref{termo1},\begin{equation}\label{dPdT}\frac{\partial P}{\partial T}=\frac{\rho+P}{T}.\end{equation}
This relation can be used to write $\dot P=s\dot T$ ($P$ is an intensive quantity in the homogeneous and isotropic universe, so $\partial_V P=0$), which can be combined with $\dot \rho=-3H(\rho+P)$ (Eq. \ref{conservationcosmo}) to obtain $\dot s/s=-3H$. Since $S\propto a^3s$,
\begin{equation*}
    \frac{\dot S}{S}=3H+\frac{\dot s}{s}=0,
\end{equation*}
showing that $S$ is a conserved quantity in the expansion of the homogeneous and isotropic universe. 

According to Eq. \ref{rhorel}, $\rho_r\propto g_*T^4$ and, according to Eq. \ref{srel}, $sa^3=const\propto g_{*s}T^3a^3\therefore T^4\propto g_{*s}^{-4/3}a^{-4}$. Consequently, $\rho_r\propto a^{-4}g_{*s}^{-4/3}g_*$. This shows that, as the universe cools, the fact that some particles become non-relativistic and $g_*$ and $g_{*s}$ decrease introduces a correction to the simpler behavior $\rho\propto a^{-4}$ (see Eq. \ref{rho(t)}), which holds for the case in which the universe is filled by particles that are eternally relativistic—this would be, for example, a universe that only contains photons. Gravitons, on the other hand, are not thermally coupled to other particles and are always relativistic, such that $\rho_{gw}\propto a^{-4}$. As a consequence,
\begin{equation*}
    \frac{\rho_{gw}}{\rho_r}\propto g_{*s}^{4/3}g^{-1}_*\therefore\frac{\rho_{gw}(t_0)}{\rho_r(t_0)}\frac{\rho_r(t_{in})}{\rho_{gw}(t_{in})}=\frac{g_{*s0}^{4/3}g^{-1}_{*0}}{g_{*s}^{4/3}(T_{in})g^{-1}_*(T_{in})}.
\end{equation*}
If a tensor mode re-enters the horizon deep in the radiation-dominated era, $\Omega_r(t_{in})=1$ and \begin{equation*}\frac{\Omega_{gw}(t_{0})}{\Omega_{gw}(t_{in})}=\Omega_{r0}\frac{g_{*s0}^{4/3}g^{-1}_{*0}}{g_{*s}^{4/3}(T_{in})g^{-1}_*(T_{in})}.\end{equation*}
As first pointed out by Ref. \cite{Watanabe:2006qe}, this implies that the transfer function $T_h$ should contain a factor of 
\begin{equation}
\label{g*}
    \frac{g_*(T_{in})}{g_{*0}}\left(\frac{g_{*s}(T_{in})}{g_{*s0}}\right)^{-4/3}.
\end{equation}
Note that, otherwise, if a mode re-entered the horizon later, it is usually the case that $g_*(T_{in})=g_{*0}$ and $g_{*s}(T_{in})=g_{*s0}$. This is because most particles decoupled from the thermal bath early on and have been non-relativistic for most of the history of the universe, such that $g_{*}$ and $g_{*s}$ have had their present values for most of the history of the universe\footnote{Neutrinos involve caveats, to which we turn a blind eye.}. Hence, in this case, that factor in the transfer function is just 1.

Finally, additional multiplicative factors may be added to the transfer function to describe the behavior of modes that enter the horizon before matter domination. Those come from numerically solving Eq. \ref{deltag} and the Friedmann equation (\ref{friedmann}) during radiation domination and the radiation-matter transition, as well as, for reheating, solving these equations in addition to Boltzmann equations describing the decay of the inflaton into radiation \cite{Nakayama:2008wy}. These solutions are then used to construct fitting functions that will be multiplicative factors in $T_h$, which are $T_1$ and $T_2$ respectively. Ref. \cite{Kuroyanagi:2014nba}, for example, determines these to be: 
\begin{align}
    \label{T1}&T_1^2(x)=1+1.57x+3.42x^2\\
    \label{T2}&T_2^2(x)=\frac{1}{1-0.22x^{3/2}+0.65x^2}.
\end{align}
These are to be evaluated at $x_t\equiv k/k_t$, where $k_t=a(t)H(t)$ is the comoving wavenumber of the gravitational wave that enters the horizon at time $t$. For radiation-matter equality, $k_{eq}=7.1\times10^{-2}\Omega_{m,0}h^2$ $\mathrm{Mpc}^{-1}$—where $h\equiv H_0/(100\;\mathrm{km\;s^{-1}\;Mpc^{-1}})$)—and, for the end of reheating, $k_R\approx1.7\times10^{14}\mathrm{\; Mpc^{-1}}[g_{*s}(T_R)/106.75]^{1/6}(T_R/10^7\mathrm{\;GeV})$ \cite{Kuroyanagi:2014nba}. Note that, for modes that re-enter the horizon later, for which $k$ is smaller, $T_1,T_2\rightarrow1$, reflecting the fact that their evolution should not be influenced by reheating and radiation domination.

Putting together Eqs. \ref{bessel}, \ref{turner}, \ref{g*}, \ref{T1}, and \ref{T2}, the transfer function is finally given by:
\begin{equation}
\label{transferfunctionfinal}
    T_h^2(k,\eta)=\Omega_{m}^2\left(\frac{3j_1(k\eta)}{k\eta}\right)^2\frac{g_*(T_{in})}{g_{*0}}\left(\frac{g_{*s}(T_{in})}{g_{*s0}}\right)^{-4/3}T_1^2(x_{eq})T_2^2(x_R).
\end{equation}

\section{Observational data concerning primordial gravitational waves}
\subsection{From the NANOGrav pulsar timing array}
As discussed in Section \ref{PTA}, PTAs search for gravitational waves by observing their effect on the time of arrival of an oscillatory signal—traditionally a pulsar's radio emission, but see Ref. \cite{alves2024artificialprecisiontimingarray} for a more general proposal—, which happens since gravitational waves will alter a photon's travel time because they stretch and contract space. In order to detect some oscillatory signal, surely observed data should capture at least one or a few complete oscillations. As a consequence, for a signal of frequency $f=1/T$, $T$ sets a rough minimum requirement for how long a successful observation campaign should last—to get an appreciable signal-to-noise ratio, one should need a duration of a few times $T$ at least. This is the reason why the NANOGrav PTA, targeting gravitational waves of frequencies as low as $f\sim 1$ nHz, has required more than a decade accumulating data to start seeing statistically significant evidence for the existence of a stochastic gravitational-wave background ($1$ $\mathrm{yr}^{-1}\approx32$ nHz) \cite{NANOGrav:2023gor}.

For its data analysis and results report, NANOGrav utilizes a power-law parametrization of the characteristic strain $h_c$ of the form
\begin{equation}
	h_c(f)=A_{GWB}\left(\frac{f}{1\;\mathrm{yr}^{-1}}\right)^{(3-\gamma)/2},
\end{equation}
where $\gamma$ is then the spectral index for the spectral density of timing residuals \cite{NANOGrav:2020bcs}. Recall from Eq. \ref{omegagw} that $\Omega_{gw}(f)\propto f^2h_c(f)^2$ and, from Eq. \ref{phinney}, that $\Omega_{gw}(f)\propto f^{2/3}$ for a stochastic background from compact object binaries—by the way, in the nHz regime, these are supermassive black holes, since lighter compact objects would be too far from merger and, hence, too early in the inspiral process, if at all in it, to produce gravitational waves of detectable strain (see e.g. Section 2 in Ref. \cite{alves2024artificialprecisiontimingarray}). Therefore, $\gamma=13/3$ for a stochastic background from supermassive black hole binaries. 

Our knowledge regarding astrophysical black holes is much better constrained by observational data than that about primordial gravitational waves. On the one hand, we have directly observed supermassive black holes in the center of galaxies with electromagnetic-wave observatories and have solid models establishing that most galaxies have one of those at their centers—and the distribution of galaxies in the universe is rather well-catalogued. Moreover, the direct detection of gravitational waves from dozens of stellar-mass compact object mergers by the LIGO-Virgo-KAGRA collaboration reassures us that numerical relativity can accurately model the emission process of binaries. On the other hand, since the observational search for primordial gravitational waves has barely reached its dawn—with completed CMB missions finding none and new CMB and gravitational-wave missions being planned for the near-future to look for them—, there is still a vast landscape of unconstrained models of the early universe, in which gravitational waves are produced with different amplitudes and spectra. As a result, binaries of supermassive black holes have been, for a sizable fraction of the community, the preferred hypothesis to explain the NANOGrav signal. For this reason, for example, Ref. \cite{NANOGrav:2023gor} provides results of a Bayesian analysis of their data with a $\gamma=13/3$ power-law prior, for which they obtain a median with 90\% credible interval of $A_{GWB}=2.4^{+0.7}_{-0.6}\times10^{-15}$. Inputting a free-spectral-index power law, they obtain $\gamma=3.2^{+0.6}_{-0.6}$ and a larger amplitude: $A_{GWB}=6.4^{+4.2}_{-2.7}\times10^{-15}$. Note that $13/3=4.\bar3$, such that $\gamma=13/3$ is outside the 90\% credible interval of the latter results. While population variance might bring $(A_{GWB},\gamma)$ for the actual gravitational-wave background from binaries closer to the results reported by NANOGrav, as shown in Fig. 11 of Ref. \cite{NANOGrav:2023gor}, these results leave the possibility of other sources for the gravitational-wave stochastic background plausible, motivating our endeavor. 

\subsection{From Big Bang nucleosynthesis}
\label{bbn}
In the 1950s and 1960s, the G. Burbidge, M. Burbidge, Fowler, and Hoyle (BBFH) paradigm attempted to explain the origin of chemical elements assuming that they were all formed in stellar interiors and supernovae. In spite of its successes, it had important shortcomings, like its prediction of a low abundance of helium, which makes up $24\%$ of the baryonic matter in the universe, an abundance that is only behind hydrogen's. 

The current paradigm, which filled in the gaps of the BBFH theory with great success, is Big Bang nucleosynthesis (BBN), which posits that light elements had a primordial origin through nuclear reactions in the hot medium that filled the radiation-dominated universe. Its ability to explain current element abundances renders it a high-precision probe of early-universe physics. Not only that, the fact that helium makes up more than a fifth of baryonic matter everywhere we look is now interpreted as strong evidence that the universe went through a hot early phase—that is, it serves as important evidence in favor of the Hot Big Bang model.

When neutrons and protons decouple, the ratio of their number densities remains approximately frozen—with small changes coming from weak-interaction processes, mainly the decay of free neutrons—at the value $n_n/n_p\approx\exp{[(m_p-m_n)/T_f]}$, where $m_n$ and $m_p$ are their masses and the freeze-out temperature $T_f$ is affected by the effective number of relativistic degrees of freedom roughly as $T_f\sim g_*^{1/6}$ \cite{2000PhR...331..283M,Kolb:1990vq}. As a result, the abundance of neutrons that will be available to form helium during nucleosynthesis is very sensitive to $g_*$.

At the BBN temperature of $\sim 1$ MeV, the relativistic standard model species are the photon, the electron, the positron, 3 neutrinos, and 3 antineutrinos, and they are in thermal equilibrium. Therefore, the effective number of relativistic degrees of freedom is $g_{*,SM}=2+(7/8)(2+2+3+3)=43/4$—note that photons, electrons, and positrons have two helicity states while neutrinos and antineutrinos have only one, justifying the presence of some factors of 2 in the determination of $g_{*,SM}$. On top of standard model particles, other relativistic species could have been present during BBN. Nevertheless, $g_*$ could not have been too far from 43/4, or the resulting abundances of light elements would disagree with the observations. $g_*$ is usually parametrized by an effective number of neutrino species by setting $g_{*}=2+(7/8)(4+2N_{eff})$.

Assuming that, besides the standard model particles, gravitons are the only relativistic species during BBN places an upper limit in their contribution to $g_{*}$ and, hence, $N_{eff}$. As worked out in Ref. \cite{2000PhR...331..283M}, the precisely measured value of $h^2\Omega_{r,0}$ can be used to write
\begin{equation}
    \label{BBNconstraint}
    \int_{f_{BBN}}^{f_{end}}\mathrm d(\ln f)h^2\Omega_{gw}(f)\leq 5.6\times10^{-6}(N_{eff}-3),
\end{equation}
where $f_{BBN}\approx1.8\times10^{-11}$ is the frequency of the tensor mode that crossed the horizon during BBN \cite{Boyle:2007zx} and, for inflationary gravitational waves, $f_{end}$ is the frequency of the tensor mode that crossed the horizon around the time of the end of inflation—more generally, $f_{end}$ is the high-frequency cutoff of the primordial gravitational-wave spectrum. Observations of $\mathrm{^4He}$ and D reveal that $N_{eff}-3\lesssim0.4$ \cite{Mossa:2020gjc,2020ApJ...896...77H,2018ApJ...855..102C,2016RvMP...88a5004C}. Despite being a constraint on the integral of $\Omega_{gw}(f)$, in the literature, this is most often used as a constraint on the value of $\Omega_{gw}(f)$ itself. These are indeed roughly equivalent in this case for $f_{BBN}<f<f_{end}$, unless $\Omega_{gw}(f)$ has some sufficiently sharp peak—and these are usually not predicted by inflationary models \cite{Boyle:2007zx}. Hence, $\Omega_{gw}(f)<5.6\times10^{-6}(N_{eff}-3)\sim2.2\times10^{-6}$. To take a closer look at why this is reasonable, consider a frequency-independent $\Omega_{gw}$. Then, \begin{equation}
    \int_{f_{BBN}}^{f_{end}}\mathrm d(\ln f)h^2\Omega_{gw}=\Omega_{gw}h^2\ln\left(\frac{f_{end}}{f_{BBN}}\right).
\end{equation} 
We know with high confidence from observations that $h\in(0.6,0.8)$ \cite{Planck:2018vyg,Riess:2019cxk}. It is also safe to assert that $\ln(f_{end}/f_{BBN})\sim O(10)-O(10^2)$ (for reference, $\ln10^2\approx4.6$ and $\ln10^{80}\approx180$). Therefore, $h^2\ln(f_{end}/f_{BBN})\sim O(1)-O(10)$. In conclusion, for the flat approximation, $\Omega_{gw}h^2\ln(f_{end}/f_{BBN})< 2.2\times10^{-6}\therefore\Omega_{gw} < 2.2\times10^{-6}/C<2.2\times10^{-6}$, where $C$ is a number whose order of magnitude is between 1 and 10.

\subsection{From ground-based gravitational-wave interferometers}
Interferometers of the LIGO-Virgo collaboration are sensitive enough to perform gravitational-wave astronomy in the $O(10)-O(1000)$ Hz band. They have concluded three observing runs, across which they have successfully detected dozens of mergers of compact-object binaries \cite{KAGRA:2021vkt,LIGOScientific:2020ibl,LIGOScientific:2018mvr}, and the fourth one is currently ongoing as of 2025. They have also been trying to search for a stochastic background of gravitational waves but, so far, no signal has been found. Given that the detector's sensitivity is known, they can place upper limits on the intensity of the possible stochastic background for some frequencies. Their results, however, are model-dependent: one has to assume some form for the background's power spectrum in order to constrain its amplitude and to assert to what range of frequencies the bound applies. For example, assuming a frequency-independent gravitational-wave background, their most updated constraint is $\Omega_{gw}\leq5.8\times10^{-9}$ in the $20-76.6$ Hz band; for a power-law in the frequency with spectral index $2/3$—expected for a superposition of compact binary coalescences as discussed in Section \ref{2/3sec}—$\Omega_{gw}(f=25\mathrm{\:Hz})\leq3.4\times10^{-9}$ in the $20-90.6$ Hz band \cite{KAGRA:2021kbb}. Ref. \cite{Kuroyanagi:2014nba} provides a formula to translate the LIGO-Virgo constraint for a flat background into one for a power-law spectrum with an arbitrary spectral index.

\section{Blue-tilted tensor spectrum and observations}
\begin{figure}
    \centering
    \includegraphics[width=0.75\linewidth]{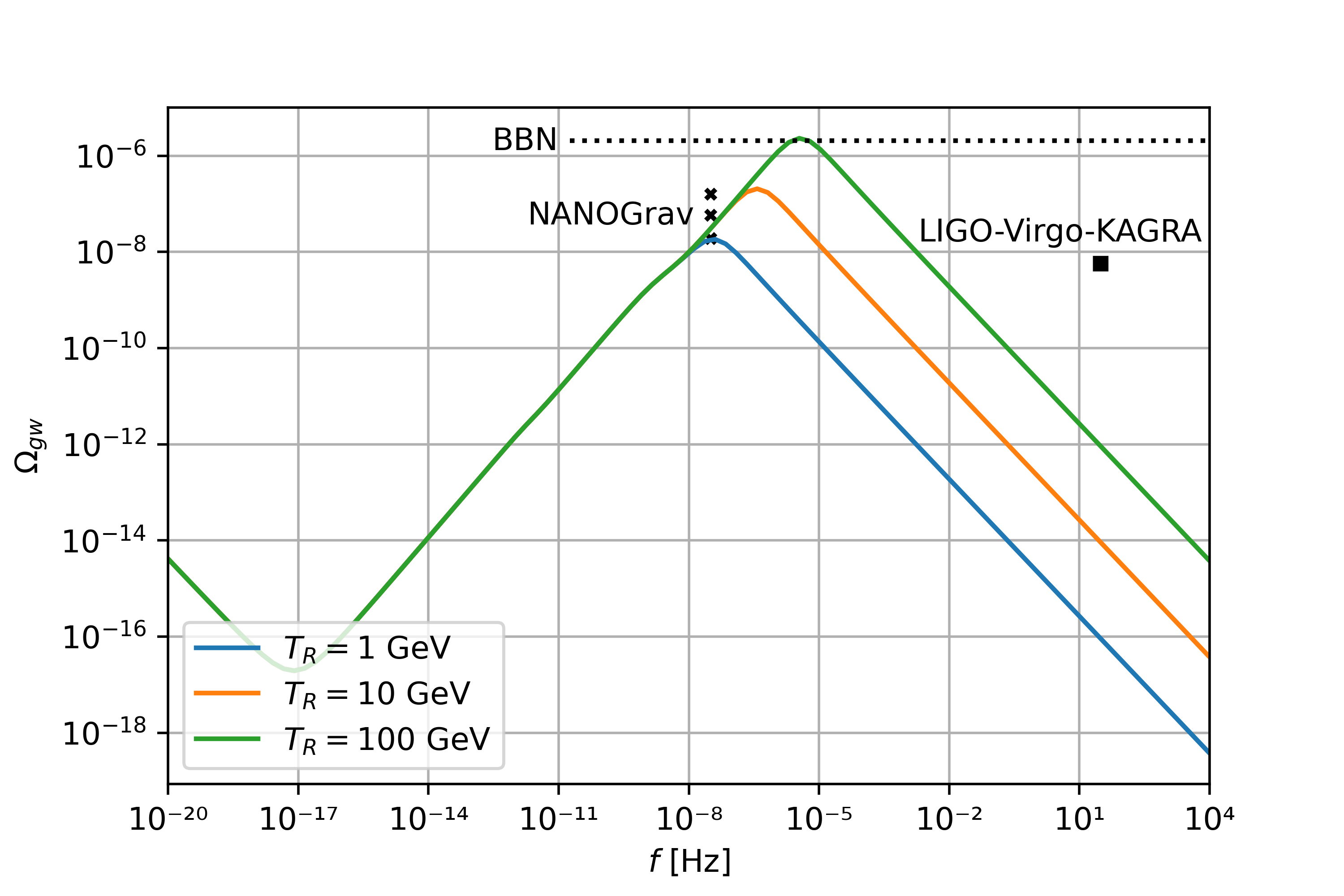}
    \caption{Density parameter $\Omega_{gw}$ as a function of frequency $f$ for inflationary gravitational waves is shown as solid curves, whose different colors label different reheating temperatures $T_R$ as indicated in the figure. Input parameters for the inflationary gravitational waves set by observation or chosen for theoretical modeling purposes are discussed in the text. The dotted line demarcates the BBN constraint, and the black square, the LIGO-Virgo-KAGRA one—models whose inflationary gravitational-wave curves lie above those are consequently ruled out by observation. The three small crosses pertaining to NANOGrav indicate the most recent median and extrema of the 90\% confidence interval for their most recent data set, analyzed leaving the $\gamma$ parameter free as discussed in the text \cite{NANOGrav:2023gor}.}
    \label{omegaT}
\end{figure}

\begin{figure}
    \centering
    \includegraphics[width=0.75\linewidth]{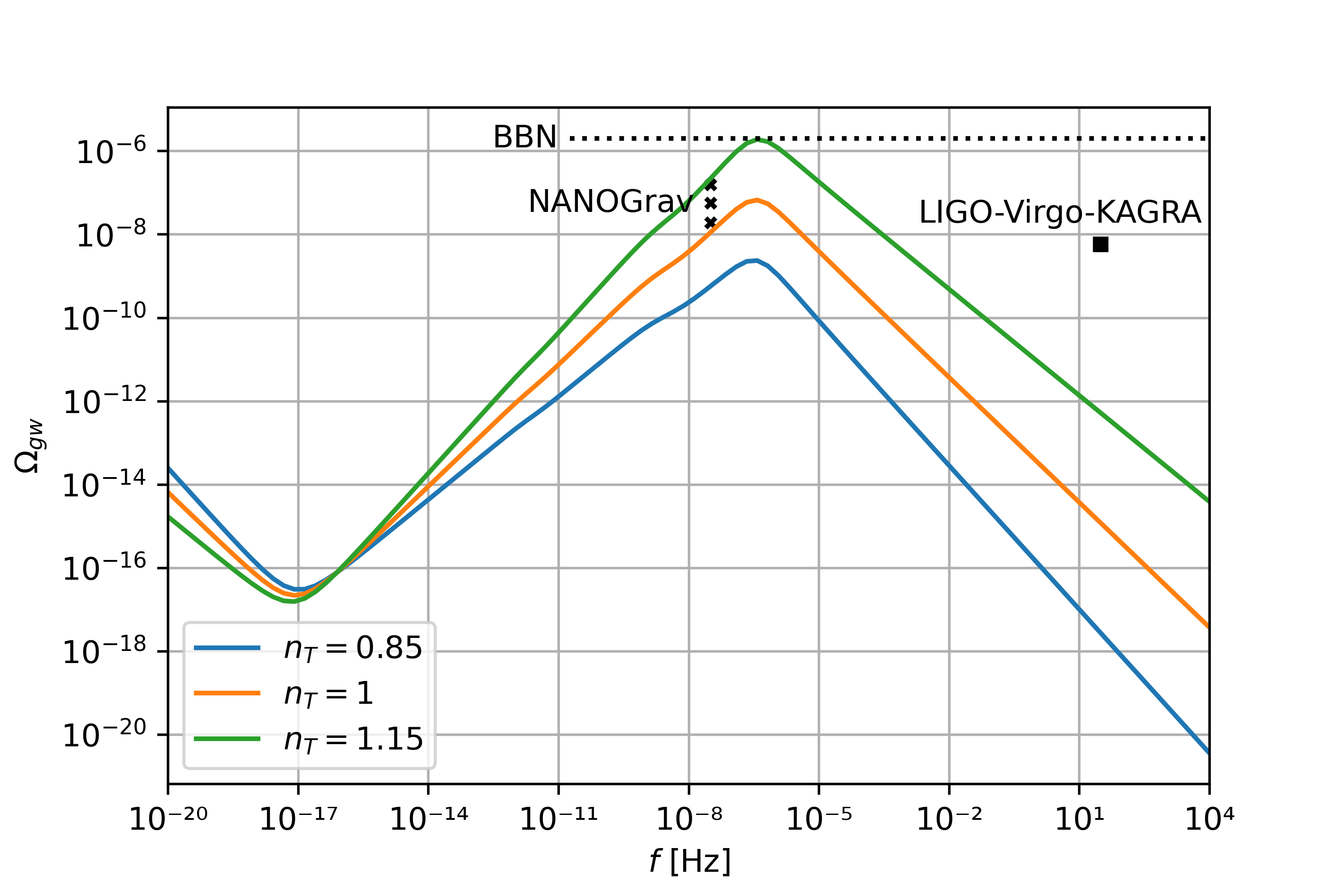}
    \caption{Recasting of Fig. \ref{omegaT} fixing $T_R=10$ GeV and varying $n_T$ for different colored curves as indicated in the figure.}
    \label{omegan}
\end{figure}

\begin{figure}
    \centering
    \includegraphics[width=0.75\linewidth]{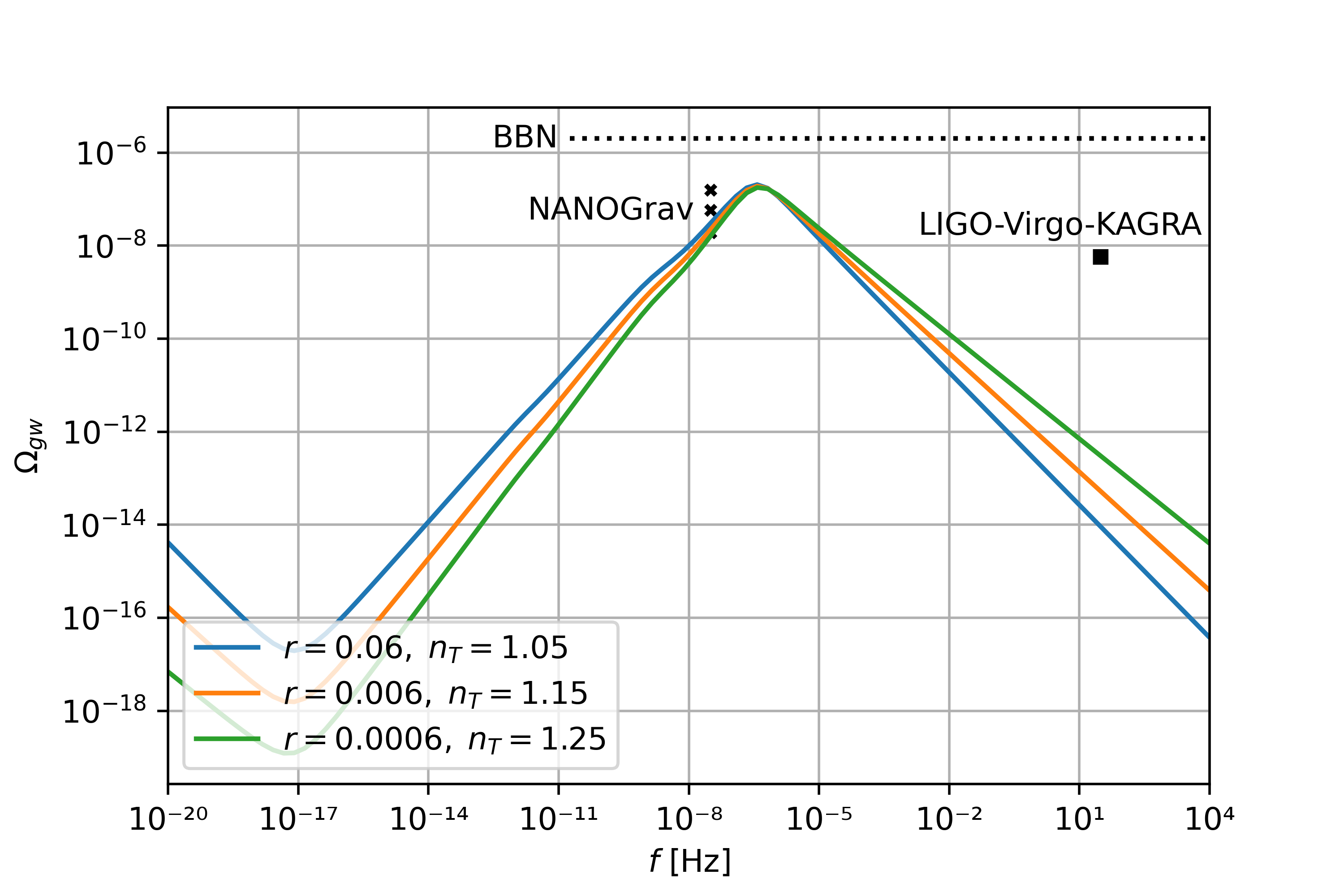}
    \caption{Recasting of Fig. \ref{omegan} letting $r$ vary alongside $n_T$ for different colored curves as indicated in the figure.}
    \label{omegar}
\end{figure}

According to Eq. \ref{slowrollnt}, slow-roll inflation produces a slightly red-tilted—that is, stronger at lower frequencies—gravitational-wave spectrum. However, as pointed out by Ref. \cite{Kuroyanagi:2020sfw}, if one modifies inflation (or early-universe physics more broadly) so as to make the spectrum blue-tilted, $\Omega_{gw}(f)$ may be consistent with the NANOGrav signal without violating the BBN and LIGO-Virgo-KAGRA constraints if the reheating temperature is low enough. This is the result we have been building towards: this dissertation has been an attempt at a self-contained explanation of Figs. \ref{omegaT}–\ref{omegar}, from which a physicist familiar with introductory relativity can learn in detail what that figure is showing. To use Eqs. \ref{ultom}, \ref{powerlawdelta}, and \ref{transferfunctionfinal} to make a plot of $\Omega_{gw}(f)$ (keeping in mind that $k=2\pi/\lambda=2\pi f/c$), we need some more observational input. From Planck's 2018 data on the CMB, we can get $\Omega_{m,0}=0.315$, $h=0.674$, $\Delta_h^2(k=0.05\mathrm{\;Mpc}^{-1})<1.26\times10^{-10}$, $\Omega_{r,0}\approx0$, and $\Omega_{\Lambda,0}\approx1-\Omega_{m,0}$ (the two latter values are important to compute $\eta_0$ using Eq. \ref{H(a)}, needed for the factor in \ref{bessel}) \cite{Planck:2018vyg}. Moreover, the particles that remain relativistic in the universe are photons and neutrinos. One can use the current value of $N_{eff}$ and the current photon and neutrino temperatures (which differ) to calculate $g_{*0}=3.36$ and $g_{*s0}=3.91$ from Eqs. \ref{gstar} and \ref{gstars} \cite{Kuroyanagi:2020sfw}. Finally, one can use, for example, the fitting formulas proposed in Ref. \cite{Kuroyanagi:2014nba} to compute $g_*(T(k_{in}))$ and $g_{*s}(T(k_{in}))$, where $T(k_{in})$ was the photon temperature when the mode with wavenumber $k$ re-entered the horizon.

Moreover, spherical Bessel functions like the one in Eq. \ref{transferfunctionfinal} may be expanded as \cite{wolframSphericalBesselJ}:
\begin{equation}
    j_v(z)=\frac{\sqrt{\pi}}{2}\left(\frac{z}{2}\right)^v \sum_{k=0}^\infty\frac{(-1)^kz^{2k}}{4^k\Gamma(k+v+\frac{3}{2})k!}.
\end{equation}
For $|z|\rightarrow\infty$—which is the case for the argument of $j_1$ in Eq. \ref{transferfunctionfinal} since observable gravitational waves are deep within the current cosmological horizon—, to $O(z^{-1})$, we have \cite{wolframSphericalBesselJ}:
\begin{equation}
    j_v(z)\propto\frac{1}{z}\sin\left(z-\frac{\pi v}{2}\right)\stackrel{v=1}{=}-\frac{\cos z}{z}.
\end{equation}
The oscillatory behavior of $\cos z/z$ can be smoothed out by replacing the trigonometric function in the numerator with its root-mean-square average, $1/\sqrt{2}$, i.e. $\cos z/z\approx1/\sqrt{2}z$.

A caveat for the BBN constraint is that the frequency of the mode crossing the horizon at the end of inflation, $f_{end}$ in Eq. \ref{BBNconstraint}, depends on the unmeasured parameters $T_R$ and $H_{end}$. Even so, with the approximation in Eq. 2.21 of Ref. \cite{Kuroyanagi:2020sfw} in mind, we assume $f_{end}>10^4$ Hz, i.e. above the LIGO-Virgo-KAGRA band, for the purpose of making Figs. \ref{omegaT}–\ref{omegar}. 

To produce our first plot, we may choose the maximum possible value for the gravitational-wave amplitude set by Planck's observations, $\Delta_h^2(k=0.05\mathrm{\;Mpc}^{-1})=1.26\times10^{-10}$. We can finally get Fig. \ref{omegaT} by choosing a fiducial blue-tilted spectral index, $n_T=1.05$, and different low—more on this soon—values of the reheating temperature, as indicated in the figure, to make $\Omega_{gw}(f)$ match the NANOGrav signal without violating the BBN and LIGO-Virgo-KAGRA constraints. To illustrate the curves' dependence on the spectral index, we fix $T_R=10$ GeV and instead vary $n_T$, with other input parameters held the same, in Fig. \ref{omegan}


To wrap up with a discussion of the gravitational-wave amplitude, we shall introduce a way to parametrize it that is commonly used, the tensor-to-scalar ratio $r$. If we were to carry out the calculations from Section \ref{gwdynamicsfrw} for scalar instead of tensorial metric perturbations, we would have found, for an exact de Sitter spacetime, in analogy with Eq. \ref{idealPT}, 
\begin{equation}
\label{idealPS}
    P_S=\left(\frac{H^2}{2\pi\dot \phi}\right)^2,
\end{equation}
i.e. a scale-invariant dimensionless power spectrum $P_S$ for scalar modes \cite{Guzzetti:2016mkm}. And, in analogy with Eq. \ref{powerlawdelta}, we would parametrize slow-roll inflation's deviation away from de Sitter with a power law:
\begin{equation}
\label{powerlawscalar}
    P_S\approx A_S\left(\frac{k}{k_*}\right)^{n_S-1},
\end{equation}
where the spectral index is $n_S-1$ purely by convention. Then, using the Friedmann equation (\ref{friedmann}), the inflaton's dynamical equation in slow roll (\ref{inflODE} for $\ddot \phi\rightarrow 0$), and the definition of the slow-roll parameter $\epsilon$ (\ref{epsilon}), we may rewrite Eq. \ref{powerlawscalar} in light of Eq. \ref{idealPS} as
\begin{equation}
\label{lmbade}
    P_S=\frac{4\pi G}{\epsilon}\left(\frac{H_*}{2\pi}\right)^2\left(\frac{k}{aH_*}\right)^{n_S-1}.
\end{equation}
Finally, we may form the ratio between the tensorial and scalar amplitudes of the dimensionless power spectrum, the tensor-to-scalar ratio $r$, usings Eqs. \ref{idealPT} and \ref{lmbade}:
\begin{equation}
\label{r16}
    r\equiv\frac{\Delta_h^2(k_*)}{P_S(k_*)}=16\epsilon.
\end{equation}
This is another important and well-known result regarding slow-roll inflation: tensorial perturbations are suppressed compared to scalar ones, and the suppression factor is proportional to the slow-roll parameter $\epsilon$. And more information of interest may be extracted from this discussion: performing algebraic manipulation, we may use Eqs. \ref{friedmann}, \ref{rhoinfl} with $\dot \phi\rightarrow0$, \ref{lmbade}, and \ref{r16} and the value of $A_S$ measured by Planck, $\ln(A_S10^{10})=3.043$ \cite{Planck:2018vyg}, to relate $r$ to the energy scale during inflation through the inflaton's potential $V$ in $c=\hbar=1$ units \cite{Guzzetti:2016mkm}:
\begin{equation}
\label{Vandr}
    V\approx(1.0\times10^{16}\mathrm{\;GeV})^4\frac{r}{0.01}.
\end{equation}
This is a good moment to put forward what it means for the reheating temperatures in Fig. \ref{omegaT} to be ``low''. As previously stated, Big Bang nucleosynthesis is arguably the earliest phase in the universe's history whose physics we know in detail. Consequently, roughly speaking, for our physical account of the history of the universe to be consistent, it is important for reheating to have happened at temperatures above $\sim10$ MeV, so that the thermal bath of relativistic standard model particles has already been produced from the inflaton's decay by the time nucleosynthesis has to have happened at temperatures around $\sim 1$ MeV. Above that lower limit, there exists a range of viable options for $T_R$ that spans many orders of magnitude. Unless one accepts or explains away a \textit{hierarchy problem}—whereby a dimensionless coupling constant is unnaturally small, where the notion of naturalness is tied to dimensionless constants in nature being $O(1)$ or so, similar in spirit to the fine-tuning problems discussed in Sections \ref{flatnessproblem} and \ref{thehorizonproblem}—, one naturally expects the Hubble parameter at the end of inflation $H_{end}\sim V_{end}^{1/4}$ not to differ by many orders of magnitude from the temperature at which reheating concludes. This is because, in many reasonable scenarios, reheating can be treated as approximately instantaneous, such that the energy density in the inflaton field will be completely furnished to the relativistic bath with no time for the expansion of the universe to dilute and cool it. Even if reheating does not instantaneously give rise to a homogeneous thermal bath of relativistic particles, e.g. if the bath takes some non-negligible time to thermalize \cite{McDonough:2020tqq}, this is usually not enough to cool $T_R$ by several orders of magnitude. As a result, if one wishes to explain NANOGrav's gravitational-wave observation with a blue-tilted primordial spectrum, one has to be careful about the balance between $T_R$ and $H_{end}$: although low reheating temperatures will enable $\Omega_{gw}(f)$ to evade existing observational constraints as proposed by Ref. \cite{Kuroyanagi:2014nba} and shown in Fig. \ref{omegaT}, realistic low-reheating-temperature models of inflation, for instance the one in Ref. \cite{Kobayashi:2019eyg}, may often be accompanied by low values of $H_{end}$—low in comparison to several simple models in which inflation occurs at an energy scale $\sim10^{15}-10^{16}$ GeV—, damping $r$ in light of Eq. \ref{Vandr} and rendering the gravitational waves too faint for observation. To conclude this discussion, in Fig. \ref{omegar}, we illustrate how different fiducial combinations of $n_T$ and $r$ may be equilibrated to have the primordial blue-tilted $\Omega_{gw}(f)$ be consistent with the observational data we have been considering. It is crucial to allow for the possibility that $r$ is lower than Figs. \ref{omegaT} and \ref{omegan}'s $r=0.06$ given that all we have empirically is an upper bound on its value.

\section{A model of the early universe with a blue-tilted tensor spectrum}
To meet our goal of understanding the content of Figs. \ref{omegaT}–\ref{omegar}, we were agnostic about what could have given rise to a blue-tilted spectrum of primordial gravitational waves, even though we explained in Section \ref{gwdynamicsfrw} why, in slow-roll inflation—a model that successfully fits the data on scalar perturbations—, one should, to the contrary, expect a slight red tilt. To conclude this dissertation with a forward-looking note, we provide a teaser of one model, among many existing ones, that does give rise to appreciable and positive values of $n_T$: the one in Ref. \cite{Tahara:2020fmn}, in which the essential ingredient is null-energy-condition (NEC) violation. The development of early-universe models with a low reheating temperature and blue-tilted gravitational-wave spectrum remains an active and exciting topic for theoretical physics research.

Energy conditions in general relativity are a posteriori constraints on the possible matter contents of spacetime. In other words, while they are not necessary, they are in line with most, if not all, types of matter we have empirically encountered, such that one might posit that they hold in general as an assumption to derive some interesting result. For example, the NEC, which we are about to state, is an assumption of Penrose's famous singularity theorem \cite{Penrose:1964wq}, according to which, under general relativity, a spacetime that contains a trapped surface, like a black hole event horizon, will inevitably contain a singularity at some point in time. Also assuming the NEC, Hawking and Penrose extended the latter's initial results, focused on black holes, to Big Bang singularities in cosmology \cite{Hawking:1970zqf}. The NEC may be stated as follows: consider spacetime is filled by matter with energy-momentum-tensor components $T_{\mu\nu}$. Then, for any null vector with components $n^\mu$, i.e. for any four-vector for which $n^\mu n_\mu=0$,
\begin{equation}
    T_{\mu\nu}n^\mu n^\nu>0.
\end{equation}

To see what the NEC means for an FLRW spacetime, let us write its line element as $ds^2=-dt^2+a^2\gamma_{ij}dx^idx^j$. Consider the vector $(1,a^{-1}\vec v)$, null by construction if $\gamma_{ij}v^iv^j=1$. Then, in the rest frame of the perfect fluid filling the universe, for which $u^0=1$ and $u^i=0$, recalling Eq. \ref{perfectfluid}, we have
\begin{equation*}
    T_{\mu\nu}n^\mu n^\nu=P+\rho-P+Pa^2\gamma_{ij}a^{-2}v^iv^j=\rho+P>0.
\end{equation*}
The NEC in FLRW is thus stating $P>-\rho$, that is, the most extreme, the most preposterous allowed negative-pressure equation of state for the content of the universe is that of dark energy. In light of Eq. \ref{conservationcosmo}, this is equivalently saying that, in an expanding universe, $\dot\rho<0$ if the content conforms with the NEC, i.e., matter that violates the NEC is not diluted but actually concentrated by the expansion of the universe. Preposterous indeed! We can draw another parallel conclusion if we take a time derivative of Friedmann's equation (\ref{friedmann}) with $k=0$ and use Eq. \ref{conservationcosmo} to write $\dot H=-4\pi G(\rho+P)$: NEC-violating matter drives cosmic dynamics with $\dot H>0$.

The model in Ref. \cite{Tahara:2020fmn} involves a Horndeski-type Lagrangian, of the form
\begin{equation}
    \mathcal L=\frac{R}{16\pi G}+G_2(\sigma,X)-G_3(\sigma,X)\square \sigma,
\end{equation}
where $\sigma$ is a scalar field and $X\equiv-g^{\mu\nu}\partial_\mu\sigma\partial_\nu\sigma/2$. Horndeski theories are notable because they have second derivatives in their Lagrangians and, yet, their equations of motion are second order—Lagrangians with second derivatives will often yield third-order dynamical equations. Because of that special feature of theirs, they can support NEC-violating solutions without pathologies \cite{Rubakov:2014jja}.

Working out the equations of motion for the metric and matter fields from the specific Horndeski Lagrangian proposed in Ref. \cite{Tahara:2020fmn} would go beyond our present scope—those equations are available in their Appendix A and the Lagrangian, in Section 3—, so we proceed with a qualitative description of what sort of early-universe history it gives rise to. In their model, usual slow-roll inflation and reheating—important, respectively, to match observational constraints on scalar and tensorial perturbations and to lead to the radiation-dominated epoch—are preceded by a NEC-violating phase. While, during slow-roll, the inflaton's field values ``slowly roll'' toward lower-potential-energy regions, the NEC-violating phase involves the inflaton field spookily ``climbing up'' toward higher potential energies, a phenomenon frequently understood as a manifestation of a negative kinetic energy. To get an intuition for why these two behaviors are related, we can consider a negative-kinetic-energy harmonic oscillator, described by the Lagrangian
\begin{equation}
    \mathcal L=-\frac{1}{2}(\dot q^2+\omega^2q^2),
\end{equation}
for which the Euler-Lagrange equation reads
\begin{equation}
    \ddot q=\omega^2q.
\end{equation}
The flipped sign turns the solution of the second-order differential equation for the usual harmonic oscillator, composed of oscillatory modes, into a combination of exponentially decaying and growing modes. This explosive behavior hints at how the physics of negative-kinetic-energy systems compares to their regular counterparts. 

The crucial result from the particle theory in Ref. \cite{Tahara:2020fmn} is
\begin{equation}
\label{NEChubble}
    H=-\frac{p}{t}
\end{equation}
during the NEC-violating phase proposed for the very early universe, where $p$ is a positive constant. Note that $\dot H=p/t^2>0$, indeed violating the NEC. If one solves Eq. \ref{deltag} using Eq. \ref{NEChubble}—which we solved in Section \ref{gwdynamicsfrw} for the simpler case of slow-roll inflation—, what one finds out is that \cite{Tahara:2020fmn}
\begin{equation}
    \Delta_h^2(k)\propto k^{3-(3p+1)/(p+1)}.
\end{equation}
Hence, $0<n_T<2$ for $p>0$, showing this NEC-violating model does succeed in producing a blue-tilted gravitational-wave spectrum. 
\chapter{\textsc{Conclusion}}

This dissertation provides the basic theoretical framework underlying the observation of gravitational waves from cosmic inflation. 

We have started with a mathematically rich discussion of what gravitational waves are and what is involved in their detection, being particularly careful about the several ways in which one can report the intensity of an observed signal. 

We then switched gears and discussed cosmology from scratch, overviewing what it means for the universe to be homogeneous and isotropic and what the implications of these special properties are for the general-relativistic description of the universe. We were then equipped to explain the essential features of the Hot Big Bang model and follow up with a survey of problems plaguing it that motivated the proposal of cosmic inflation in the 1980s as an adjustment to the Big Bang model. 

Putting things together, we studied perturbations in the inflationary universe and focused on the tensorial modes of perturbation, which are the gravitational waves, working out the spectrum produced during inflation to be observed in the present universe. We discussed existing observational constraints that a primordial spectrum of gravitational waves should respect, as well as the gravitational-wave signal for which NANOGrav has been building statistical significance. We ended by showing how a certain set of features (a low reheating temperature alongside a blue-tilted gravitational-wave spectrum) would allow the early universe to produce the signal observed by NANOGrav while not violating those constraints and, then, by providing a brief overview of a model that actualizes one of the central required features (a blue-tilted spectrum).

The origin of the universe remains as astonishing as hard to explain. But, hopefully, this dissertation will leave one with the sense that our knowledge about the evolution of the cosmos is already beautifully and encouragingly rich, and that the continued development of gravitational-wave astronomy might be the key to unprecedentedly profound discoveries.
\addto{\captionsenglish}{\renewcommand{\bibname}{\centering REFERENCES}}
\selectlanguage{english}
\bibliographystyle{unsrt}
\bibliography{biblio}
\end{document}